\documentclass[journal]{IEEEtran}

\usepackage{graphicx}
\usepackage[noend]{algpseudocode}
\usepackage{placeins}
\usepackage{float}
\usepackage{tabularx,colortbl}
\usepackage{amsthm}
\usepackage{amsmath}
\usepackage{mathtools}
\usepackage{amssymb}
\usepackage{cleveref}
\usepackage{algorithm}
\usepackage{verbatim}
\usepackage{cite}
\usepackage{url}
\usepackage{cases}
\usepackage{amsmath}
\usepackage{amsfonts}
\usepackage{amssymb}
\usepackage{verbatim}
\usepackage{multirow}
\usepackage{mathtools}
\usepackage{multicol}
\usepackage{xcolor}
\usepackage{MnSymbol}
\usepackage{booktabs}
\usepackage[all,cmtip]{xy}

\usepackage{newfloat}
\DeclareFloatingEnvironment[
    fileext=los,
    listname={List of Boxes},
    name=Box,
    placement=tbhp,
    within=section,
]{cr_box}

\definecolor{BoxBackground}{RGB}{235,235,255}

\IEEEoverridecommandlockouts

\begin{document}
\title{Analog to Digital Cognitive Radio:\\ Sampling, Detection and Hardware}
\author{
\thanks{
This work was funded by the European Union's Horizon 2020 research and innovation program under grant no. 646804-ERC-COG-BNYQ, and by the Israel Science Foundation under grant no. 335/14.
}
Deborah Cohen, Shahar Tsiper, Yonina C. Eldar \\
Technion -- Israel Institute of Technology, Haifa, Israel \\
\{debby@tx, tsiper@tx, yonina@ee\}.technion.ac.il}

\maketitle

\global\long\def\dint#1#2#3#4{\int\limits _{#1}^{#2}#3\,\text{d}#4}
\global\long\def\v#1{\mathbf{#1}}
\global\long\def\sinc{\text{sinc}}
\global\long\def\hz#1{\,\text{#1Hz}}

\section{Introduction}

The radio spectrum is the radio frequency (RF) portion of the electromagnetic spectrum. These spectral resources are traditionally allocated to licensed or primary users (PUs) by governmental organizations. As can be observed in ``RF Spectral Resources", most of the frequency bands are already allocated to one or more PUs. Consequently, new users can hardly find free frequency bands. Following the ever-increasing demand from new wireless communication applications, this issue has become critical over the past few years.

\begin{figure}[ht]
  \begin{center}
    \includegraphics[width=1\columnwidth]{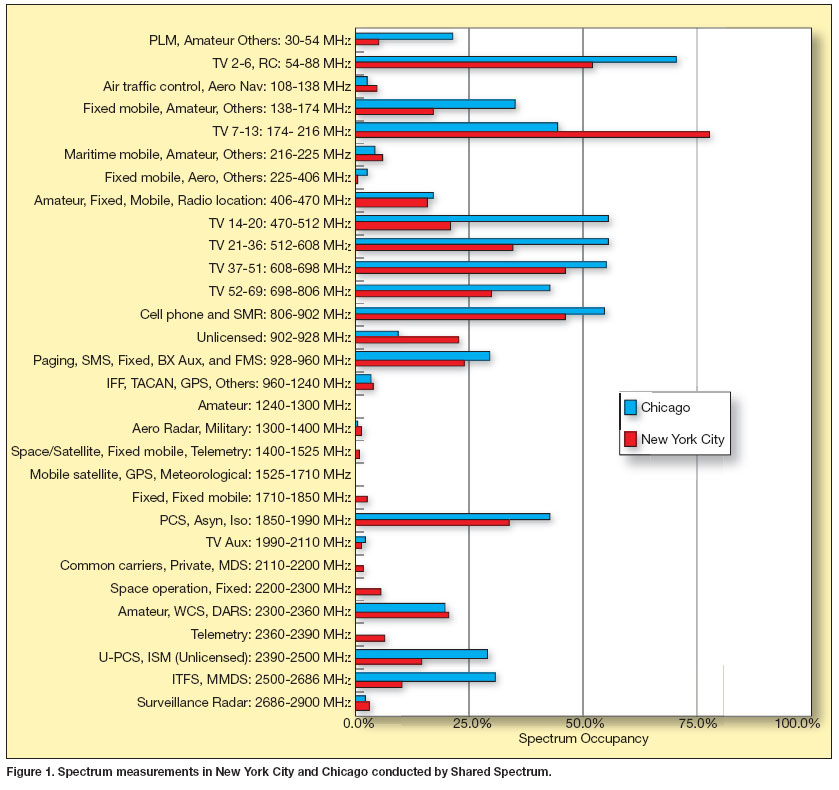}
    \caption{Spectrum occupancy measurements in New York City, New York from Aug. 30th.-Sept. 3rd 2004 and Chicago, Illinois from Nov. 16th-18th 2005 (figure courtesy of SSC~\cite{SpecMeasChicNY}).}
    \label{fig:ny_chi}
  \end{center}
\end{figure}

Various studies~\cite{Study1, Study2, study3} have shown that this over-crowded spectrum is usually significantly underutilized in frequency, time and space.
A 2002 FCC report shows large temporal and geographic variations in terms of the spectrum usage, from $15\%$ to $85\%$ utilization. Figure~\ref{fig:ny_chi} presents the spectrum occupancy measurements conducted by the Shared Spectrum Company (SSC) in two major metropolitans, New York and Chicago. We observe that the occupancy of most frequency bands does not exceed $50\%$, and half of them are even below $10\%$~\cite{thesis_liu}. This shows that spectrum shortage is often due to the inflexibility of the frequency allocation, resulting in low utilization efficiency.

In order to bridge between spectrum scarcity and sparsity and make better use of spectral resources, the cognitive radio (CR)~\cite{Mitola, Haykin, MitolaMag} paradigm has been proposed. The idea of adaptive learning for spectrum sensing can be traced back to Shannon~\cite{Shannon}. This technology, which is under development, allows the spectrum to be used more efficiently by granting secondary users opportunistic access to licensed spectral bands when the corresponding PU is not active. A CR transceiver scans for unused bands and changes its transmission and reception parameters to different frequencies depending on the spectrum utilization.

CR faces many issues at various levels, and challenges traditional analog, digital and network processing techniques to meet its specific radio sensitivity requirements and wideband frequency agility~\cite{Cabric_CS}. The CR cycle includes two major functionalities: spectrum sensing and spectrum access. Through spectrum sensing, CRs collect information about the status of the surrounding spectrum's occupancy, that is the PUs' activity, allowing for adapted exploitation of the vacant spectral bands.
In order to minimize the interference caused to PUs, the spectrum sensing task performed by a CR should be reliable and fast~\cite{cognitive1, cognitive2, WidebandMishali} and obey the requirements of the IEEE 802.22 protocol, described in ``IEEE 802.22 Standard for WRAN''.

For the past ten years, CR and its challenges have been thoroughly reviewed in the literature. Several works~\cite{cog, crReview, wangReview, maReview, zengReview, yucekReview} focus on spectrum sensing and survey sensing techniques, along with their performance and limitations. These techniques are essentially energy detection, matched filter and cyclostationary detection. Collaborative CR networks, where different users share their sensing results and cooperatively decide on the licensed spectrum occupancy, have also been proposed to overcome practical propagation issues such as path loss, channel fading and shadowing. Other works~\cite{maReview, hossainReview} deal with spectrum access, which uses the information gathered from spectrum sensing to plan the spectrum exploitation by unlicensed users. This functionality includes spectrum analysis, spectrum access, and spectrum mobility~\cite{CR_Andrea}. It ensures coexistence with PUs and other CRs, by minimizing interference to the former and sharing spectrum with the latter.

In this article, we focus on the issue of spectrum sensing from the analog to digital interface point of view, which has received little attention so far. Our motivation stems from one of the main challenges of spectrum sensing in the context of CR, which is the sampling rate bottleneck. This issue arises since CRs typically deal with wideband signals with prohibitively high Nyquist rates. Sampling at this rate requires very sophisticated and expensive analog to digital converters (ADCs), leading to a torrent of samples. Therefore, the classic spectrum sensing methods described above, which are traditionally performed on Nyquist rate samples, are difficult to implement in practice on wideband signals. Our main goal is to provide an analog to digital CR framework including an analog preprocessing and sub-Nyquist sampling front-end, and subsequent low rate digital processing.

We first survey recent methods for spectrum sensing at sub-Nyquist rates, paving the way to efficient sensing with low computational and power requirements. We then review spectrum sensing strategies that aim at overcoming other diverse challenges in the context of CR, such as coping with low signal to noise ratio (SNR) regimes, channel fading and shadowing effects. Throughout the paper, we consider both theoretical and practical aspects and present hardware implementation of theoretical concepts, demonstrating real-time wideband spectrum sensing for CR from low rate samples.

\begin{figure*}[t]\fboxsep1em
\colorbox{BoxBackground}{\begin{minipage}{1\textwidth}\begin{multicols*}{2}
\section*{RF Spectral Resources}

Spectral resources are managed by governmental organizations that allocate them to fixed users. In the United States, regulatory responsibility for the radio spectrum is divided between the Federal Communications Commission (FCC), which administers spectrum for non-Federal use, and the National Telecommunications and Information Administration (NTIA), responsible for the Federal use.


In Europe, the European Telecommunications Standards Institute (ETSI) and the Electronic Communications Committee (ECC) cooperate on aspects related to the regulatory environment for radio spectrum. Figure~\ref{fig:us_map} shows the US frequency allocation chart. As can be seen, all of the bands spanning the VHF, UHF and SHF frequency ranges (0.003--30GHz) are pre-allocated to one or even several licensed PUs. This poses inherent difficulties in introducing new technologies that require occasional usage of this spectral range.

\end{multicols*}
 \begin{center}
    \includegraphics[width=1\columnwidth]{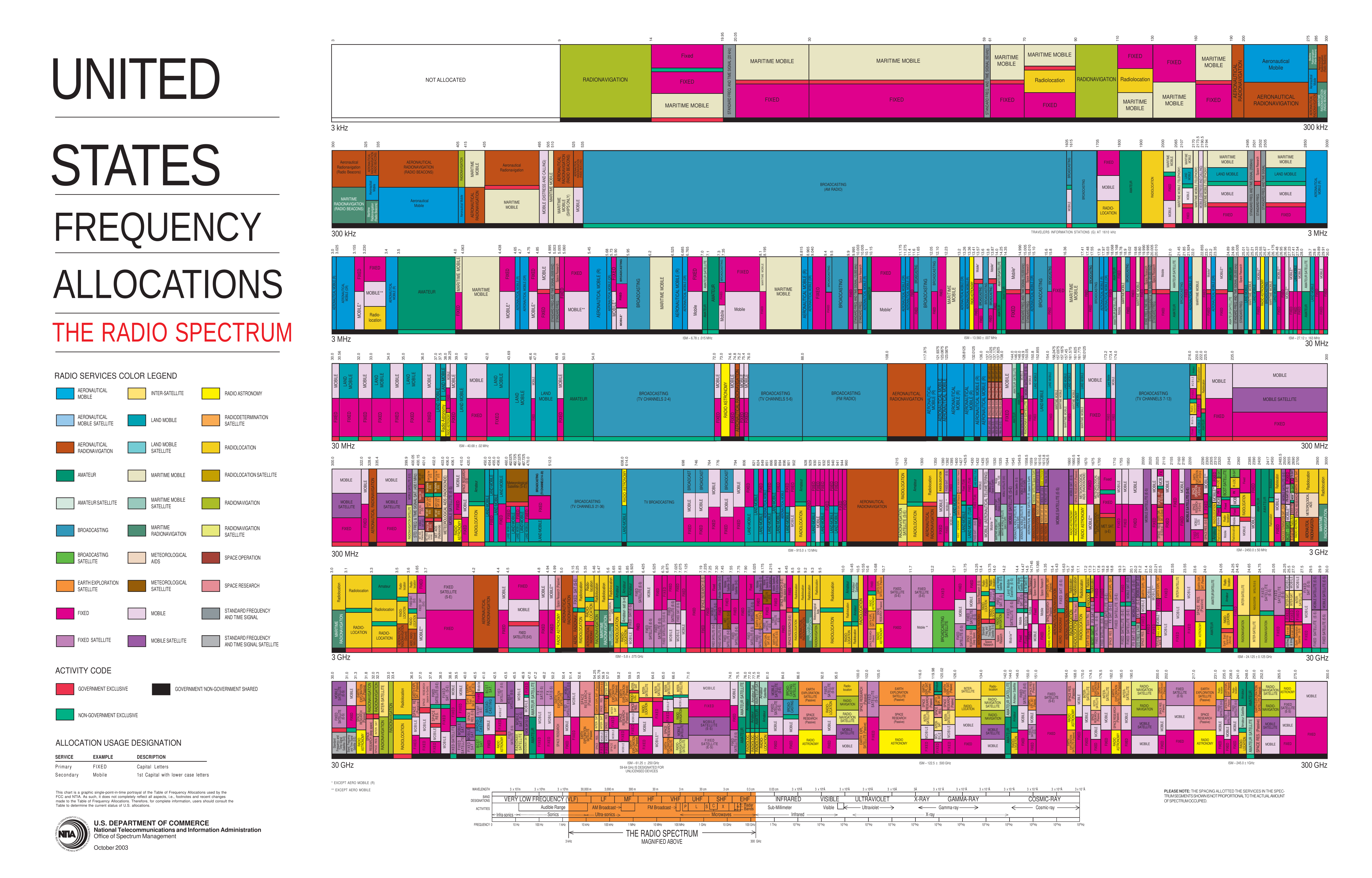}
    \caption{United States frequency allocation chart (courtesy of the U.S. Department of Commerce, National Telecommunications and Information Administration, Office of Spectrum Management - January 2016).}
    \label{fig:us_map}
 \end{center}


\end{minipage}}\end{figure*}

\section{CR challenges}

In this article, we focus on CR spectrum sensing. In practice, the information gathered from spectrum sensing is used to plan spectrum access by the unlicensed users. For completeness, we first quickly review the main components and challenges of spectrum access.

Spectrum analysis or management, which directly follows spectrum sensing, ensures coexistence with PUs and other CRs. The ambient RF environment is analyzed in order to characterize the behavior of PUs and the properties of the detected spectrum holes in terms of interference, duration of availability and more. Then, spectrum access can be optimized to maximize the CR throughput while maintaining interference caused to the licensed users below a target threshold~\cite{hossainReview}. Several techniques have been proposed to minimize interference to PUs as well as ensure proper reception of secondary signals, such as waveform design and multi-carrier approaches~\cite{maReview, hossainReview}. These are regrouped under the term of spectrum sculpting~\cite{maReview}. Besides minimizing interference to the PUs, spectrum sharing needs to be coordinated within the CR network. Various power control and resource allocation schemes that deal with this issue are reviewed in~\cite{wangReview, hossainReview}. Spectrum access further requires synchronization between the CR transmitter and receiver~\cite{hossainReview}.

The function of spectrum mobility ensures adaptation to changes in the spectrum occupancy. When a licensed user starts accessing the channel currently being used by a CR, the latter has to vacate the band and switch to another free spectral band. This operation is referred to as hand-off~\cite{wangReview, hossainReview}. The multi-carrier transmission approach evoked above allows to maintain uninterrupted communication in such scenarios~\cite{maReview}. Additional issues of security against malicious users and various attacks to the network are discussed in~\cite{wangReview, yucekReview, hossainReview}.

We now focus on spectrum sensing, which is the fundamental enabler to spectrum access. In order to increase the chance of finding an unoccupied spectral band, CRs have to sense a wide band of spectrum. Nyquist rates of wideband signals are high and can even exceed today's best ADCs front-end bandwidths. Besides, such high sampling rates generate a large number of samples to process, affecting speed and power consumption. To overcome the rate bottleneck, several sampling methods have been proposed that leverage the a priori known received signal's structure, enabling sampling reduction. These include the random demodulator~\cite{RandomDemodulator1, RandomDemodulator2}, multi-rate sampling~\cite{MultiRate}, multicoset sampling and the modulated wideband converter (MWC)~\cite{WidebandMishali, Mishali_multicoset, Mishali_theory, Xampling}.

The CR then performs spectrum sensing on the acquired samples to detect the present of PUs' transmissions. The simplest and most common spectrum sensing approach is energy detection~\cite{Urkowitz_energy}, which does not require any a priori knowledge on the input signal. Unfortunately, energy detection is very sensitive to noise and performs poorly in low SNRs. This becomes even more critical in sub-Nyquist regimes since the sensitivity of energy detection is amplified due to aliasing of the noise~\cite{Castro}. Therefore, this scheme fails to meet CR performance requirements in low SNRs. In contrast, matched filter (MF) detection~\cite{MF1, MF2}, which correlates a known waveform with the input signal to detect the presence of a transmission, is the optimal linear filter for maximizing SNR in the presence of additive stochastic noise. However, this technique requires perfect knowledge of the potential received transmission. When no a priori knowledge can be assumed on the received signals' waveform, MF is difficult to implement. A compromise between both methods is cyclostationary detection~\cite{Gardner_review, Napo_review}. This strategy is more robust to noise than energy detection but at the same time only assumes that the signal of interest exhibits cyclostationarity, which is a typical characteristic of communication signals. Consequently, cyclostationary detection is a natural candidate for spectrum sensing from sub-Nyquist samples in low SNRs.

Besides noise, the task of spectrum sensing for CRs is further complicated as a result of path loss, fading and shadowing~\cite{collaborative_balak, collaborative_brodersen, collaborative_ghasemi}. These phenomena are due to the signal's propagation, that can be affected by obstacles and multipath, and result in the attenuation of the signal's power. To overcome these practical issues, collaborative CR networks have been considered, where different users share their sensing results and cooperatively decide on the licensed spectrum occupancy. Cooperative spectrum sensing may be classified into three catagories based on the way the data is shared by the CRs in the network: centralized, distributed and relay-assisted. In each of these settings, two options of data fusion arise: decision fusion, or hard decision, where the CRs only report their binary local decisions, and measurement fusion, or soft decision, where they share their samples~\cite{collaborative_balak}. Cooperation has been shown to improve detection performance and relax sensitivity requirements by exploiting spatial diversity~\cite{collaborative_ghasemi, collaborative_poor}.
At the medium access control (MAC) level, cooperation introduces the need for a tailored communication protocol and a control channel~\cite{cog, crReview}, which can be implemented as a dedicated frequency channel or as an underlay ultra-wideband (UWB) channel~\cite{Cabric_CS}. These CR communication challenges are outside the scope of this paper.

Finally, CRs may require, or at least benefit from joint spectrum sensing and direction of arrival (DOA) estimation. DOA recovery enhances CR performance by allowing exploitation of vacant bands in space in addition to the frequency domain. For example, a spectral band occupied by a PU situated in a certain direction with respect to the CR, may be used by the latter for transmission to the opposite direction, where receivers do not sense the PU's signal. In order to estimate jointly the carrier frequencies and DOAs of the received transmissions, arrays of sensors have been considered and DOA recovery techniques such as MUSIC~\cite{pisarenko73, schmidt86}, ESPRIT~\cite{PAUL1986} or compressed sensing (CS)~\cite{CSBook} based approaches, may then be adapted to the joint carrier and DOA estimation problem. We review several algorithms that jointly estimate the carrier frequencies and DOAs of the received transmissions, from the same low rate samples obtained from an array of sensors.

This article focuses on the spectrum sensing challenges for CR outlined above. We first review sub-Nyquist sampling methods for multiband signals and then consider different aspects of spectrum sensing performed on low rate samples, including cyclostationary detection, collaborative spectrum sensing and joint carrier frequency and DOA estimation. Our emphasis is on practical low rate acquisition schemes and tailored recovery that can be implemented in real CR settings. In particular, we examine the analog to digital interface of CRs, as opposed to the review in~\cite{bjornReview}, which is concerned with the digital implications of compressive spectrum sensing. This allows to demonstrate the realization of the theoretical concepts on hardware prototypes. We focus in particular on the implementation of one sampling scheme reviewed here, the MWC, and show how the same low rate samples can be used in the different extensions of spectrum sensing described above.

We note that CRs are faced with additional fundamental difficulties, such as the hidden-node problem and potential interference from secondary users to existing primary communication links. These are beyond the scope of this review but need to be addressed as well for CRs to become practical.

\begin{figure*}[ht!]\fboxsep1em
\colorbox{BoxBackground}{\begin{minipage}{1\textwidth}\begin{multicols*}{2}
\section*{IEEE 802.22 Standard for WRAN}
\label{box:protocol}

During the past decade, the use of analog TV radio bands has been slowly decaying, mainly due to the introduction of digital TV broadcasts based on new transmitting media such as cable and satellite.
The once crowded VHF and UHF frequency range (between $54$ and $860\hz M$) is less occupied today, particularly in remote rural areas.
Concurrently, a new need has emerged in these areas for fast wireless data networks with long range capabilities, up to 30 km.
For these operating range requirements, the VHF/UHF radio spectrum, able to traverse these distances with relatively low transmitting power, is a natural choice.
Combining the need for available frequencies in this range with the fact that it is mostly underutilized today, has laid the foundation for designing efficient spectrum sharing techniques, including CR.\@

IEEE 802.22 is the first international standard especially designed to achieve this goal.
It describes a wireless regional area network (WRAN) relying on new CR technologies, that exploits the vacant white space in the VHF/UHF range, while still reserved to licensed TV bands.
Managing the spectrum and transmitting broadband communications only in vacant frequency slots enables fast internet access in hard-to-reach areas with low population density, and therefore has high potential for extensive worldwide usage

To exploit analog TV bands' white spaces, an opportunistic approach is selected. Both base stations (BS) and clients, referred to as customer-premises equipment (CPE), are spread apart geographically, as seen in Fig.~\ref{fig:WRAN-Diagram}. In order to enable a true CR network, the standard defines necessary capabilities from both the BSs and CPEs to enable \emph{cognitive sensing} and adapt to possibly rapid changes incurred by incumbent users.

The physical (PHY) layer is the lowest network layer. It contains the circuitry used to transmit and receive analog communication signals. 
For implementing CR technologies, the PHY has to rapidly adapt to spectral changes, and be both agile and flexible enough to jump between carrier frequencies without losing information. It is required to constantly listen to the operating band during designated quiet times, in order to make sure the incumbent user does not wake up and require its use.
The media access control (MAC), situated above the PHY layer in the network hierarchy, is responsible for CR communication management. Its specifications include fast and dynamic adaptation to changes in the environment by constantly sensing the spectrum.
The requirements for each of the different network layers are particularly demanding when compared to other RF standards from the 802.xx family.




To conclude, the 802.22 WRAN standard holds great promise in delivering fast broadband connections to remote rural areas, with broadcast ranges of up to 30 km, by smartly sharing the already allocated VHF and UHF spectrum. Spectrum coexistence is made possible by CR tools, which require advanced technologies. These introduce tremendous challenges in terms of the necessary hardware to meet such harsh demands, while preserving small form factors and low power levels. Sub-Nyquist methods for spectral sensing and reconstruction offer a way to alleviate part of this burden by exploiting the signals' spectral structure, effectively making CR technology more accessible.

\begin{center}
	\includegraphics[width=1\columnwidth]{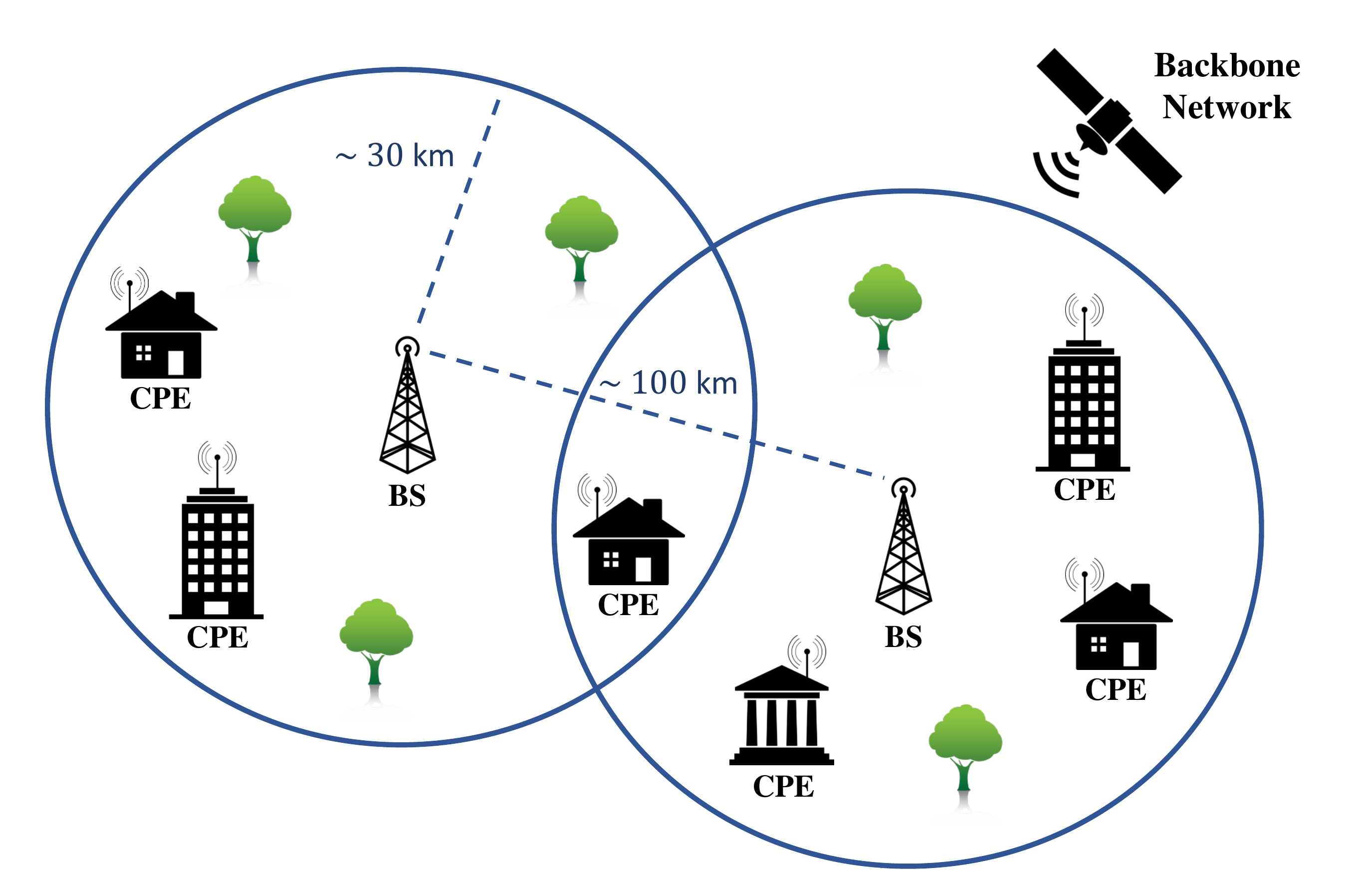}
	\caption{CR network illustration, including BSs and CPEs.}
	\label{fig:WRAN-Diagram}
\end{center}


\end{multicols*}\end{minipage}}\end{figure*}

\section{Sub-Nyquist Sampling for CR}

\begin{figure*}[ht!]\fboxsep1em
\colorbox{BoxBackground}{\begin{minipage}{1\textwidth}\begin{multicols*}{2}
\section*{Compressed Sensing Recovery}
\label{box:cs}

Compressed sensing (CS)~\cite{CSBook, SamplingBook} is a framework for simultaneous sensing and compression of finite-dimensional vectors, which relies on linear dimensionality reduction. In particular, the field of CS focuses on the following recovery problem
\begin{equation}
\bf z = Ax,
\end{equation}
where $\bf x$ is a $N\times 1$ sparse vector, namely with few non zero entries, and $\bf z$ is a vector of measurements of size $M<N$. CS provides recovery conditions and algorithms to reconstruct $\bf x$ from the low-dimensional vector $\bf z$.

The different spectrum sensing applications described in this paper deal primarily with analog signals and sampling techniques, while CS inherently defines a digital framework. We will discuss how the analog approaches of low rate sampling use CS as a tool for recovery and adapt it to the analog setting. To that end, we describe here two CS greedy recovery algorithms, that solve the optimization problem
\begin{equation}
\mathbf{\hat{x}}= \arg \min_\mathbf{x} ||\mathbf{x}||_0 \quad \text{s.t. } \mathbf{z = Ax},
\end{equation}
where $||\cdot ||_0$ denotes the $\ell_0$-norm.
The first algorithm we consider belongs to the family of matching pursuit (MP) methods~\cite{MallatMP}. The orthogonal matching pursuit (OMP) algorithm iteratively proceeds by finding the column of $\bf A$ most correlated to the signal residual $\bf r$,
\begin{equation}
\label{eq:omp1}
i = \arg \max |\mathbf{A}^H \mathbf{r}|,
\end{equation}
where the absolute value is computed element-wise.
The residual is obtained by subtracting the contribution of a partial estimate $\mathbf{\hat{x}}_{\ell}$ of the signal at the $\ell$th iteration, from $\bf z$, as follows
\begin{equation}
\label{eq:omp2}
\mathbf{r}=\mathbf{z}-\mathbf{A}\mathbf{\hat{x}}_{\ell}.
\end{equation}
Once the support set is updated by adding the index $i$, the coefficients of $\mathbf{\hat{x}}_{\ell}$ over the support set are updated, so as to minimize the residual error.

Other greedy techniques include thresholding algorithms. We focus here on the iterative hard thresholding (IHT) method proposed in~\cite{DaviesIHT}. Starting from an initial estimate $\mathbf{\hat{x}}_0 =0$, the algorithm iterates a gradient descent step with step size $\mu$ followed by hard thresholding, i.e.
\begin{equation}
\label{eq:iht}
\mathbf{\hat{x}}_{\ell} = \mathcal{T}(\mathbf{\hat{x}}_{\ell-1} + \mu \mathbf{A}^H (\mathbf{z}-\mathbf{A}\mathbf{\hat{x}}_{\ell-1}), k)
\end{equation}
until a convergence criterion is met. Here $\mathcal{T}(\mathbf{x}, k)$ denotes a thresholding operator on $\bf x$ that sets all but the $k$ entries of $\bf x$ with the largest magnitudes to zero, and $k$ is the sparsity level of $\bf x$, assumed to be known.

These two greedy algorithms and other CS recovery techniques can be adapted to further settings, such as multiple measurement vectors (MMV), where the measurements $\bf z$ and sparse objective $\bf x$ become matrices, infinite measurement vectors (IMV), block sparsity and more, as we will partially discuss in the paper. Further details on CS recovery conditions and techniques can be found in~\cite{CSBook, SamplingBook}.


\end{multicols*}\end{minipage}}\end{figure*}

\begin{figure}[H]
  \begin{center}
    \includegraphics[width=0.9\columnwidth]{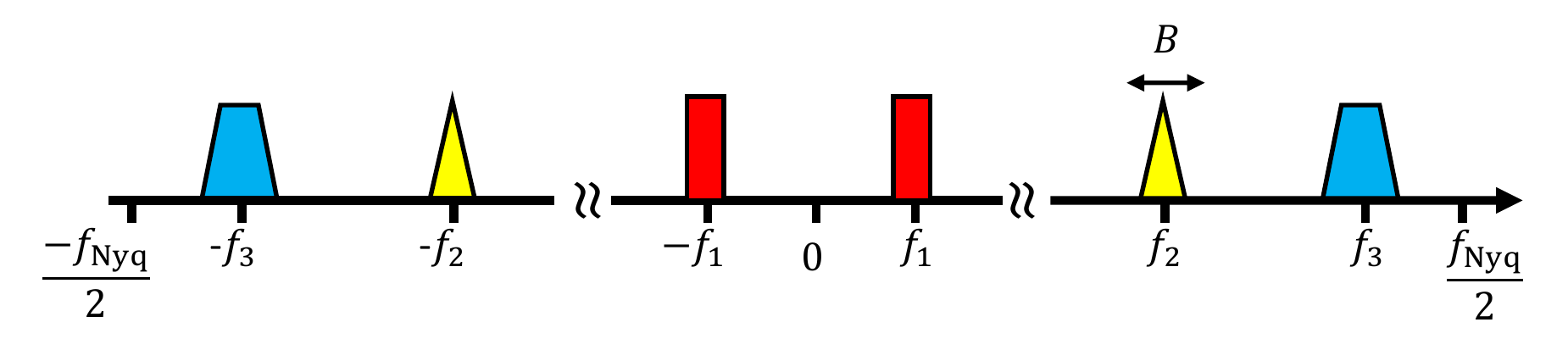}
    \caption{Multiband model with $K=6$ bands. Each band does not exceed the bandwidth $B$ and is modulated by an unknown carrier frequency $|f_i| \leq f_{\text{Nyq}}/2$, for $i=1,2,3$.}
    \label{fig:multiband}
  \end{center}
\end{figure}

CR receivers sense signals composed of several transmissions with unknown support, spread over a wide spectrum. Such sparse wideband signals belong to the so-called multiband model~\cite{Mishali_multicoset, Mishali_theory, SamplingBook}. An example of a multiband signal $x(t)$ with $K$ bands is illustrated in Fig.~\ref{fig:multiband}, with individual bandwidths not greater than $B$, centered around unknown carrier frequencies $|f_i| \leq f_{\text{Nyq}}/2$, where $f_{\text{Nyq}}$ denotes the signals' Nyquist rate and $i$ indexes the transmissions. Note that, for real-valued signals, $K$ is an even integer due to spectral conjugate symmetry and the number of transmissions is $N_{\text{sig}}=K/2$.

When the frequency support of $x(t)$ is known, classic sampling methods such as demodulation, undersampling ADCs and interleaved ADCs (see~\cite{SamplingBook, MagazineMishali} and references therein) reduce the sampling rate below Nyquist.
Here, since the frequency location of the transmissions are unknown, classic processing first samples $x(t)$ at its Nyquist rate $f_{\text{Nyq}}$, which may be prohibitively high. Sensing such wideband signals may be performed by splitting them into several frequency bands and processing each band separately. This can be done either in series or in parallel. However, this increases the latency in the first case and the hardware complexity in the second.

To overcome the sampling rate bottleneck, several blind sub-Nyquist sampling and recovery schemes have been proposed that exploit the signal's structure and in particular its sparsity in the frequency domain, but do not require knowledge of the carrier frequencies. It has been shown in~\cite{Mishali_multicoset} that the minimal sampling rate for perfect blind recovery in multiband settings is twice the Landau rate~\cite{LandauCS}, that is twice the occupied bandwidth.
This rate can be orders of magnitude lower than Nyquist. In the remainder of this section, we survey several sub-Nyquist methods, that theoretically achieve the minimal sampling rate. Two main limitations of sub-Nyquist sampling are noise enhancement~\cite{Castro}, which is discussed later on and may be partially overcome by techniques reviewed below such as cyclostationary detection and collaborative spectrum sensing, and the inherent assumption that the channel is sparse and contains spectral holes.

\subsection{Multitone Model and the Random Demodulator}
\label{sec:multitone}

\begin{figure}[H]
  \begin{center}
    \includegraphics[width=0.9\columnwidth]{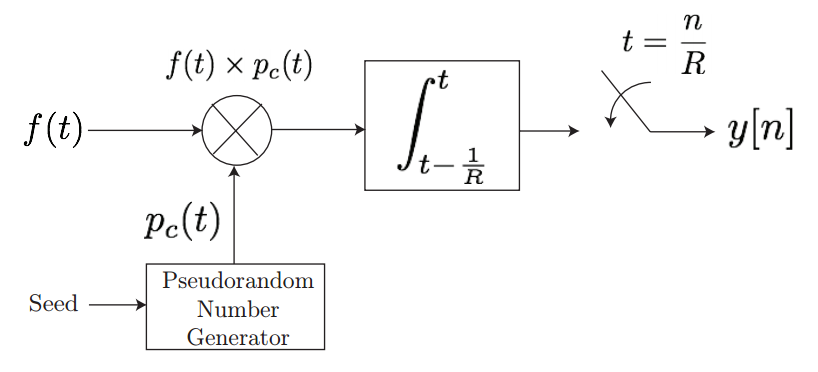}
    \caption{Block diagram for the random demodulator, including a random number generator, a mixer, an accumulator and a sampler~\cite{RandomDemodulator2}.}
    \label{fig:randomD}
  \end{center}
\end{figure}

Tropp et al.~\cite{RandomDemodulator1, RandomDemodulator2} consider a discrete multitone model and suggest sampling using the random demodulator, depicted in Fig.~\ref{fig:randomD}. Multitone functions are composed of $K$ active tones spread over a bandwidth $W$, such that
\begin{equation}
f(t) = \sum\limits_{\omega \in \Omega} b_{\omega} e^{-2 \pi i \omega t}, \quad t \in [0,1).
\end{equation}
Here, $\Omega$ is composed of $K$ normalized frequencies, or tones, that are a subset of the integers in the interval $[-W/2, W/2]$, and $b_{\omega}$, for $\omega \in \Omega$, are a set of complex-valued amplitudes. The number of active tones $K$ is assumed to be much smaller than the bandwidth $W$. The goal is to recover both the tones $\omega$ and the corresponding amplitudes $b_{\omega}$.

To sample the signal $f(t)$, it is first modulated by a high-rate sequence $p_c(t)$ created by a pseudo-random number generator. It is then integrated and sampled at a low rate, as shown in Fig.~\ref{fig:randomD}. The random sequence used for demodulation is a square wave, which alternates between the levels $\pm 1$ with equal probability.
The $K$ tones present in $f(t)$ are thus aliased by the pseudorandom sequence.
The resulting demodulated signal $y(t) = f(t) p_c(t)$ is then integrated over a period $1/R$ and sampled at the low rate $R$. This integrate-and-dump approach results in the following samples
\begin{equation}
y_m = R \int_{m/R}^{(m+1)/R} y(t) \mathrm{d}t, \quad m = 0,1,\dots, R-1.
\end{equation}

%

The samples $y_m$ acquired by the random demodulator can be written as a linear combination of the $W \times 1$ sparse amplitude vector $\bf b$ that contains the coefficients $b_{\omega}$ at the corresponding locations $\omega$~\cite{RandomDemodulator1, RandomDemodulator2}. In matrix form, we write
\begin{equation}
\bf y = \Phi b,
\end{equation}
where $\bf y$ is the vector of size $R$ that contains the samples $y_m$ and $\bf \Phi$ is the known sampling matrix that describes the overall action of the system on the vector of amplitudes $\bf b$, namely demodulation and filtering (see~\cite{RandomDemodulator2} for more details).
Capitalizing on the sparsity of the vector $\bf b$, the amplitudes $b_{\omega}$ and their respective locations $\omega$ can be recovered from the low rate samples $\bf y$ using CS~\cite{CSBook} techniques such as those discussed in ``Compressed Sensing Recovery'', in turn allowing for the recovery of $f(t)$.
The minimal required number of samples $R$ for perfect recovery of $f(t)$ in noiseless settings is $2K$~\cite{CSBook}.

The random demodulator is one of the pioneer and innovative attempts to extend the inherently discrete and finite CS theory to analog signals.
However, truly analog signals, as those we consider here, require a prohibitively large number of harmonics to approximate them well within the discrete model.
When attempting to approximate signals such as those from the multiband model, the number of tones $W$ is of the order of the Nyquist rate and the number of samples $R$ is a multiple of $KB$.
This in turn renders the reconstruction computationally prohibitive and very sensitive to the grid choice (see~\cite{MagazineMishali} for a detailed analysis).
Furthermore, the time-domain approach precludes processing at a low rate, even for multitone inputs since interpolation to the Nyquist rate is an essential ingredient of signal reconstruction.
In terms of hardware and practical implementation, the random demodulator requires accurate modulation by a periodic square mixing sequence and accurate integration, which may be challenging when using analog signal generators, mixers and filters.

In contrast to the random demodulator, which adopts a discrete multitone model, the rest of the approaches we focus on treat the analog multiband model, illustrated in Fig.~\ref{fig:multiband}, which is of interest in the context of CR.\@

\subsection{Multi-Rate Sampling}
\label{sec:multirate}

\begin{figure}[ht]
  \begin{center}
    \includegraphics[width=0.9\columnwidth]{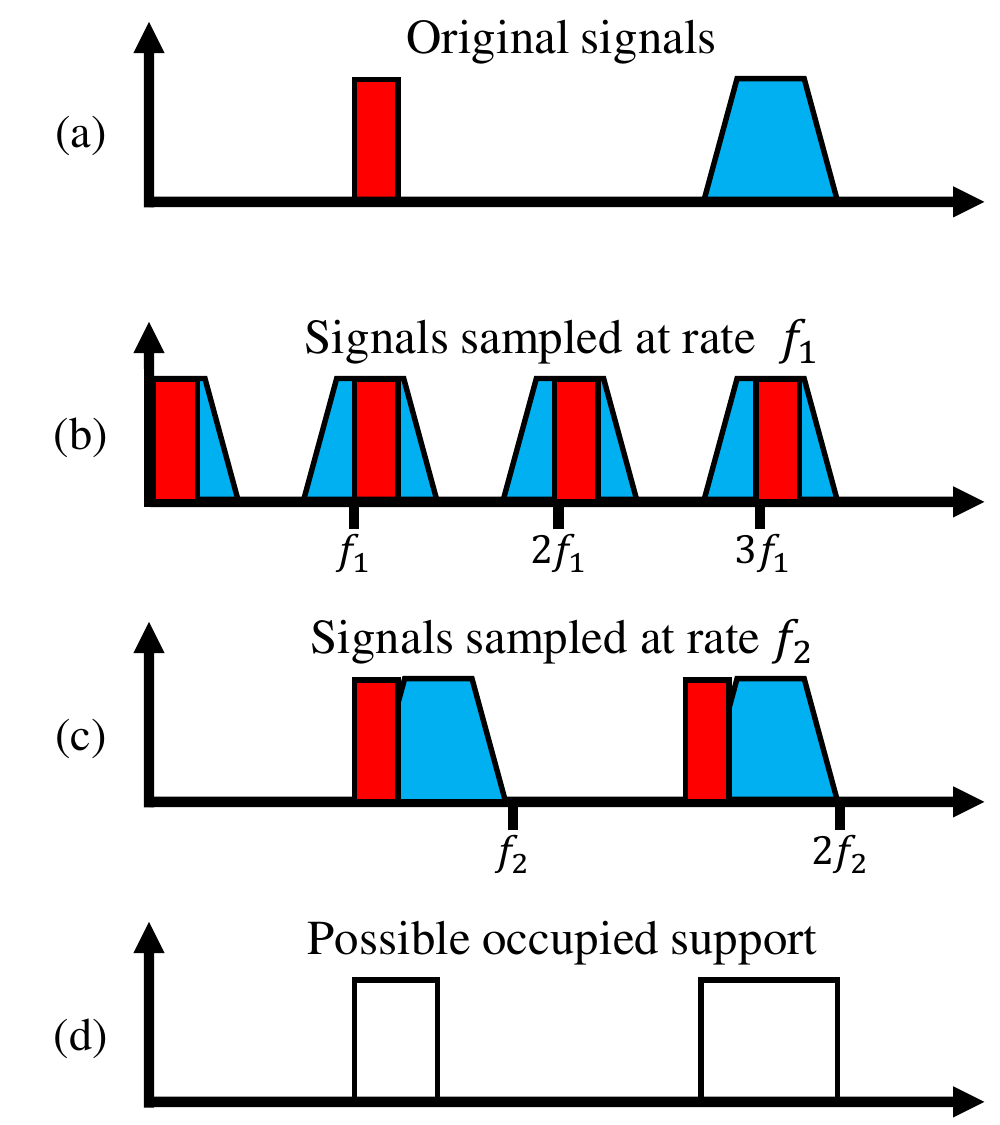}
    \caption{Action of the SMRS on a multiband signal: (a) The input signal with $K=2$ bands, (b) signals sampled at rate $F_1$ in channel 1, (c) signals sampled at rate $F_2$ in channel 2, and (d) possible support which is the intersection of the supports in channel 1 and 2~\cite{MultiRate}.}
    \label{fig:multirate_scheme}
  \end{center}
\end{figure}

An alternative sampling approach is based on the synchronous multi-rate sampling (SMRS)~\cite{MultiRate} scheme, which has been proposed in the context of electro-optical systems to undersample multiband signals. The SMRS samples the input signal at $P$ different sampling rates $F_i$, each of which is an integer multiple of a basic sampling rate $\Delta f$. This procedure aliases the signal with different aliasing intervals, as illustrated in Fig.~\ref{fig:multirate_scheme}. The Fourier transform of the undersampled signals is then related to the original signal through an underdetermined system of linear equations,
\begin{equation}
\label{eq:multirate_sys}
\mathbf{z}(f)= \mathbf{Q} \mathbf{x}(f).
\end{equation}
Here, $\mathbf{x}(f)$ contains frequency slices of size $\Delta f$ of the original signal $x(t)$ and $\mathbf{z}(f)$ is composed of the Fourier transform of the sampled signal. Each channel contributes $M_i = F_i/\Delta f$ equations to the system (\ref{eq:multirate_sys}), which concatenates the observation vector of all the channels. The measurement matrix $\bf Q$ has exactly $P$ non-zero elements in each column, that correspond to the locations of the spectral replica in each channel baseband $[0, F_i]$.

This approach assumes that either the signal or the sampling time window are finite. The continuous variable $f$ is then discretized to a frequency resolution of $\Delta f$. Since $x(t)$ is sparse in the frequency domain, the vector $\mathbf{x}(f)$ is sparse and can be recovered from (\ref{eq:multirate_sys}) using CS techniques, for each discrete frequency $f$. An alternative recovery method, referred to as the reduction procedure and illustrating in Fig.~\ref{fig:multirate_scheme}, consists of detecting baseband frequencies in which there is no signal, by observing the samples. These frequencies are assumed to account for the absence of signals of interest in all the frequencies that are down-converted to that baseband frequency, which allows to reduce the number of sampling channels. This assumption does not hold in the case where two or more frequency components cancel each other due to aliasing, which happens with probability zero. Once the corresponding components are eliminated from (\ref{eq:multirate_sys}), the reduced system is inverted using the Moore-Penrose pseudo-inverse to recover $\mathbf{x}(f)$.

There are several drawbacks to the SMRS that limit its performance and potential implementation. First, the discretization process affects the SNR since some of the samples are thrown out. Furthermore, spectral components down-converted to off the grid frequencies are missed. In addition, the first recovery approach requires a large number of sampling channels, proportional to the number of active bands $K$, whereas the reduction procedure does not ensure a unique solution and the inversion problem is ill-posed in many cases. Another difficulty is that, in practice, synchronization between channels sampling at different rates is challenging. Finally, this scheme samples wideband signals using low rate samplers. Practical ADCs introduce an inherent bandwidth limitation, modeled by an anti-aliasing low pass filter (LPF) with cut-off frequency determined by the sampling rate, which distorts the samples. Thus, the SMRS implementation requires low rate samplers with large analog bandwidth.

\subsection{Multicoset sampling}
\label{sec:multico}
A popular sampling scheme for sampling wideband signals at the Nyquist rate is multicoset or interleaved ADCs~\cite{Bresler,Mishali_multicoset,MagazineMishali} in which several channels are used, each operating at a lower rate.
We now discuss how such systems can be used in the sub-Nyquist regime.

\begin{figure}[ht]
  \begin{center}
    \includegraphics[width=0.9\columnwidth]{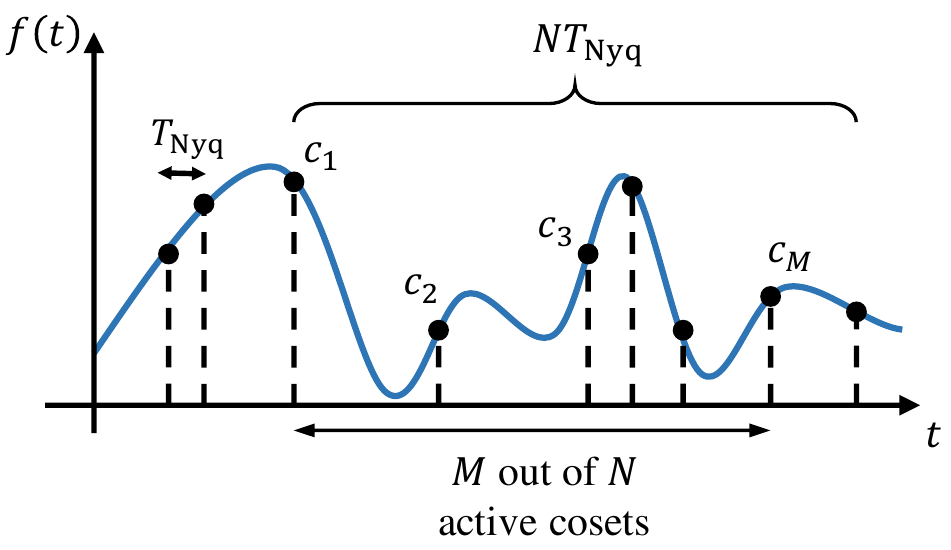}
    \caption{Illustration of multicoset sampling.}
    \label{fig:multico_signal}
  \end{center}
\end{figure}

\begin{figure}[ht]
  \begin{center}
    \includegraphics[width=0.9\columnwidth]{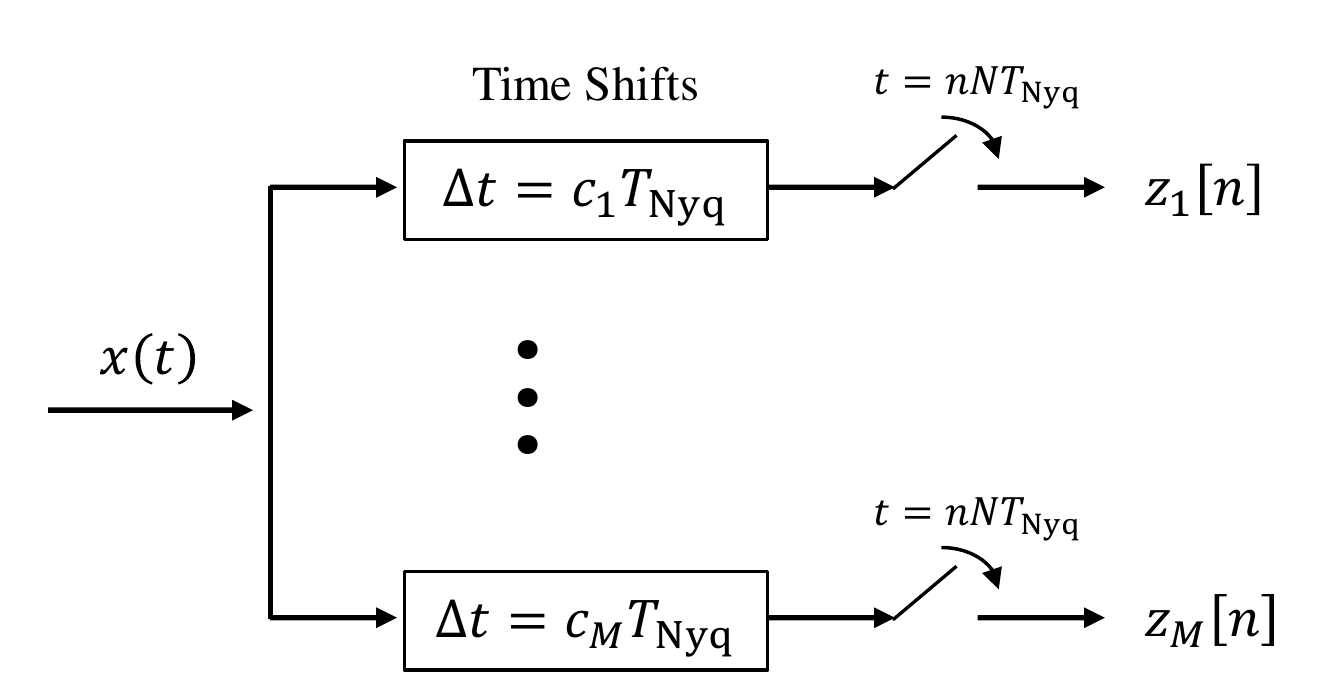}
    \caption{Schematic implementation of multicoset sampling. The input signal $x(t)$ is inserted into the multicoset sampler that splits the signal into $M$ branches, and delays each one by a fixed coefficient $c_i T_{\text{Nyq}} $. Every branch is then sampled at the low rate $1/(NT_{\text{Nyq}})$, and then digitally processed to perform spectrum sensing and signal reconstruction.}
    \label{fig:multico_scheme}
  \end{center}
\end{figure}

Multicoset sampling may be described as the selection of certain samples from the uniform Nyquist grid, as shown in Fig.~\ref{fig:multico_signal}, where $T_{\text{Nyq}}=1/f_{\text{Nyq}}$ denotes the Nyquist period.
More precisely, the uniform grid is divided into blocks of $N$ consecutive samples, from which only $M<N$ are kept. Mathematically, the $i$th sampling sequence is defined as
\begin{equation}
\label{eq:multico_samples}
x_{c_i}[n]= \left\{ \begin{array}{ll}
x(nT_{\text{Nyq}}), & n=mN+c_i, m \in \mathbb{Z} \\
0, & \text{otherwise},
\end{array} \right.
\end{equation}
where the cosets $c_i$ are ordered integers so that $0 \leq c_1 < c_2 < \cdots < c_M < N$.
A possible implementation of the sampling sequences (\ref{eq:multico_samples}) is depicted in Fig.~\ref{fig:multico_scheme}.
The building blocks are $M$ uniform samplers at rate $1/NT_{\text{Nyq}}$, where the $i$th sampler is shifted by $c_i T_{\text{Nyq}}$ from the origin.
When sampling at the Nyquist rate, $M=N$ and $c_i=(i-1)$.

\begin{figure}[ht]
	\begin{center}
		\includegraphics[width=1\columnwidth]{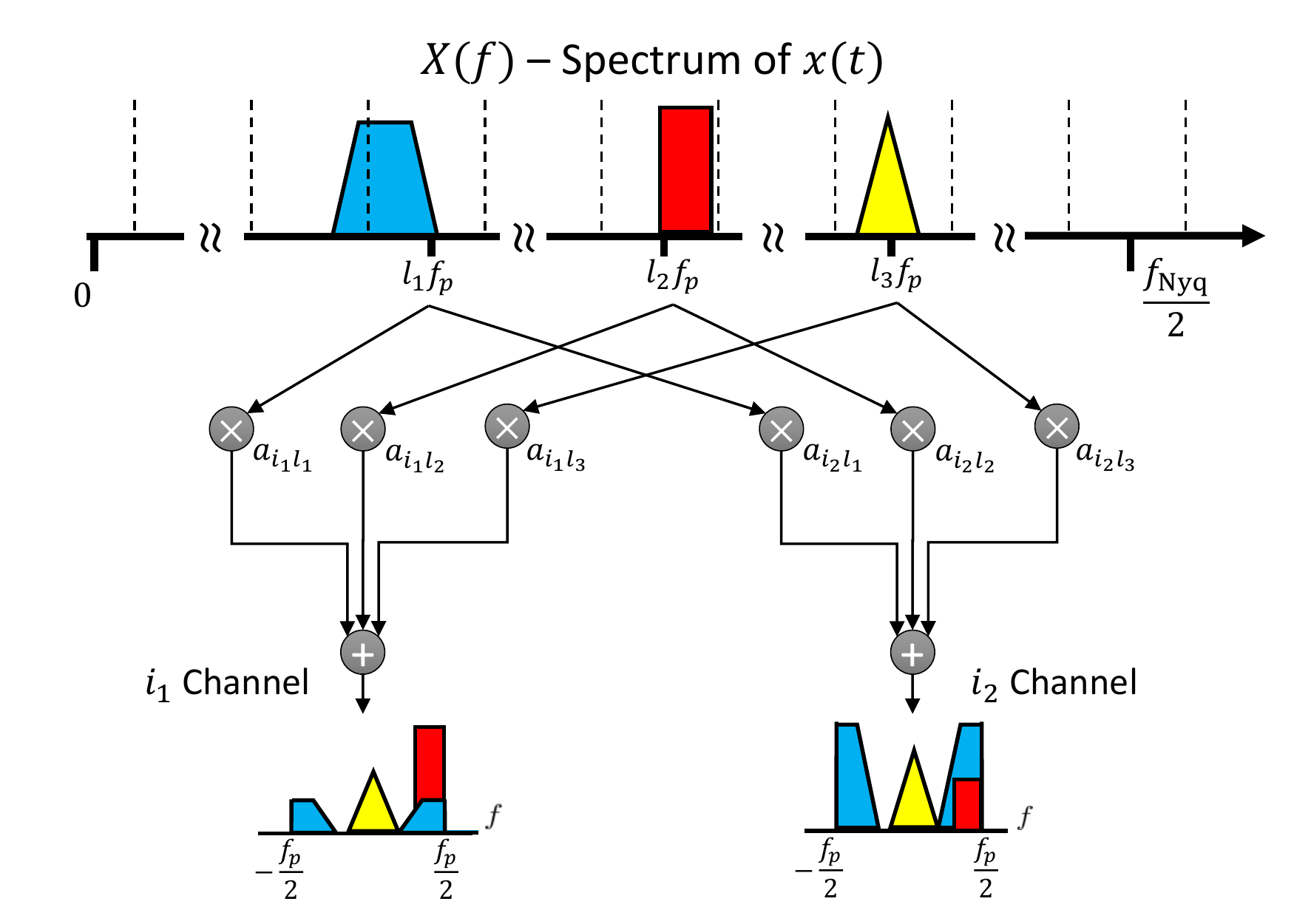}
		\caption{The spectrum slices of the input signal $\mathbf{x}(f)$ are shown here to be multiplied by the coefficients $a_{il}$ of the sensing matrix $\mathbf{A}$, resulting in the measurements $z_i$ for the $i$th channel. Note that in multicoset sampling, only the slices' complex phase is modified by the coefficients $a_{il}$. In the MWC sampling described below, both the phases and amplitudes are affected.}
		\label{fig:zAx}
	\end{center}
\end{figure}

The samples in the Fourier domain can then be written as linear combinations of spectrum slices of $x(t)$, such that~\cite{Mishali_multicoset}
\begin{equation}
\mathbf{z}(f) = \mathbf{A} \mathbf{x}(f), \qquad f \in \mathcal{F}_s.
\label{eq:multico}
\end{equation}
Here, $\mathbf{x}(f)$ denotes the spectrum slices of $x(t)$, $f_s = \frac{1}{NT_{\text{Nyq}}} \ge B$ is the sampling rate of each channel and $\mathcal{F}_s=[-f_s/2, f_s/2]$. This relation is illustrated in Fig.~\ref{fig:zAx}. The $M \times N$ sampling matrix $\mathbf{A}$ is determined by the selected cosets $c_i$.
The recovery processing described below is performed in the time domain, where we can write
\begin{equation}
\mathbf{z}[n] = \mathbf{A} \mathbf{x}[n], \qquad n \in \mathbb{Z},
\label{eq:multico_time}
\end{equation}
The vector $\mathbf{z}[n]$ collects the measurements at $t=n/f_s$ and $\mathbf{x}[n]$ contains the samples sequences corresponding to the spectrum slices of $x(t)$.

The goal can then be stated as the recovery of $\mathbf{x}[n]$ from the samples $\mathbf{z}[n]$.
The system (\ref{eq:multico_time}) is underdetermined due to the sub-Nyquist setup and known as infinite measurement vectors (IMV) in the CS literature~\cite{CSBook, SamplingBook}. With respect to these two properties, the digital reconstruction algorithm encompasses the following three stages~\cite{Mishali_multicoset} that we explain in more detail below:
\begin{enumerate}
\item The continuous-to-finite (CTF) block constructs a finite frame (or basis) from the samples.
\item The support recovery formulates an optimization problem whose solution's support is identical to the support $S$ of $\mathbf{x}[n]$, that is the active slices.
\item The signal is digitally recovered by reducing (\ref{eq:multico_time}) to the support of $\mathbf{x}[n]$.
\end{enumerate}

The recovery of $\mathbf{x}[n]$ for every $n$ independently is inefficient and not robust to noise. Instead, the CTF method, developed in~\cite{Mishali_multicoset}, exploits the fact that the bands occupy continuous spectral intervals so that $\mathbf{x}[n]$ are jointly sparse, that is they have the same spectral support $S$ over time.
The CTF produces a finite system of equations, called multiple measurement vectors (MMV)~\cite{CSBook, SamplingBook} from the infinite number of linear systems described by (\ref{eq:multico_time}). The samples are first summed as
\begin{equation}\label{eq:CTF0}
\mathbf{Q}= \sum_n \mathbf{z}[n] \mathbf{z}^H[n],
\end{equation}
and then decomposed to a frame $\bf V$ such that $\mathbf{Q=VV}^H$. Clearly, there are many possible ways to select $\bf V$. One option is to construct it by performing an eigendecomposition of $\bf Q$ and choosing $\bf V$ as the matrix of eigenvectors corresponding to the non zero eigenvalues. The finite dimensional MMV system
\begin{equation} \label{eq:CTF}
\bf V= A U,
\end{equation}
is then solved for the sparsest matrix $\bf U$ with minimal number of non-identically zero rows using CS techniques~\cite{CSBook, SamplingBook}. The key observation of this recovery strategy is that the indices of the non zero rows of $\bf U$ coincide with the active spectrum slices of $\mathbf{z}[n]$~\cite{Mishali_multicoset}. These indices are referred to as the support of $\mathbf{z}[n]$ and are denoted by $S$.

Once the support $S$ is known, $\mathbf{x}[n]$ is recovered by reducing the system of equations (\ref{eq:multico_time}) to $S$. The resulting matrix $\mathbf{A}_S$, that contains the columns of $\bf A$ corresponding to $S$, is then inverted as follows
\begin{equation}
\mathbf{x}_S[n]= \mathbf{A}_S^{\dagger} \mathbf{z}[n].
\label{eq:Reconstruct}
\end{equation}
Here, $\mathbf{x}_S[n]$ denotes the vector $\mathbf{x}[n]$ reduced to its support. The remaining entries of $\mathbf{x}[n]$ are equal to zero.

The overall sampling rate of the multicoset system is
\begin{equation}
f_{tot}=Mf_s=\frac{M}{N}f_{\text{Nyq}}.
\end{equation}
The minimal number of channels is dictated by CS results~\cite{CSBook} such that $M \geq 2K$ with $f_s\geq B$ per channel. The sampling rate can thus be as low as $2KB$, which is twice the Landau rate~\cite{LandauCS}.

Although this sampling scheme seems relatively simple and straightforward, it suffers from several practical drawbacks~\cite{MagazineMishali}. First, as in the SMRS approach, multicoset sampling requires low rate ADCs with large analog bandwidth. Another issue arises from the time shift elements, since maintaining accurate time delays between the ADCs on the order of the Nyquist interval $T_{\text{Nyq}}$ is difficult. Finally, the number of channels $M$ required for recovery of the active bands can be prohibitively high. The MWC, presented in the next section, uses similar recovery techniques while overcoming these practical sampling issues.

\subsection{MWC sampling}
The MWC~\cite{Mishali_theory} exploits the blind recovery ideas developed in~\cite{Mishali_multicoset} and combines them with the advantages of analog RF demodulation.
To circumvent the analog bandwidth issue in the ADCs, an RF front-end mixes the input signal $x(t)$ with periodic waveforms.
This operation imitates the effect of delayed undersampling used in the multicoset scheme and results in folding the spectrum to baseband with different weights for each frequency interval. These characteristics of the MWC enable practical hardware implementation, which will be described later on.

More specifically, the MWC is composed of $M$ parallel channels. In each channel, $x(t)$ is multiplied by a periodic mixing function $p_i(t)$ with period $T_p=1/f_p$ and Fourier expansion
\begin{equation}
p_i(t) =\sum_{l=-\infty}^{\infty} a_{il} e^{j\frac{2\pi}{T_p} lt}.
\end{equation}
The mixing process aliases the spectrum, such that each band appears in baseband.
The signal then goes through a LPF with cut-off frequency $f_s/2$ and is sampled at rate $f_s \ge f_p $. The analog mixture boils down to the same mathematical relation between the samples and the $N=f_{\text{Nyq}}/f_s$ frequency slices of $x(t)$ as in multicoset sampling, namely (\ref{eq:multico}) in frequency and (\ref{eq:multico_time}) in time, as shown in Fig.~\ref{fig:zAx}. Here, the $M \times N$ sampling matrix $\mathbf{A}$ contains the Fourier coefficients $a_{il}$ of the periodic mixing functions. The recovery conditions and algorithm are identical to those described for multicoset sampling. 

Choosing the channels' sampling rate $f_s$ to be equal to the mixing rate $f_p$ results in a similar configuration as the multicoset scheme in terms of the number of channels. In this case, the minimal number of channels required for the recovery of $K$ bands is $2K$. The number of branches dictates the total number of hardware devices and thus governs the implementation complexity. Reducing the number of channels is a crucial challenge for practical implementation of a CR receiver. The MWC architecture presents an interesting flexibility that permits trading channels for sampling rate, allowing to drastically reduce the number of channels, even down to a single channel.

\begin{figure}
\centering
\includegraphics[width=0.9\columnwidth]{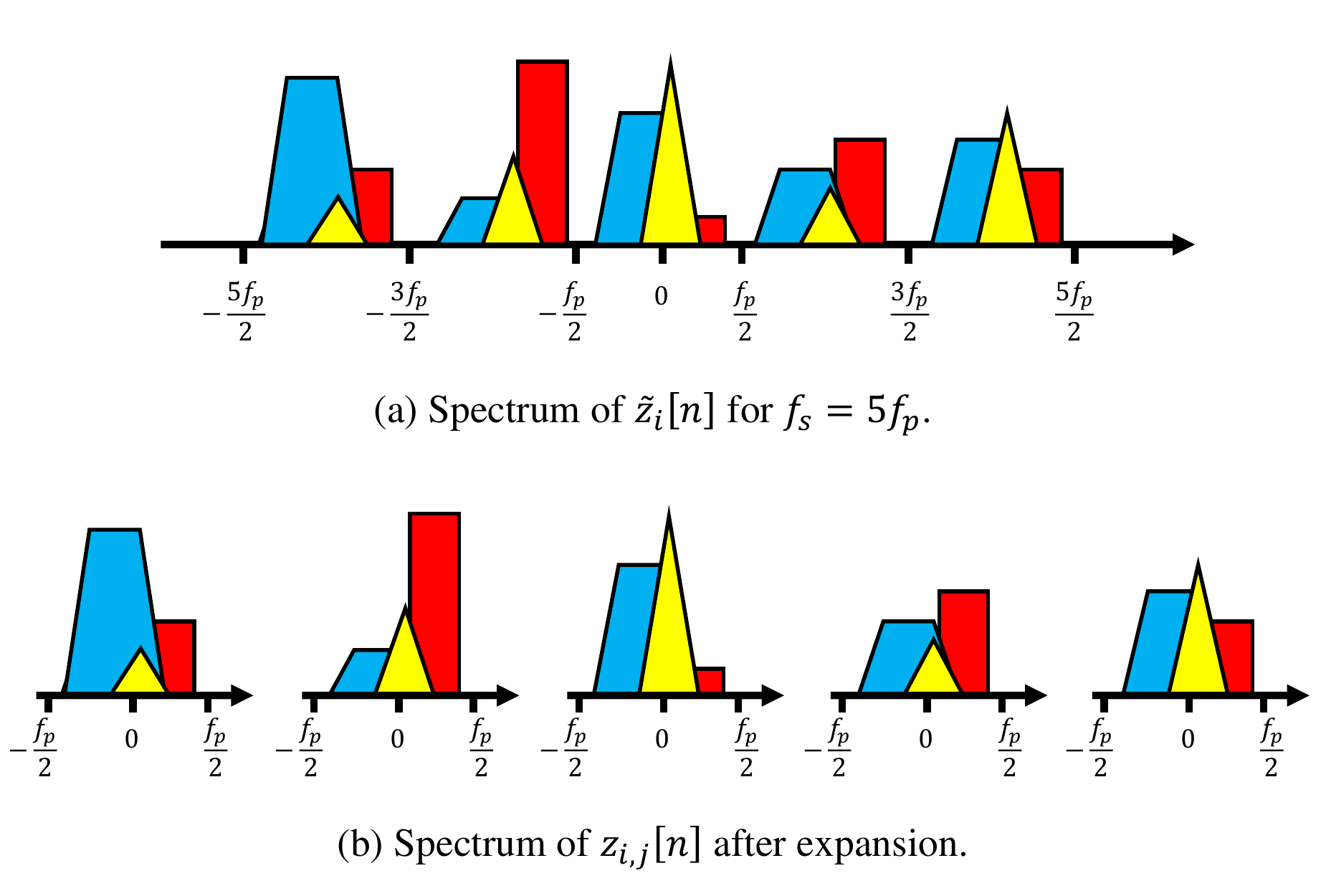}
\caption{Illustration of the expander configuration for $q=5$. (a) Spectrum of the output $\tilde{z}_i[n]$ of the physical $i$th channel, (b) spectrum of the samples $z_{i,j}[n]$ of the $q=5$ equivalent virtual channels, for $j=1,\dots,5$, after digital expansion.\label{fig:Expander}}
\end{figure}

Consider a configuration where $f_s=qf_p$, with odd $q$. In this case, the $i$th physical channel provides $q$ equations over $\mathcal{F}_p=[-f_p/2,f_p/2]$, as illustrated in Fig.~\ref{fig:Expander}. Conceptually, $M$ physical channels sampling at rate $f_s=qf_p$ are then equivalent to $Mq$ channels sampling at $f_s=f_p$. The number of channels is thus reduced at the expense of higher sampling rate $f_s$ in each channel and additional digital processing. The output of each of the $M$ physical channels is digitally demodulated and filtered to produce samples that would result from $Mq$ equivalent virtual branches. This happens in the so-called expander module, directly after the sampling stage and before the digital processing described above, in the context of multicoset sampling.
At its brink, this strategy allows to collapse a system with $M$ channels to a single branch with sampling rate $f_s=Mf_p$ (further details can be found in~\cite{Mishali_theory, mwc_hardware, TsiperCalib}).

\begin{figure*}[ht]
	\begin{center}
		\includegraphics[width=1\textwidth]{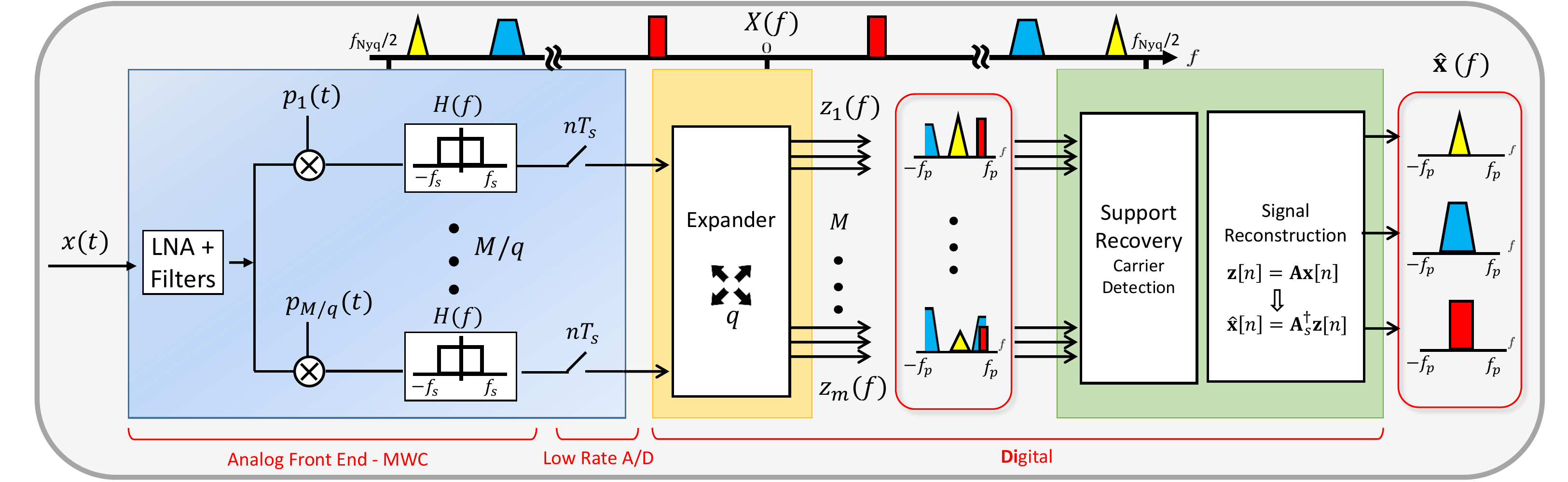}
		\caption{Schematic implementation of the MWC analog sampling front-end and digital signal recovery from low rate samples.}
		\label{fig:HighLevel}
	\end{center}
\end{figure*}

The MWC sampling and recovery processes are illustrated in Fig.~\ref{fig:HighLevel}.
This approach results in a hardware-efficient sub-Nyquist sampling method that does not suffer from the practical limitations described in previous sections, in particular, the analog bandwidth limitation of low rate ADCs. In addition, the number of MWC channels can be drastically reduced below $2K$ to as few as one, using higher sampling rate $f_s$ in each channel and additional digital processing. This tremendously reduces the burden on hardware implementation. However, the choice of appropriate periodic functions $p_i(t)$ to ensure correct recovery is challenging. Some guidelines are provided in~\cite{mishali2009expected, gan2013deterministic, SamplingBook}.

\subsection{Uniform Linear Array based MWC}\label{sec:alt_mwc}

\begin{figure}
  \begin{center}
    \includegraphics[width=1\columnwidth]{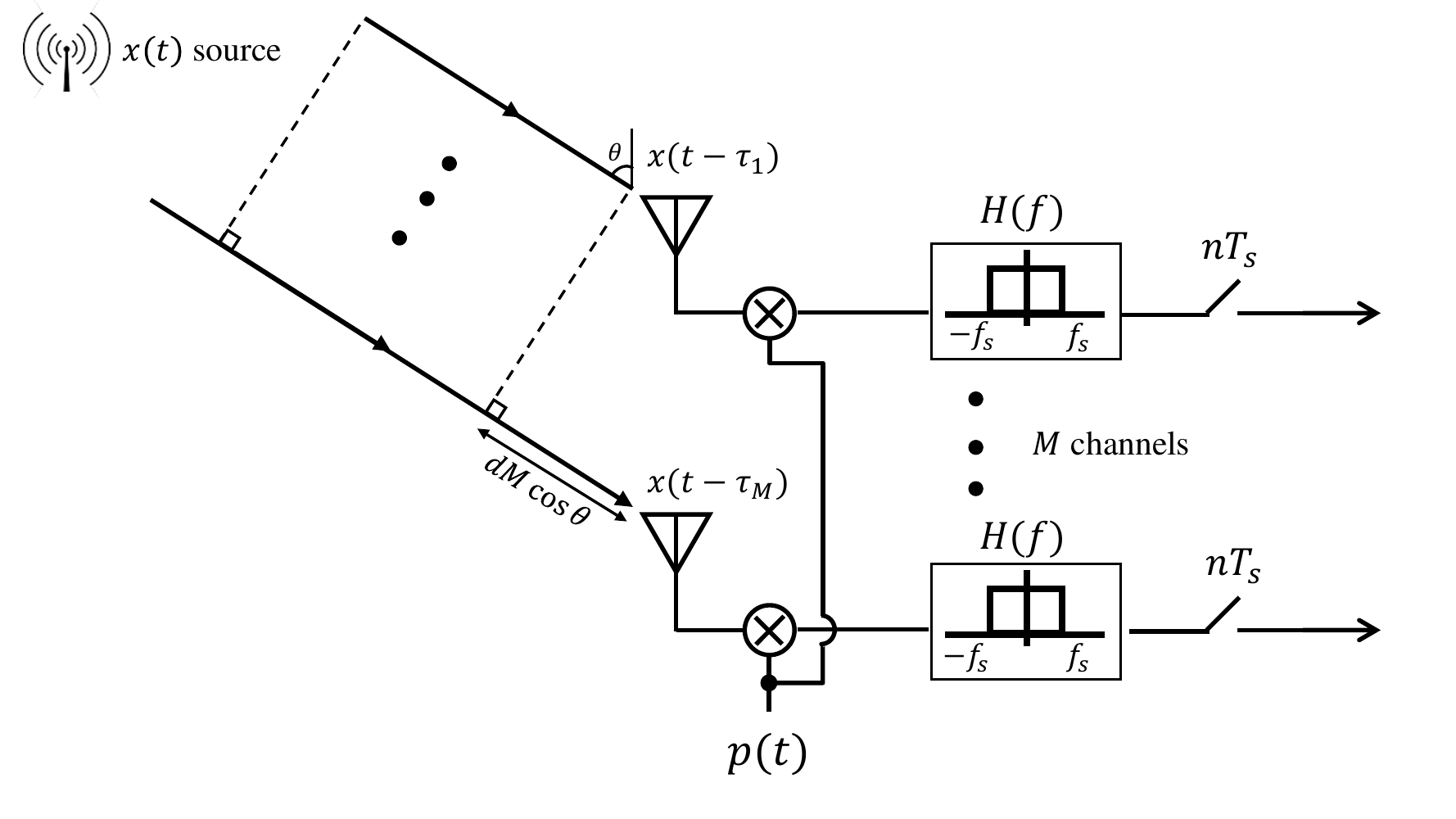}
    \caption{ULA configuration with $M$ sensors, with distance $d$ between two adjacent sensors. Each sensor includes an analog front-end composed of a mixer with the same periodic function $p(t)$, a LPF and a sampler, at rate $f_s$.}
    \label{fig:ULAfig}
  \end{center}
\end{figure}

An alternative sensing configuration, composed of a uniform linear array (ULA) and relying on the sampling paradigm of the MWC, is presented in~\cite{cohen_doa}. The sensing system consists of a ULA composed of $M$ sensors, with two adjacent sensors separated by a distance $d$, such that $d<c/(|\cos(\theta)|f_{\text{Nyq}})$, where $c$ is the speed of light and $\theta$ is the angle of arrival (AOA) of the signal $x(t)$. This system, illustrated in Fig.~\ref{fig:ULAfig}, capitalizes on the different accumulated phases of the input signal between sensors, given by
$
e^{j2\pi f_i \tau_m}
$, where
\begin{equation} \label{eq:delay}
\tau_{m}=\frac{dm}{c}\cos(\theta)
\end{equation}
is the delay at the $m$th sensor with respect to the first one.
Each sensor implements one channel of the MWC, that is the input signal is mixed with a periodic function, low-pass filtered and then sampled at a low rate.

This configuration has three main advantages over the standard MWC.\@
First, it allows for a simpler design of the mixing functions which can be identical in all sensors.
The only requirement on $p(t)$, besides being periodic with period $T_p$, is that none of its Fourier series coefficients within the signal's Nyquist bandwidth is zero.
Second, the ULA based system outperforms the MWC in terms of recovery performance in low SNR regimes. Since all the MWC channels belong to the same sensor, they are all affected by the same additive sensor noise.
In the ULA architecture, each channel belongs to a different sensor with uncorrelated sensor noise between channels.
This alternative architecture benefits from the same flexibility as the standard MWC in terms of collapsing the channels, which translates here into reducing the antennas.
This leads to a trade-off between hardware complexity, governed by the number of antennas, and SNR.\@
Finally, as will be shown later on, the modified system can be easily extended to enable joint spectrum sensing and DOA estimation.

\begin{figure}
\begin{center}
\includegraphics[width = 0.3\textwidth]{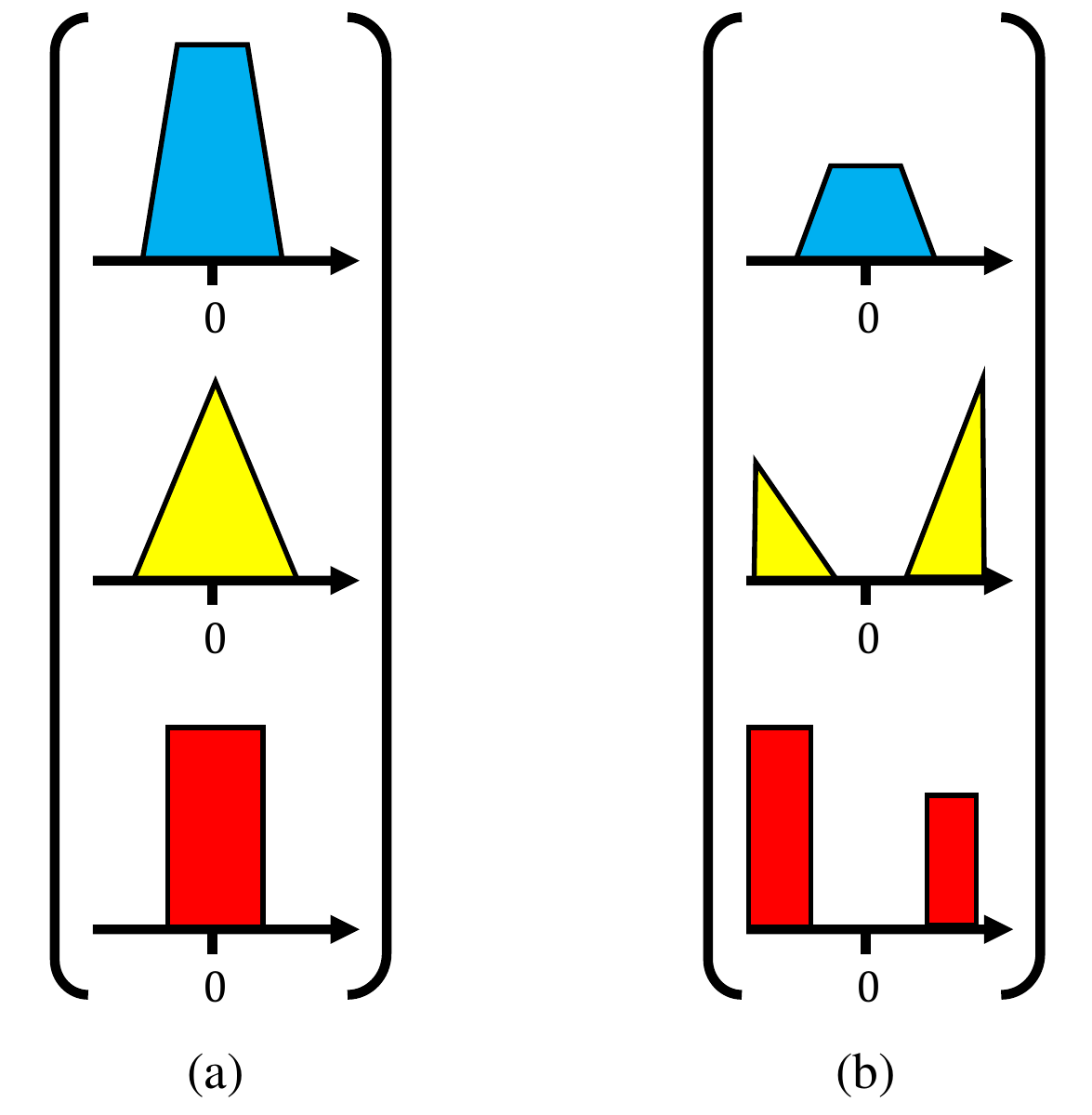}
\end{center}
\caption{The left pane shows the original source signals at baseband (before modulation). The right pane presents the output signals at baseband
$\mathbf{x}(f)$ after modulation, mixing, filtering and sampling.\label{fig:alt_mwc_sig}}
\end{figure}

Similarly to the previous sampling schemes, the samples $\mathbf{z}(f)$ can be expressed as a linear transformation of the unknown vector of slices $\mathbf{x}(f)$, such that
\begin{equation}
\label{eq:alternative_mwc}
\mathbf{z}(f)=\mathbf{A} \mathbf{x}(f), \qquad f \in \mathcal{F}_s.
\end{equation}
Here, $\mathbf{x}(f)$ is a non sparse vector that contains cyclic shifted, scaled and sampled versions of the active bands, as shown in Fig.~\ref{fig:alt_mwc_sig}.
In contrast to the previous schemes, in this configuration, the matrix $\mathbf{A}$, defined by
\begin{equation} \label{eq:A}
\mathbf{A}=
\left(\begin{matrix}e^{j2\pi f_{1}\tau_{1}} & \cdots & e^{j2\pi f_{N}\tau_{1}}\\
\vdots &  & \vdots\\
\\
e^{j2\pi f_{1}\tau_{M}} & \cdots & e^{j2\pi f_{N}\tau_{M}}
\end{matrix}\right),
\end{equation}
depends on the unknown carrier frequencies.

Two approaches are presented in~\cite{cohen_doa} to recover the carrier frequencies of the transmissions composing the input signal. The first is based on CS algorithms and assumes that the carriers lie on a predefined grid.
In this case, the resulting sensing matrix, which extends $\bf A$ with respect to the grid, is known and the expanded vector $\mathbf{x}(f)$ is sparse. This leads to a similar system as (\ref{eq:multico}) or (\ref{eq:multico_time}) which can be solved using the recovery paradigm from~\cite{Mishali_multicoset}, described in the context of multicoset sampling. In the second technique, the grid assumption is dropped and ESPRIT~\cite{PAUL1986} is used to estimate the carrier frequencies. This approach first computes the sample covariance of the measurements $\mathbf{z}[n]$ as
 \begin{equation} \label{eq:Rest}
\mathbf{R}=\sum_{n}\mathbf{z}[n] \mathbf{z}^H[n],
\end{equation}
and performs a singular value decomposition (SVD). The non zero singular values correspond to the signal's subspace and the carrier frequencies are then estimated from these. Once the carriers are recovered, the signal itself is reconstructed by inverting the sampling matrix $\bf A$ in (\ref{eq:alternative_mwc}).

The minimal number of sensors required by both reconstruction methods in noiseless settings is $M=2K$, with each sensor sampling at the minimal rate of $f_{s}=B$ to allow for perfect signal recovery~\cite{cohen_doa}.
The proposed system thus achieves the minimal sampling rate $2KB$ derived in~\cite{Mishali_multicoset}. We note that the expander strategy proposed in the context of the MWC can be applied in this configuration as well.

\section{MWC Hardware}

\subsection{MWC Prototype}

\begin{figure*}
		\includegraphics[width=1\textwidth]{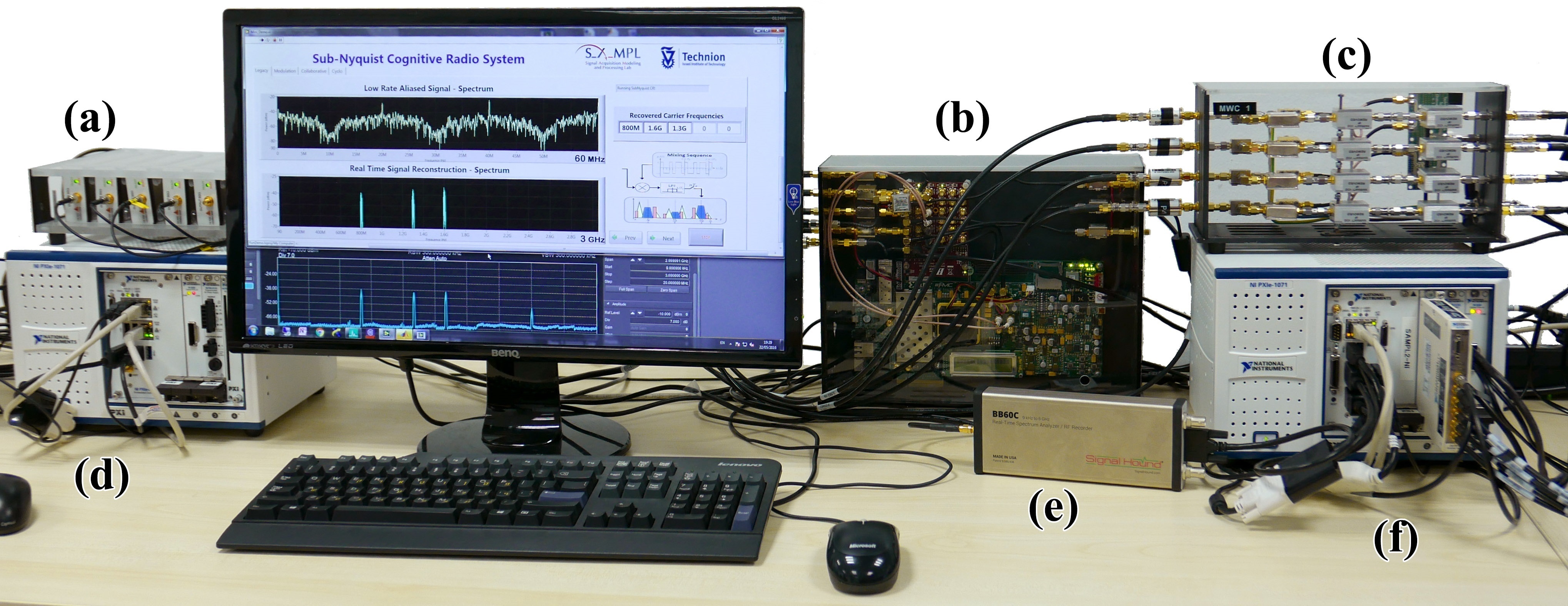}
		\caption{MWC CR system prototype: (a) vector signal Generators, (b) FPGA mixing sequences generator, (c) MWC analog front-end board, (d) RF combiner, (e) spectrum analyzer, (f) ADC and DSP.\label{fig:Prototype}}
\end{figure*}

One of the main aspects that distinguish the sub-Nyquist MWC from other sampling schemes is its practical implementation~\cite{mwc_hardware}, proving the feasibility of sub-Nyquist sampling even under distorting effects of analog components and physical phenomena.
A hardware prototype, shown in Fig.\ref{fig:Prototype}, was developed and built according to the block diagram in Fig.~\ref{fig:HighLevel}. In particular, the system receives an input signal with Nyquist rate of $6\,\text{GHz}$ and spectral occupancy of up to $200\,\text{MHz}$, and samples at an effective rate of $480\,\text{MHz}$, that is only $8\%$ of the Nyquist rate and $2.4$ times the Landau rate.
This rate constitutes a relatively small oversampling factor of $20\%$ with respect to the theoretical lower sampling bound.
This section describes the different elements of the hardware prototype, shown in Fig.~\ref{fig:Prototype}, explaining the various considerations that were taken into account when implementing the theoretical concepts on actual analog components.

\begin{figure}[H]
	\begin{center}
		\includegraphics[width=1\columnwidth]{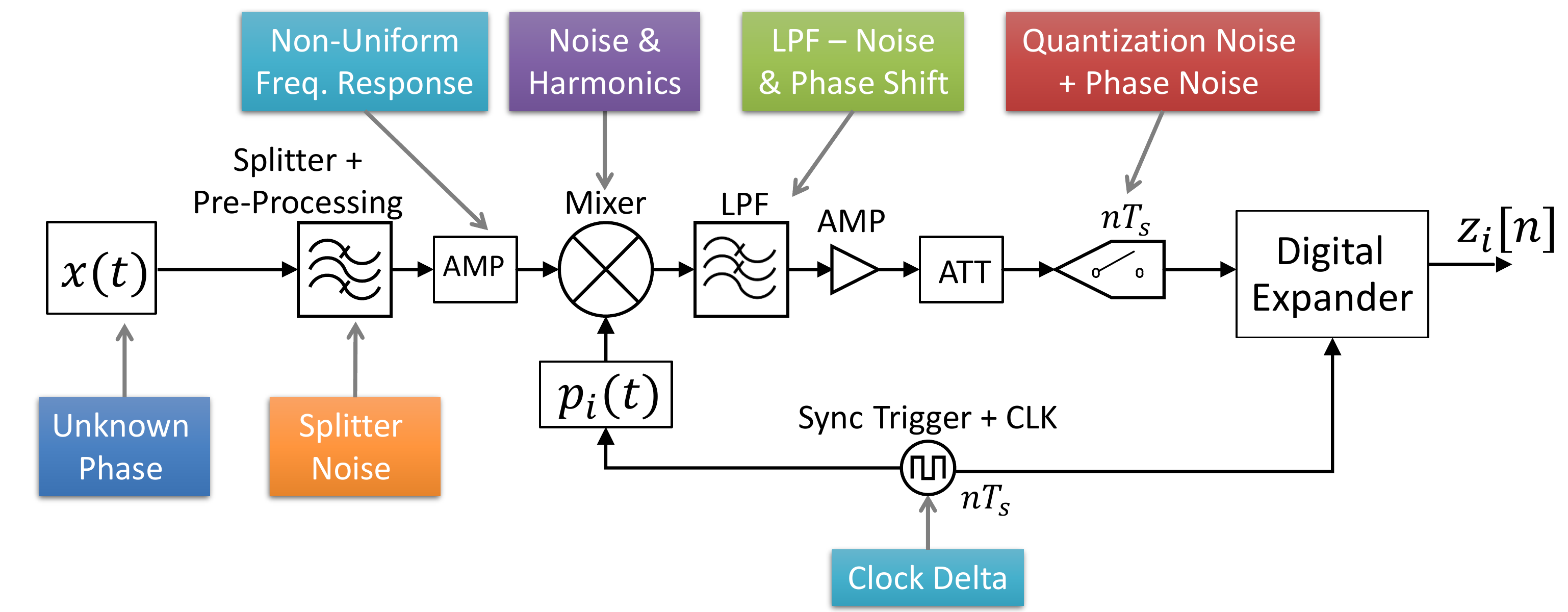}
		\caption{Hardware RF chain detailed schematics, including amplifiers, attenuators, filters, mixers, samplers and synchronization signals required for precise and accurate operation. The distortions induced by each component are indicated as well.}
		\label{fig:OneChannelRF}
	\end{center}
\end{figure}

At the heart of the system lies the proprietary MWC board~\cite{mwc_hardware} that implements the sub-Nyquist analog front-end. The card uses a high speed 1-to-4 analog splitter that duplicates the wideband signal to $M=4$ channels, with an expansion factor of $q=5$, yielding $Mq=20$ virtual channels after digital expansion. Then, an analog preprocessing step, composed of preliminary equalization, impedance corrections and gain adjustments, aims at maintaining the dynamic range and fidelity of the input in each channel. Indeed, the signal and mixing sequences must be amplified to specific levels before entering the analog mixers to ensure proper behavior emulating mathematical multiplication with the mixing sequences. The entire analog path of the multiband input signal can be seen in Fig.~\ref{fig:OneChannelRF}.

The modulated signal next passes through an analog anti-aliasing LPF.\@
The anti-aliasing filter must be characterized by both an almost linear phase response in the pass band between 0 to $50\,\text{MHz}$ and an attenuation of more than $20\,\text{dB}$ at $f_s/2=60\,\text{MHz}$.
A Chebyshev LPF of 7th order with cut-off frequency ($-3\,\text{dB}$) of $50\,\text{MHz}$ was chosen for the implementation. After impedance and gain corrections, the signal now has a spectral content limited to $50\,\text{MHz}$, that contains a linear combination of the occupied bands with different amplitudes and phases, as seen in Fig.~\ref{fig:zAx}.
Finally the low rate analog signal is sampled by a National Instruments\textsuperscript{\textcopyright} ADC operating at $120\,\text{MHz}$, leading to a total sampling rate of $480\,\text{MHz}$.

\begin{table}[t]
  \begin{centering}
    \begin{tabular}{>{\centering}p{0.9cm}>{\centering}p{1.3cm}>{\centering}m{4.2cm}l}
      & \textbf{Value}  & \textbf{Notes}  \tabularnewline
      \toprule
      $f_{s}$  & $120\hz M$  & $\left(q+1\right)f_{p}$ - Sampling rate  \tabularnewline
      $f_{p}$  & $20\hz M$  & $1/T_{p}$  \tabularnewline
      $q$  & 5  & Expansion factor  \tabularnewline
      $M/q$  & 4  & \# Hardware Channels  \tabularnewline
      $f_{\mbox{max}}$  & $3\hz G$  & $f_{\mbox{max}}=f_{\mbox{Nyq}}/2$  \tabularnewline
      $B$  & $18.5\hz M$  & Bandwidth on each carrier  \tabularnewline
      $M_p$  & 305  & Number of $\pm1$ intervals in each period of $p_{i}(t)$  \tabularnewline
    \end{tabular}
    \par\end{centering}
  \caption[MWC Specs]{The MWC parameters used in the setup of Fig.~\ref{fig:Prototype}.}
  \label{tab:mwc}
\end{table}

The mixing sequences that modulate the signal play an essential part in signal recovery.
They must have low cross-correlations with each other, while spanning a large bandwidth determined by the Nyquist rate of the input signal, and yet be easy enough to generate with relatively cheap, off-the-shelf hardware.
The sequences $p_i(t)$, for $i=1,\dots,4$, are chosen as truncated versions of Gold Codes~\cite{gold1967optimal}, which are commonly used in telecommunication (CDMA) and satellite navigation (GPS).\@
Mixing sequences based on Gold codes were found to give good results in the MWC system~\cite{Mishali_sequences}, primarily due to their small cross-correlations.

Since Gold codes are binary, the mixing sequences are restricted to alternating $\pm1$ values. This fact allows to digitally generate the sequences on a dedicated FPGA.\@
Alternatively, they can be implemented on a small chip with very low power and complexity. The added benefit of producing the mixing sequences on such a platform is that the entire sampling scheme may be synchronized and triggered using the same FPGA with minimally added phase noise and jitter, keeping a closed synchronization loop with the samplers and mixers.
A XiLinX VC707 FPGA acts as the central timing unit of the entire sub-Nyquist CR setup by generating the mixing sequences and the synchronization signals required for successful operation.
It is crucial that both the mixing period $T_p$ and the low rate samplers operating at $(q+1)f_p$ (due to intended oversampling) are fully synchronized, in order to ensure correct modeling of the entire system, and consequently guarantee accurate support detection and signal reconstruction.

The digital back-end is implemented using a National Instruments PXIe-1065 computer with DC coupled ADC.\@
Since the digital processing is performed at the low rate $f_s$, very low computational load is required in order to achieve real time recovery. MATLAB\textsuperscript{\textregistered}and LabVIEW\textsuperscript{\textregistered} environments are used for implementing the various digital operations and provide an easy and flexible research platform for further experimentations, as discussed in the next sections.
The sampling matrix $\bf A$ is computed once off-line, using the calibration process outlined in~\cite{TsiperCalib} and described in ``Hardware Calibration of the MWC''.\@

\subsection{Support Recovery}

The prototype is fed with RF signals composed of up to 5 carrier transmissions with an unknown total bandwidth occupancy of up to $200\,\text{MHz}$, and Nyquist rate of $6\,\text{GHz}$. An RF input $x(t)$ is generated using vector signal generators (VSG), each producing one modulated data channel with individual bandwidth of up to $20\,\text{MHz}$.
The input transmissions then go through an RF combiner, resulting in a dynamic multiband input signal.
This allows to test the system's ability to rapidly sense the input spectrum and adapt to changes, as required by modern CR standards (see ``IEEE 802.22 Standard for WRAN'' for details).
In addition, the described setup can simulate more complex scenarios, including collaborative spectrum sensing~\cite{coll_dist,coll_cent}, joint DOA estimation~\cite{cohen_doa}, cyclostationary based detection~\cite{cohen_cyclo} and various modulation schemes such as PSK, OFDM and more, for verifying sub-Nyquist data reconstruction capabilities.

\begin{figure}
	\begin{center}
		\includegraphics[width=1\columnwidth]{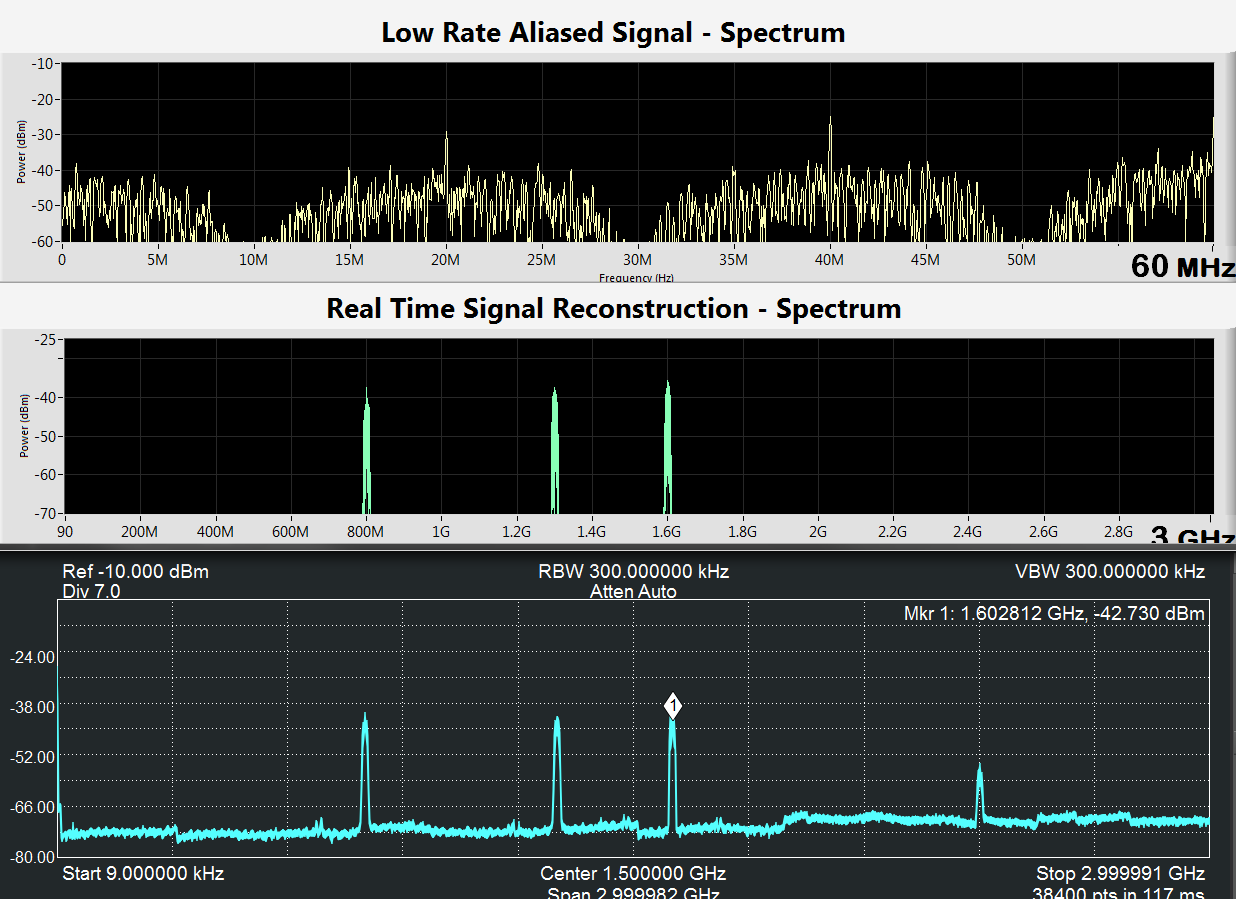}
		\caption{Screen shot from the MWC recovery software: low rate samples acquired from one MWC channel at rate $ 120\,\text{MHz}$ (top), digital reconstruction of the entire spectrum, performed from sub-Nyquist samples (middle), true input signal $x(t)$ showed using a fast spectrum analyzer (bottom).}
		\label{fig:LegacySim}
	\end{center}
\end{figure}

Support recovery is digitally performed on the low rate samples, as presented above in the context of multicoset sampling.
The prototype successfully recovers the support of the transmitted bands, when SNR levels are above $15\,\text{dB}$, as demonstrated in Fig.~\ref{fig:LegacySim}. More sophisticated detection schemes, such as cyclostationary detection, allow to achieve perfect support recovery from the same sub-Nyquist samples in lower SNR regimes of $0-10\,\text{dB}$, as seen in Fig.~\ref{fig:Cyclo1}, and will be further discussed below.

\begin{figure}
	\begin{center}
		\includegraphics[width=1\columnwidth]{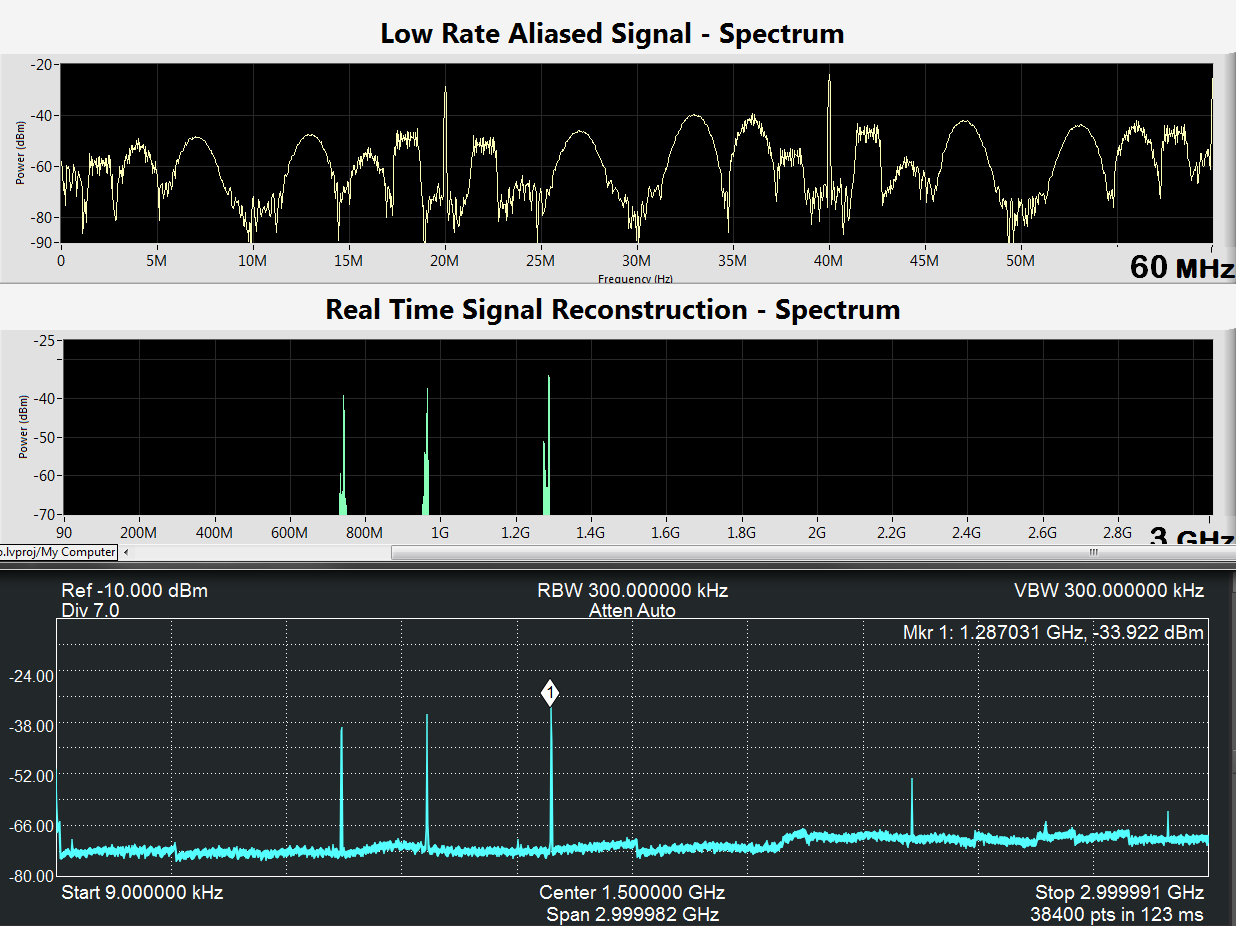}
		\caption{The setup is identical to Fig.~\ref{fig:LegacySim}. In this case, the individual transmissions have low bandwidth, highlighting the structure of the signal when folding to baseband.\label{fig:LegacyLowBW1}}
	\end{center}
\end{figure}

The main advantage of the MWC is that sensing is performed in real-time for the entire spectral range, even though the operation is performed solely on sub-Nyquist samples, which results in substantial savings in both computational and memory complexity. In additional tests, it is shown that the bandwidth occupied in each band can also be very small without impeding the performance, as shown in Fig.~\ref{fig:LegacyLowBW1}, where the support of signals with very low bandwidth (just 10\% occupancy within the $20\,\text{MHz}$ band) is correctly detected.

\subsection{Signal Reconstruction}

Once the support is recovered, the data is reconstructed from the sub-Nyquist samples, by applying~\eqref{eq:Reconstruct}. This step is performed in real-time, reconstructing the signal bands $\mathbf{z}[n]$ one sample at a time, with low complexity due to the small dimensions of the matrix-vector multiplication.
We note that reconstruction does not require interpolation to the Nyquist grid.
The active transmissions are recovered at the low rate of $20\,\text{MHz}$, corresponding to the bandwidth of the slices $\mathbf{z}(f)$.

\begin{figure}
	\begin{center}
		\includegraphics[width=1\columnwidth]{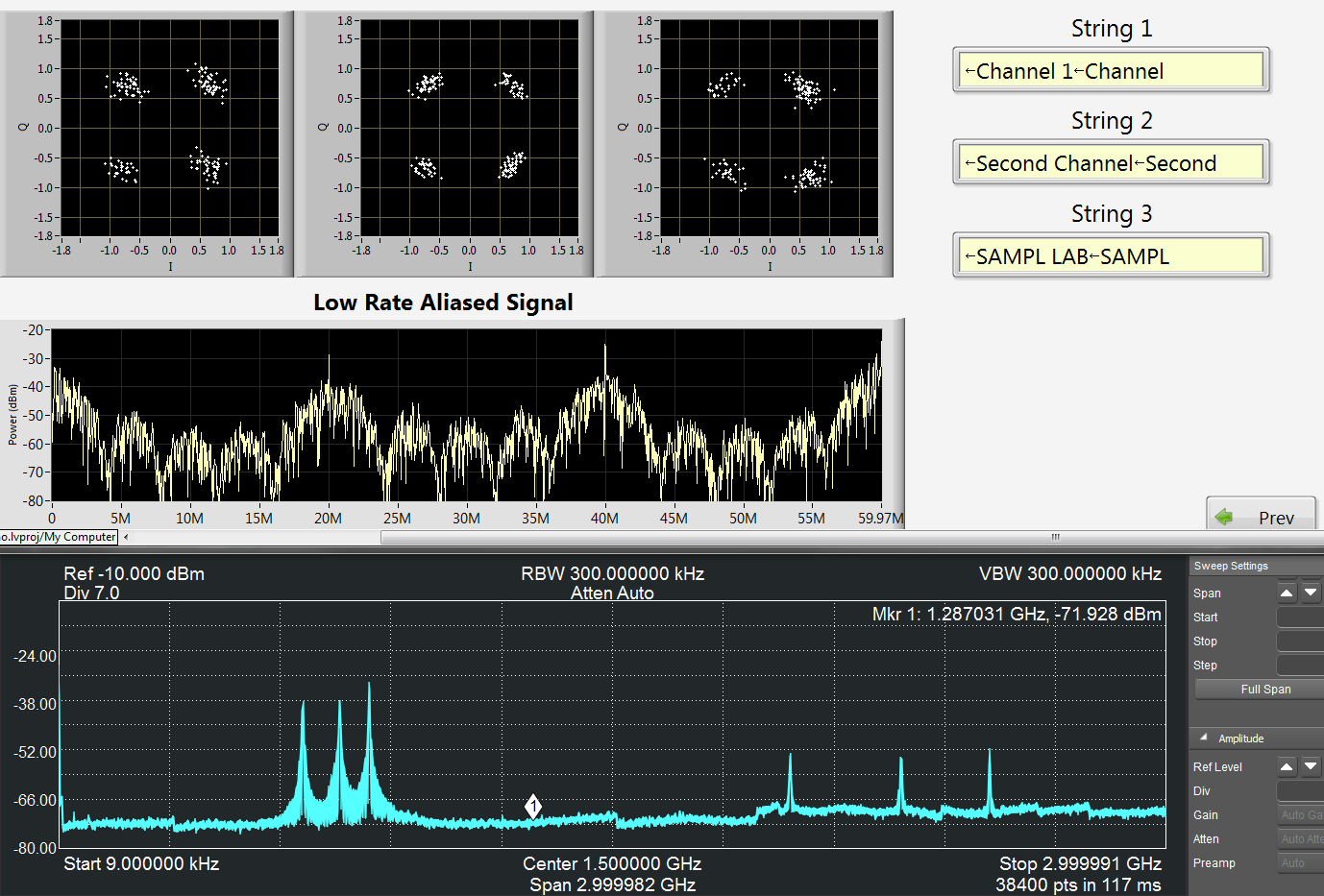}
		\caption{Demodulation, reconstruction and detection of $N_{\text{sig}}=3$ inputs from sub-Nyquist samples using the MWC CR prototype: I/Q phase diagrams, showing the modulation pattern of the transmitted bands after reconstruction from the low rate samples (top left), sub-Nyquist samples from an MWC channel $z_i[n] $ in the Fourier domain (middle), signal sampled by an external spectrum analyzer showing the entire bandwidth of 3\hz{G} (bottom),  information sent on each carrier, proving successful reconstruction (top right).\label{fig:ModulationSim}}
	\end{center}
\end{figure}

The prototype's digital recovery stage is further expanded to support decoding of common communication modulations, including BPSK, QPSK, QAM and OFDM.\@
An example for the decoding of three QPSK modulated bands is presented in Fig.~\ref{fig:ModulationSim}, where the I/Q constellations are shown after reconstructing the original transmitted signals $\v{x}_S$ using~\eqref{eq:Reconstruct}, from their low-rate and aliased sampled signals $\v{z}_n$.
The I/Q constellations of the baseband signals is displayed, each individually decoded using a general QPSK decoder.
In this example, the user broadcasts text strings, that are then deciphered and displayed on screen.
There are no restrictions regarding the modulation type, bandwidth or other parameters, since the baseband information is exactly reconstructed regardless of its respective content.
Therefore, any digital modulation method, as well as analog broadcasts, can be transmitted and deciphered without loss of information, by applying any desirable decoding scheme directly on the sub-Nyquist samples.

By combining both spectrum sensing and signal reconstruction, the MWC prototype serves as two separate communication devices.
The first is a state-of-the-art CR that performs real time spectrum sensing at sub-Nyquist rates, and the second is a receiver able to decode multiple data transmissions simultaneously, regardless of their carrier frequencies, while adapting to spectral changes in real time.
In cases where the support of the potential active transmissions is a priori known (e.g.\ potential cellular carriers), the MWC may be used as an RF demodulator that efficiently acquires several frequency bands simultaneously.

Other schemes would require a dedicated demodulation channel for each potentially active band.
In this case, the mixing sequences should be designed so that their Fourier coefficients are non zero only in the bands of interest, increasing SNR, and the support recovery stage is not needed~\cite{SequenceBlocker_Murmann}.

\begin{figure*}[t]\fboxsep1em
\colorbox{BoxBackground}{\begin{minipage}{1\textwidth}\begin{multicols*}{2}

\section*{Hardware Calibration of the MWC}

In the sampling system described above, the system matrix $\bf A$ is theoretically known and contains the Fourier series coefficients of the mixing sequences, such that
\begin{equation}
\label{eq:AtheoryCoeffs}
\mathbf{A}_{il}=c_{il}=\frac{1}{T_{p}}\dint 0{T_{p}}{p_{i}\left(t\right)e^{-j\frac{2\pi}{T_{p}}lt}}t\,.
\end{equation}
When calculating the matrix coefficients using~\eqref{eq:AtheoryCoeffs}, perfect support recovery and signal reconstruction is guaranteed both theoretically and by numerical verification performed in software simulations.
However, in practice, analog and physical distortions and imperfections affect the mixing and sampling process, and some modeling assumptions, that describe the system matrix in theory, no longer hold.
The main effects that distort the transfer function are:
\begin{itemize}
	\item The mixing procedure introduces nonlinearities. In general, mixers are intended
	to modulate narrowband signals with one sine carrier, as opposed
	to our mixing sequences that effectively contain over a hundred different harmonics.
	\item The analog filters have a non-flat frequency response, both in amplitude and phase.
	\item The actual design uses amplifiers and attenuators. These components exhibit
	non-linear frequency response.
	\item The phase noise and jitter, due to variations in components, cables and
	clock deltas lead to unknown and varying time shifts between the mixing and sampling stages.
\end{itemize}
An accurate method for estimating the effective $\v A$ is crucial for successful support recovery and signal reconstruction.
An adaptive calibration scheme is proposed in~\cite{TsiperCalib}.
The calibration procedure estimates the elements of $\v A$ with no prior knowledge on the mixing series $p_{i}(t)$ except for their period length $T_{p}$.

Since our system is not time invariant (e.g.\ samplers), nor linear (e.g.\ mixers), one cannot find the entries of the system matrix by simply measuring its response to an impulse. To circumvent this difficulty, the system's response is investigated for every frequency band of the spectrum by injecting known sinusoidal inputs sequentially.
In each iteration, the following input
\begin{equation}
x_{l}\left(t\right)=\alpha_{l}\sin\left[2\pi\left(lf_{p}+f_{0}\right)t \right],\ l\in\left[0,1,\ldots,L_{0}\right],
\end{equation}
is fed to the system. Here, $f_0=0.1f_p$ was heuristically chosen and the amplitudes $\alpha_l$ increase with $l$ to compensate for the attenuation of the Fourier coefficients of the mixing sequences at high frequencies. Every sine wave corresponds to a specific spectral band and translates to a relevant column of the matrix $\bf A$. The sub-Nyquist samples are then digitally processed to estimate the system matrix coefficients, column by column. Performance of the calibrated system is illustrated in Fig.~\ref{fig:CalibSNR}.

\begin{center}
	\includegraphics[width=1\columnwidth]{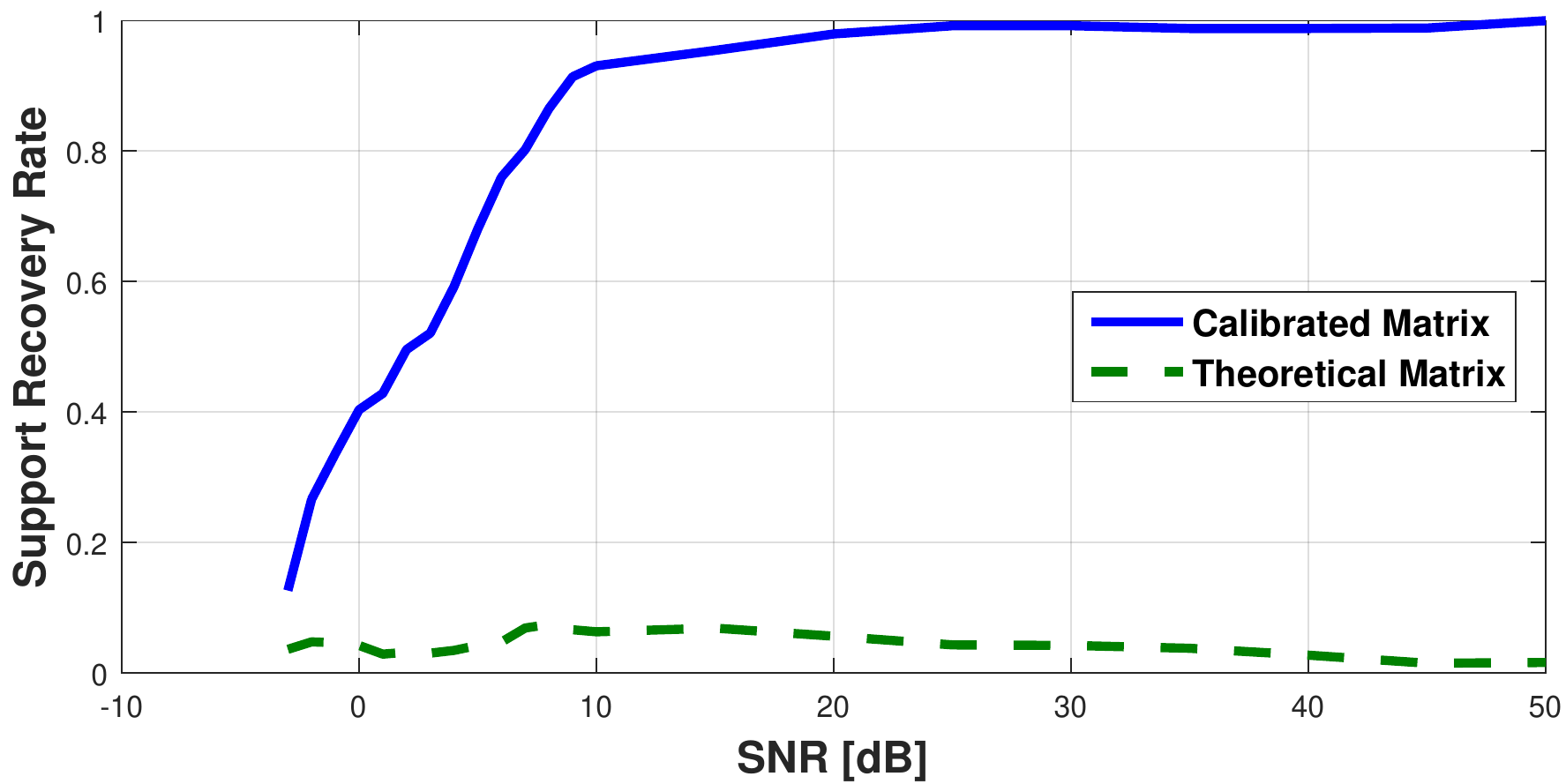}
	\caption[SNR Results]{Hardware reconstruction success rate of the calibrated matrix versus the theoretical matrix.}
	\label{fig:CalibSNR}
\end{center}

\end{multicols*}\end{minipage}}\end{figure*}

\section{Statistics Detection}

In the previous section, we reviewed recent sub-Nyquist sampling methods that reconstruct a multiband signal, such as a CR signal, from low rate samples.
However, the final goal of CRs often only requires detection and not necessarily perfect reconstruction of the PUs' transmissions.
In this regard, several works have proposed performing detection on second-order signal statistics, which share the same frequency support as the original signal.
In particular, power and cyclic spectra have been considered for stationary and cyclostationary~\cite{Gardner_review} (see ``Cyclostationarity'') signals, respectively.
Instead of recovering the signal from the low rate samples, its statistics are reconstructed and its support is estimated~\cite{Davies, Davies2, Leus, Leus_m, geert_ita, wang, cohen2013cognitive, TianFeature, TianFeatureJ, cabric_cyc, TianLeus, LeusCyclo, cohen_cyclo}.

Recovering second-order statistics rather than the signal itself benefits from two main advantages.
First, it allows to further reduce the sampling rate, as we will discuss below.
Intuitively, statistics have less degrees of freedom than the signal itself, requiring less samples for their reconstruction.
This follows from the assumption that the signal of interest is either stationary or cyclostationary.
Going one step further, the sparsity constraint can even be removed in this case and the power/cyclic spectrum of non sparse signals can be recovered from samples obtained below the Nyquist rate~\cite{Davies, Davies2, wang, cohen2013cognitive, TianLeus, cohen_cyclo}.
This is useful for CRs operating in less sparse environments, in which the lower bound of twice the Landau rate may exceed the Nyquist rate.
Second, robustness to noise is increased due to the averaging performed to estimate statistics.
This is drastically improved in the case of cyclostationary signals in the presence of stationary noise.
Indeed, exploiting cyclostationarity properties exhibited by communication signals allows to separate them from stationary noise, leading to better detection in low SNR regimes~\cite{GardnerBook}.
In this section, we first review power spectrum detection techniques in stationary settings and then extend these to cyclic spectrum detection of cyclostationary signals.

\subsection{Power Spectrum Based Detection}

In the statistical setting, the signal $x(t)$ is modeled as the sum of uncorrelated wide-sense stationary transmissions.
The stationarity assumption is key to further reducing the sampling rate.
In frequency, stationarity is expressed by the absence of correlation between distinct frequency components.
Specifically, as shown in~\cite{Papoulis}, the Fourier transform of a wide-sense stationary signal is a nonstationary white process, such that
\begin{equation}
\label{eq:papou}
\mathbb{E} [X(f_1) X^*(f_2) ] = S_x(f_1) \delta(f_1 -f_2).
\end{equation}
Here, the power spectrum $S_x(f)$ of $x(t)$ is the Fourier transform of its autocorrelation $r_x(\tau)$. Thus, obviously, the support of $S_x(f)$ is identical to that of $X(f)$. In this case, we will see that the autocorrelation matrix of the $N$ spectrum frequency slices of $x(t)$ contained in $\mathbf{x}(f)$ is diagonal and presents only $N$ degrees of freedom, which is key to sampling rate reduction.

Another intuitive interpretation can be given in the time domain. The autocorrelation of stationary signals $r_x(\tau) = \mathbb{E} \left[ x(t)x(t-\tau) \right]$ is a function only of the time lags $\tau$ and not of the sampling times $t$.
Therefore, the autocorrelation recovery capabilities of a sampling set is determined by the density of the associated difference set, namely the set that contains the time lags. The cardinality of the difference set, which depends on the choice of sampling times, is, in the worst case, equal to twice that of the original set. This corresponds for example to a uniform sampling set.
As a consequence, when the sampling scheme is not tailored to power spectrum recovery, the minimal required sampling rate in the blind setting is the Landau rate~\cite{cohen2013cognitive}. However, for an appropriate choice of sampling times~\cite{Leus, pp_nested}, the cardinality of the difference set can be on the order of the square of that of the original sampling set. In this case, under certain conditions, the density of the different set can be arbitrarily high even if the density of the sampling set goes to zero. With appropriate design, the autocorrelation or power spectrum may thus be estimated from samples with arbitrarily low average sampling rate~\cite{Davies, Davies2, Leus, pp_nested, pp_coprime, tar} at the expense of increased latency.

We first review power spectrum recovery techniques that do not exploit any specific design.
We then present methods that further reduce the sampling rate by adapting the sampling scheme to the purpose of autocorrelation or power spectrum estimation.



\subsubsection{Power Spectrum Recovery}

In this section, we first focus on sampling with generic MWC or multicoset schemes without specific design of the mixing sequences or cosets, respectively.

To recover $S_x(f)$ from the low rates samples $\mathbf{z}(f)$ obtained via multicoset sampling or the MWC~\cite{cohen2013cognitive}, consider the correlation matrix of the latter $\mathbf {R_z}(f) = \mathbb{E} [\mathbf{z}(f) \mathbf{z}^H(f) ]$. From (\ref{eq:multico}), $\mathbf {R_z}(f)$ can be related to correlations between the slices $\mathbf{x}(f)$, that is $\mathbf {R_x}(f) = \mathbb{E} [\mathbf{x}(f) \mathbf{x}^H(f) ]$, as follows
\begin{equation}
\mathbf{R_z}(f) = \mathbf{A} \mathbf{R_x}(f) \mathbf{A}^H, \qquad f \in \mathcal{F}_s.
\label{eq:autoco_analog}
\end{equation}
From (\ref{eq:papou}), the correlation matrix $\mathbf {R_x}(f)$ is diagonal and contains the power spectrum $S_x(f)$ at the corresponding frequencies, as
\begin{equation}
\mathbf{R}_{\mathbf{x}_{(i,i)}}(f)=S_x \left( f+ i f_s-\frac{f_{\text{Nyq}}}{2} \right), \qquad f \in \mathcal{F}_s.
\end{equation}
Recovering the power spectrum $S_x(f)$ is thus equivalent to recovering the matrix $\mathbf {R_x}(f)$.
Exploiting the fact that $\mathbf {R_x}(f)$ is diagonal and denoting by $\mathbf{r_x}(f)$ its diagonal, (\ref{eq:autoco_analog}) can be reduced to
\begin{equation}
\mathbf{r_z}(f) = \mathbf{(\bar{A} \odot A)} \mathbf{r_x}(f),
\label{eq:rzrx}
\end{equation}
where $\mathbf{r_z}(f)=\text{vec}(\mathbf{R_z}(f))$ concatenates the columns of $\mathbf{R_z}(f)$. The matrix $\mathbf{\bar{A}}$ is the conjugate of $\bf A$ and $\odot $ denotes the Khatri-Rao product~\cite{khatri}.

Generic choices of the sampling parameters, either mixing sequences or cosets, which are only required to ensure that $\bf A$ is full spark, are investigated in~\cite{cohen2013cognitive}. The Khatri-Rao product $\mathbf{(\bar{A} \odot A)}$ is full spark as well if $M> N/2$, that is the number of rows of $\bf A$ is at least half the number of slices $N$. The minimal sampling rate to recover $\mathbf{r_x}(f)$, and consequently $S_x(f)$, from $\mathbf{r_z}(f)$ in (\ref{eq:rzrx}) is thus equal to the Landau rate $KB$, namely half the rate required for signal recovery~\cite{cohen2013cognitive}.
The recovery of $\mathbf{r_x}(f)$ is performed using the procedure presented in the context of signal recovery on (\ref{eq:rzrx}), that is CTF, support recovery and power spectrum reconstruction (rather than signal reconstruction).

The same result for the minimal sampling rate is valid for non sparse signals, for which $KB$ is on the order of $f_{\text{Nyq}}$~\cite{cohen2013cognitive}. The power spectrum of such signals can be recovered at half their Nyquist rate. This means that even without any sparsity constraints on the signal in crowded environments, a CR can retrieve its power spectrum by exploiting stationarity. In this case, the system (\ref{eq:rzrx}) is overdetermined and $\mathbf{r_x}(f)$ is obtained by a simple pseudo-inverse operation.

Obviously, in practice, we do not have access to $\mathbf{R_z}(f)$, which thus needs to be estimated. The overall sensing time is divided into $N_f$ frames or windows of length $N_s$ samples. In~\cite{cohen2013cognitive}, different choices of $N_f$ and $N_s$ are examined for a fixed sensing time. In order to estimate the autocorrelation matrix $\mathbf{R_z}(f)$ in the frequency domain, estimates of $\mathbf{z}_i(f), 1 \le i \le M$, denoted by $\hat{\mathbf{z}}_i(f)$, are first computed using the fast Fourier transform (FFT) on the samples $z_i[n]$ over a finite time window. The elements of $\mathbf{R_z}(f)$ are then obtained by averaging the correlations between $\hat{\mathbf{z}}_i(f)$ over the frames.
In practice, the number of samples dictates the number of discrete Fourier transform (DFT) coefficients in the frequency domain and therefore the resolution of the reconstructed power spectrum. Once $\mathbf{\hat{r}_x}(f)$ is reconstructed, several detection statistics can be computed such as power or eigenvalue based test statistics~\cite{arts2015analytical}.

\subsubsection{Sampling Scheme Design}

Sampling approaches specifically designed for estimating the autocorrelation of stationary signals at much finer lags than the sample spacings have been studied recently in detail~\cite{Leus, pp_nested, pp_coprime, romero2015compression}.
These rely on the observation that the autocorrelation is a function of the lags only, namely the differences between pairs of sampling times. Thus, the correlation may be estimated at time lags contained in the difference set, also referred to as the difference co-array, which is the set composed of all differences between pairs of sampling times.
Since the size of the difference set may be greater than that of the original sampling set, depending on the choice of sampling times, we may need less sampling times for autocorrelation recovery than for signal recovery. Therefore, the sampling times should be carefully chosen so as to maximize the cardinality of the difference set.

The first approach we present, for autocorrelation recovery at sub-Nyquist rates, adopts multicoset sampling and designs the cosets in (\ref{eq:multico_samples}) to obtain a maximal number of differences.
In the previous section, the results were derived for any coset selection.
Here, we show that the sampling rate may be lower if the cosets are carefully chosen.
When using multicoset sampling, the sampling matrix $\bf A$ in~\eqref{eq:autoco_analog} or~\eqref{eq:rzrx}, is a partial Fourier matrix with $(i,k)$th element $e^{j\frac{2\pi}{N} c_i k}$.
A typical element of $\mathbf{(\bar{A} \odot A)}$ is then $e^{j\frac{2\pi}{N} (c_i-c_j) k}$.
If all cosets are distinct, then the size of the difference set over one period is greater than or equal to $2M-1$.
This bound corresponds to a worst case scenario, as discussed in the previous section and leads to a sampling rate of at least half Nyquist in the non sparse setting and at least Landau for a sparse signal with unknown support.
This happens for example is we select the first or last $M$ cosets or if we keep only the even or odd cosets.

\begin{figure}[t]
\begin{center}
\includegraphics[width=0.8\columnwidth]{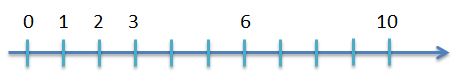}
\caption{Minimal sparse ruler of order $M=6$ and length $N=10$.\label{fig:min}}
\end{center}
\end{figure}

To maximize the size of the difference set and increase the rank of $\mathbf{(\bar{A} \odot A)}$, the cosets can be chosen~\cite{Leus, romero2015compression} using minimal linear and circular sparse rulers~\cite{leech1956representation}.
A linear sparse ruler is a set of integers from the interval $[0,N]$, such that the associated difference sets contains all integers in $[0,N]$. Intuitively, it may be viewed as a ruler with some marks erased but still able to measure all integer distances between $0$ and its length. For example, consider the minimal sparse ruler of length $N=10$. This ruler requires $M=6$ marks, as shown in Fig.~\ref{fig:min}. Obviously, all the lags $0 \leq \tau \leq 10$ on the integer grid are identifiable. 
There is no closed form expression for the maximum compression ratio $M/N$ that is achievable using a sparse ruler; however, the following bounds hold
\begin{equation} \label{eq:ruler_bound}
\frac{\sqrt{\tau(N-1)}}{N} \leq \frac{M}{N} \leq \frac{\sqrt{3 (N-1)}}{N},
\end{equation}
where $\tau \approx 2.4345$~\cite{romero2015compression}. A circular or modular sparse ruler extends this idea to include periodicity. Such designs that seek minimal sparse rulers, that is rulers with minimal number of marks $M$, allow to achieve compression ratios $M/N$ on the order of $\sqrt{N}$. As $N$ increases, the compression ratio may be arbitrarily low.

Two additional sampling techniques specifically designed for autocorrelation recovery are nested arrays~\cite{pp_nested} and co-prime sampling~\cite{pp_coprime}, presented in the context of autocorrelation estimation as well as beamforming and DOA estimation applications. In nested and co-prime structures, similarly to multicoset, the corresponding co-arrays have more degrees of freedom than those of the original arrays, leading to a finer grid for the time lags with respect to the sampling times. We now quickly review both sampling structures and their corresponding difference co-arrays and show how the autocorrelation of an arbitrary stationary signal can be recovered on the Nyquist grid from low rate samples.

In its simplest form, the nested array~\cite{pp_nested} structure has two levels of sampling density. The first level samples are at the $N_1$ locations ${\{\ell T_{\text{Nyq}}\}}_{1 \leq \ell \leq N_1}$ and the second level samples are at the $N_2$ locations ${\{(N_1+1)kT_{\text{Nyq}}\}}_{1 \leq k \leq N_2}$. This nonuniform sampling is then repeated with period $(N_1+1)N_2T_{\text{Nyq}}$.
Since there are $N_1+N_2$ samples in intervals of length $(N_1+1)N_2T_{\text{Nyq}}$, the average sampling rate of a nested array sampling set is given by
\begin{equation}
\label{eq:nested_density}
f_s = \frac{N_1+N_2}{(N_1+1)N_2T_{\text{Nyq}}} \equiv \frac{1}{N_1T_{\text{Nyq}}}+\frac{1}{N_2T_{\text{Nyq}}}.
\end{equation}
This rate can be arbitrarily low since $N_1$ and $N_2$ may be as large as we choose, at the expense of latency.

Now, consider the difference co-array which has contribution from the cross-differences and the self-differences. The non negative cross-differences, normalized by $T_{\text{Nyq}}$ for clarity, are given by
\begin{equation} \label{eq:nested_diff}
n = (N_1+1)k-\ell, \quad 1 \leq k \leq N_2, 1 \leq \ell \leq N_1.
\end{equation}
All differences in the range $1 \leq n \leq (N_1+1)N_2-1$ are covered, except for multiples of $N_1+1$.
These are are precisely the self differences among the second array.
As a result, the difference co-array is a filled array composed of all integers $-[(N_1+1)N_2-1] \leq n \leq [(N_1+1)N_2-1]$.
Going back to our autocorrelation estimation problem, this result shows that by proper averaging, we can estimate $R(\tau)$ at any lag $\tau$ on the Nyquist grid from nested array samples with arbitrarily low sampling rate, as
\begin{equation}
\hat{R}[n]= \begin{cases} 
\frac{1}{Q} \sum\limits_{q=0}^{Q-1} x(\tilde{N}(k+q)) x^*(\ell+\tilde{N}q), & n=\tilde{N}k-\ell, \\
\frac{1}{Q} \sum\limits_{q=0}^{Q-1} x(\tilde{N}(k+q)) x^*(\tilde{N}(\ell+q)), & n=\tilde{N}(k-\ell)\,,
\end{cases}
\end{equation}
where $Q$ is the number of snapshots used for averaging, and $\tilde{N}=N_1+1$. Here $k$ and $\ell$ are such that (\ref{eq:nested_diff}) holds.

Another sampling technique designed for autocorrelation recovery is co-prime sampling. It involves two uniform sampling sets with spacing $N_1T_{\text{Nyq}}$ and $N_2T_{\text{Nyq}}$ respectively, where $N_1$ and $N_2$ are co-prime integers. The average sampling rate of such a sampling set, given by
\begin{equation}
\label{eq:coprime_density}
f_s = \frac{1}{N_1T_{\text{Nyq}}} + \frac{1}{N_2T_{\text{Nyq}}},
\end{equation}
can again be made arbitrarily small compared to the Nyquist rate $1/T_{\text{Nyq}}$ by choosing arbitrarily large co-prime numbers $N_1$ and $N_2$. The associated difference set, normalized by $T_{\text{Nyq}}$, is composed of elements of the form $n=N_1k-N_2\ell$. Since $N_1$ and $N_2$ are co-prime, there exist integers $k$ and $\ell$ such that the above difference achieves any integer value $n$. Therefore, the autocorrelation may be estimated by proper averaging, as
\begin{equation}
\hat{R}[n]=\frac{1}{Q} \sum_{q=0}^{Q-1} x(N_1(k+N_2 q)) x^*(N_2(\ell + N_1q)),
\end{equation}
where $k$ and $\ell$ are such that $n=N_1k-N_2\ell$.

The main drawback of both techniques, besides the practical issue of analog bandwidth similarly to multicoset sampling, is the added latency required for averaging. Furthermore, in practice, synchronizing ADCs with different sampling rates can be challenging. Finally, nested array sampling still requires one sampler operating at the Nyquist rate. Thus, there is no saving in terms of hardware, but only in memory and computation.

\subsection{Cyclostationary Detection}

\begin{figure*}[ht!]\fboxsep1em
\colorbox{BoxBackground}{\begin{minipage}{1\textwidth}\begin{multicols*}{2}
\section*{Cyclostationarity}\label{box:cyclo}

Cyclostationary processes have statistical characteristics that vary periodically with time.
Examples of periodic phenomena that give rise to random data abound in engineering and science. In particular, in communications, periodicity arises from the underlying data modulation mechanisms, such as carrier modulation, periodic keying or pulse modulation~\cite{GardnerTS}.
A characteristic function of such processes, referred to as the cyclic spectrum, extends the traditional power spectrum to a two dimensional map, with respect to two frequency variables, angular and cyclic. The cyclic spectrum exhibits spectral peaks at certain frequency locations, the cyclic frequencies, which are determined by the signal's parameters, particularly the carrier frequency and symbol rate~\cite{GardnerBook}.

A process $x(t)$ is said to be wide-sense cyclostationary with period $T_0$ if its mean $\mu_x(t)=\mathbb{E}[x(t)]$ and autocorrelation $R_s(t,\tau) = \mathbb{E}[x(t)x(t+\tau)]$ are both periodic with period $T_0$~\cite{Gardner_review}, that is
\begin{equation}
\mu_x(t+T_0)= \mu_x(t), \qquad R_x(t+T_0,\tau) = R_x(t,\tau),
\end{equation}
for all $t \in \mathbb{R}$.
Given a wide-sense cyclostationary random process, its autocorrelation $R_x(t,\tau)$ can be expanded in a Fourier series
\begin{equation}
R_x(t,\tau) = \sum_{\alpha}R_x^{\alpha}(\tau) e^{j 2\pi \alpha t},
\end{equation}
where the sum is over integer multiples of the fundamental frequency $1/T_0$ and the Fourier coefficients $R_x^{\alpha}(\tau)$ are referred to as cyclic autocorrelation functions.
The cyclic spectrum is obtained by taking the Fourier transform of the cyclic autocorrelation functions with respect to $\tau$, namely
\begin{equation}
\label{eq:SCF} S_x^{\alpha}(f)=\int_{-\infty}^{\infty} R_x^{\alpha}(\tau) e^{-j 2\pi f \tau}\mathrm{d}\tau,
\end{equation}
where $\alpha$ is referred to as the cyclic frequency and $f$ is the angular frequency~\cite{Gardner_review}. If there is more than one fundamental frequency $1/T_0$, then the process $x(t)$ is said to be polycyclostationary in the wide sense. In this case, the cyclic spectrum contains harmonics (integer multiples) of each of the fundamental cyclic frequencies~\cite{GardnerBook}. These cyclic frequencies are governed by the transmissions' carrier frequencies and symbol rates as well as modulation types.

An alternative and more intuitive interpretation of the cyclic spectrum expresses it as the cross-spectral density $S_x^{\alpha}(f)=S_{uv}(f)$ of two frequency-shifted versions of $x(t)$, $u(t)$ and $v(t)$, such that
\begin{equation}
u(t)  \triangleq  x(t) e^{-j \pi \alpha t}, \quad
v(t)  \triangleq  x(t) e^{+j \pi \alpha t}.
\end{equation}
Then, from~\cite{Papoulis}, it holds that
\begin{equation}
\label{eq:scf2}
S_x^{\alpha}(f)=S_{uv}(f)=\mathbb{E} \left[ X \left(f+\frac{\alpha}{2} \right) X^*\left(f-\frac{\alpha}{2}\right)\right].
\end{equation}
As expressed in (\ref{eq:scf2}), the cyclic spectrum $S_x^{\alpha}(f)$ measures correlations between different spectral components of $x(t)$. Stationary signals, which do not exhibit spectral correlation between distinct frequency components, appear only at $\alpha =0$. This property is the key to robust detection of cyclostationary signals in the presence of stationary noise, in low SNR regimes.

The support region in the $(f,\alpha)$ plane of the cyclic spectrum of a bandpass cyclostationary signal is composed of four diamonds, as shown in Fig.~\ref{fig:diamonds}. Thus, the cyclic spectrum $S_x^{\alpha}(f)$ of a multiband signal with $K$ uncorrelated transmissions is supported over $4K$ diamond-shaped areas. Figure~\ref{fig:cyc_ex} illustrates the cyclic spectrum of two modulation types, AM and BPSK.\@

  \begin{center}
    \includegraphics[width=0.5\columnwidth]{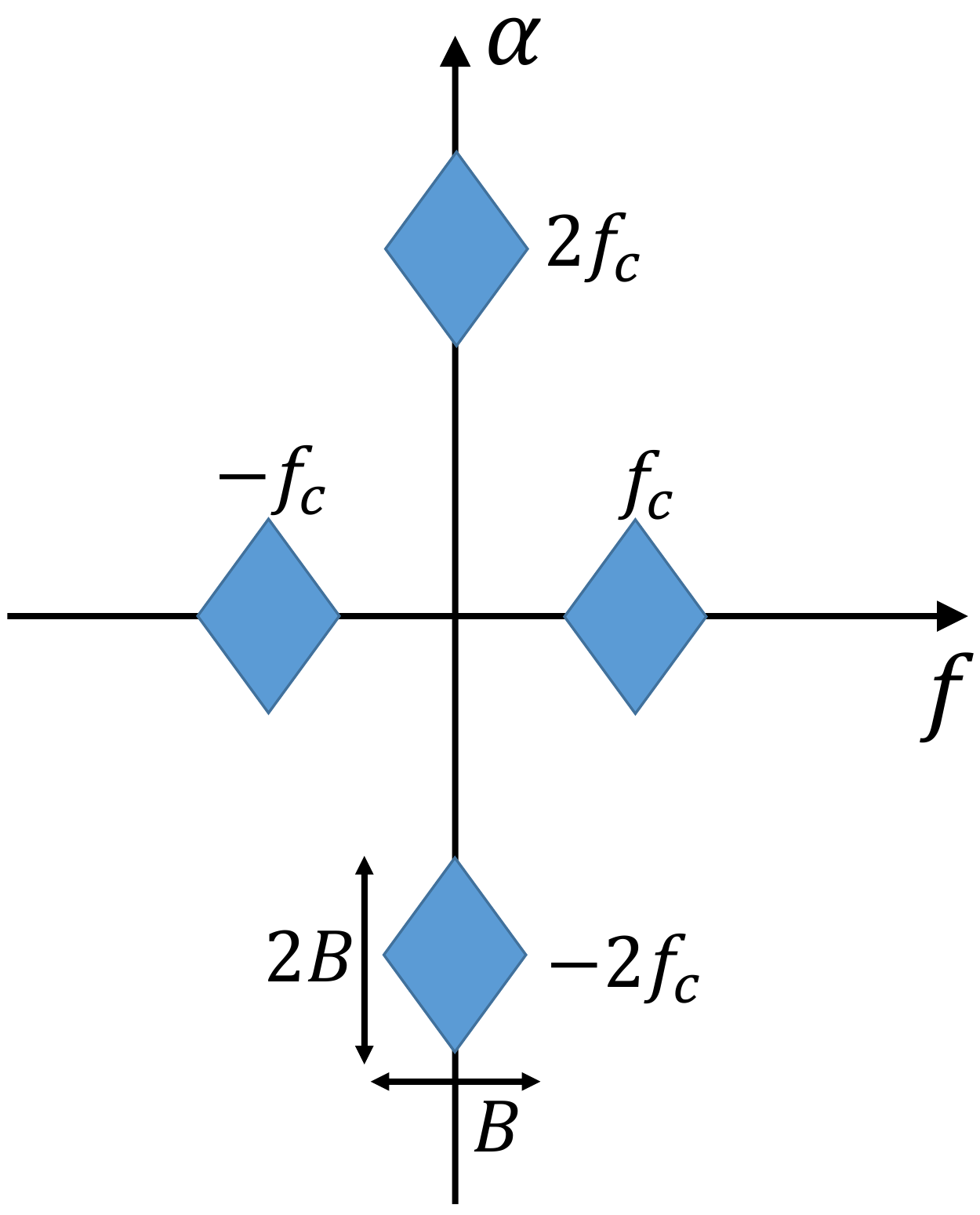}
    \caption{Support region of the cyclic spectrum of a bandpass cyclostationary signal with carrier frequency $f_c$ and bandwidth $B$.\label{fig:diamonds}}
  \end{center}

\begin{center}
    \includegraphics[width=1\columnwidth]{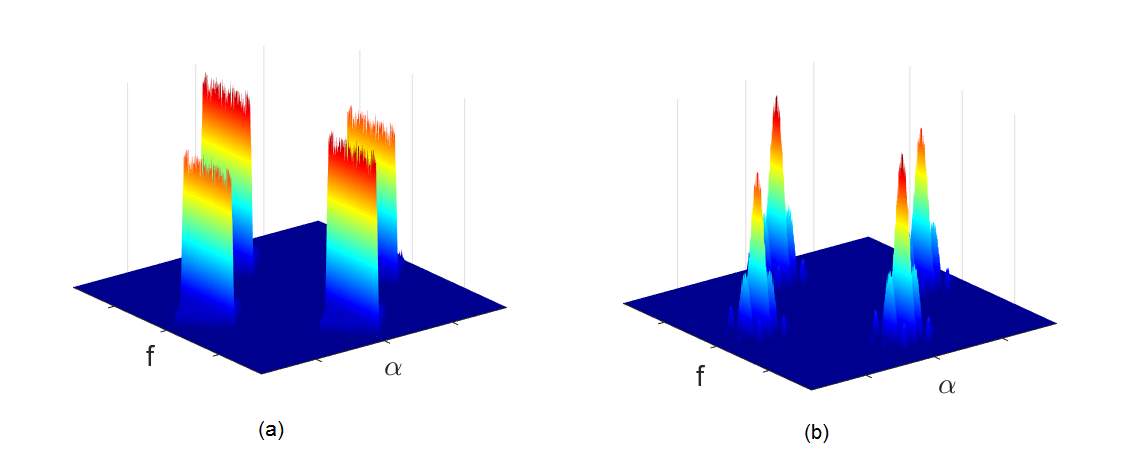}
    \caption{Cyclic spectrum magnitude of bandpass signals with (a) AM modulation and (b) BPSK modulation.\label{fig:cyc_ex}}
\end{center}

\end{multicols*}\end{minipage}}\end{figure*}

Communication signals often exhibit statistical periodicity, due to modulation schemes such as carrier modulation or periodic keying~\cite{GardnerTS}. Therefore, such signals are cyclostationary processes~\cite{GardnerBook}.
A characteristic function of these processes, the cyclic spectrum $S_x^{\alpha}(f)$, exhibits spectral peaks at the cyclic frequencies, which are determined by the signal's parameters of periodicity, such as the carrier frequency and symbol rate~\cite{GardnerBook}.
The formal definition of the cyclic spectrum is presented in ``Cyclostationarity''.
Cyclostationary based detection distinguishes between the signals of interest, assumed to be cyclostationary, and stationary noise and interference by measuring spectral correlation~\cite{GardnerBook}.
Since stationary noise and interference exhibit no spectral correlation, as shown in (\ref{eq:papou}), such detectors are highly robust to noise.
With noise enhancement being one of the main limitations of sub-Nyquist sampling based techniques, cyclostationary detection performed on the reconstructed cyclic spectrum from compressive measurements is a natural candidate for improving sub-Nyquist detection.
In this section, we first provide some general background on cyclostationarity and then review sub-Nyquist cyclostationary detection approaches.

\subsubsection{Cyclic Spectrum Recovery}

In the previous section, we showed how the power spectrum $S_x(f)$ can be reconstructed from correlations $\mathbf{R_z}(f)$ between the samples obtained using the MWC or multicoset sampling.
To that end, we first related $S_x(f)$ to the slices' correlation matrix $\mathbf{R_x}(f)$ and then recovered the latter from $\mathbf{R_z}(f)$.
Here, this approach is extended to the cyclic spectrum $S_x^{\alpha}(f)$.
We first show how it is related to shifted correlations between the slices, namely $\mathbf{R}_\mathbf{x}^a(\tilde{f}) = \mathbb{E} \left[ \mathbf{x}(\tilde{f}) \mathbf{x}^H(\tilde{f}+a) \right]$, for $a \in [0, f_s]$ and $\tilde{f} \in [0,f_s-a]$.
Next, similarly to power spectrum recovery, $\mathbf{R}_\mathbf{x}^a(\tilde{f})$ is reconstructed from shifted correlations of the samples $\mathbf{R}_\mathbf{z}^a(\tilde{f})  = \mathbb{E} \left[\mathbf{z}(\tilde{f}) \mathbf{z}^H(\tilde{f}+a) \right]$.
Once the cyclic spectrum $S_x^{\alpha}(f)$ is recovered, the transmissions' carriers and bandwidth may be estimated by locating its peaks.
Since the cyclic spectrum of stationary noise $n(t)$ is zero for $\alpha \neq 0$, cyclostationary detection is more robust to noise than stationary detection.

The alternative definition of the cyclic spectrum (\ref{eq:scf2}), presented in the cyclostationary box, implies that the elements in the matrix $\mathbf {R}_\mathbf{x}^a(\tilde{f})$ are equal to $S_x^{\alpha}(f)$ at the corresponding $\alpha$ and $f$.
Indeed, it can easily be shown~\cite{cohen_cyclo} that
\begin{equation}
\label{eq:mapping}
\mathbf{R}_\mathbf{x}^a{(\tilde{f})}_{(i,j)} = S_x^{\alpha}(f),   \\
\end{equation}
for
\begin{eqnarray}
\label{eq:mapping_param}
\alpha&=&(j-i)f_s+a \nonumber \\ f&=&-\frac{f_{\text{Nyq}}}{2} + \tilde{f}- \frac{f_s}{2}+ \frac{(j+i)f_s}{2}+\frac{a}{2}.
\end{eqnarray}
Here $\mathbf{R}_\mathbf{x}^a{(\tilde{f})}_{(i,j)}$ denotes the $(i,j)$th element of $\mathbf{R}_\mathbf{x}^a(\tilde{f})$.
This means that each entry of the cyclic spectrum $S_x^{\alpha}(f)$ can be mapped to an element from one of the correlation matrices $\mathbf{R}_{\mathbf{x}}^a(\tilde{f})$, and vice versa.
From (\ref{eq:multico}) and similarly to (\ref{eq:autoco_analog}) in the context of power spectrum recovery, we relate the shifted correlation matrices of $\mathbf{x}(f)$ and $\mathbf{z}(f)$ as
\begin{equation}
\mathbf{R}_\mathbf{z}^a(\tilde{f}) = \mathbf{A} \mathbf{R}_\mathbf{x}^a(\tilde{f}) \mathbf{A}^H, \quad \tilde{f} \in \left[0, f_s-a \right],
\label{eq:autoco2}
\end{equation}
for all $a \in [0, f_s]$.

Recall that, in the context of stationary signals, $\mathbf{R_x}(f)$ is diagonal. Here, understanding the structure of $\mathbf{R}_\mathbf{x}^a(\tilde{f})$ is more involved. In~\cite{cohen_cyclo}, it is shown that $\mathbf{R}_\mathbf{x}^a(\tilde{f})$ contains non zero elements on its $0$, $1$ and $-1$ diagonals and anti-diagonals. Besides the non zero entries being contained only in the three main and anti-diagonals, additional structure is exhibited, limiting to two the number of non zero elements per row and column of the matrix $\mathbf{R}_\mathbf{x}^a(\tilde{f})$. The above pattern follows from two main considerations. First, each frequency component, namely each entry of $\mathbf{x}(f)$, is correlated to at most two frequencies from the shifted vector of slices $\mathbf{x}(\tilde{f}+a)$, one from the same frequency band and one from the symmetric band. Second, the correlated component can be either in the same/symmetric slice or in one of the adjacent slices.

\begin{figure}
  \begin{center}
    \includegraphics[width=0.9\columnwidth]{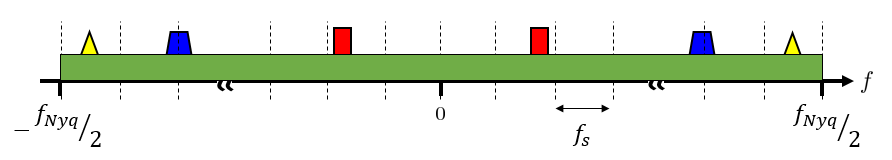}
    \caption{Original spectrum $X(f)$. The cyclostationary transmissions are shown in yellow, blue and red, buried in stationary noise in green.}
    \label{fig:schema_x}
  \end{center}
\end{figure}

\begin{figure}
  \begin{center}
    \includegraphics[width=0.85\columnwidth]{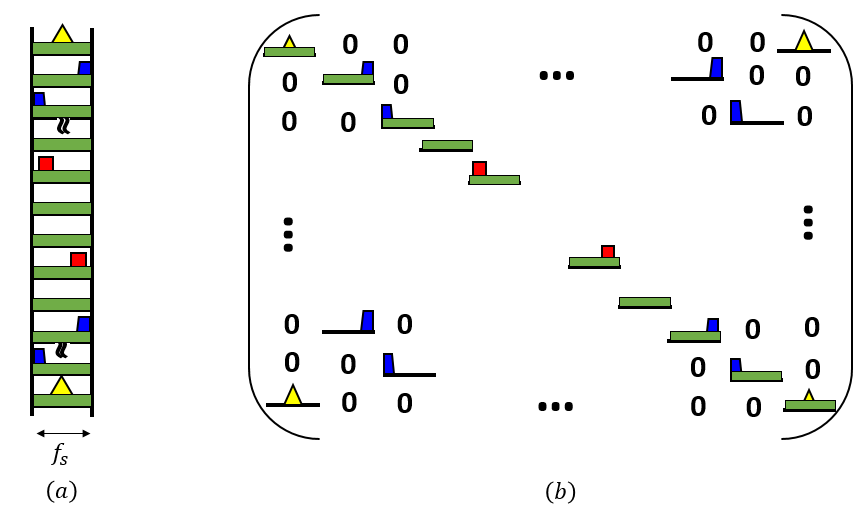}
    \caption{(a) Spectrum slices vector $\mathbf{x}(\tilde{f})$, (b) correlated slices of $\mathbf{x}(\tilde{f})$ in the matrix $\mathbf{R}_\mathbf{x}^0(\tilde{f})$.}
    \label{fig:schema_0}
  \end{center}
\end{figure}

\begin{figure}
  \begin{center}
    \includegraphics[width=0.85\columnwidth]{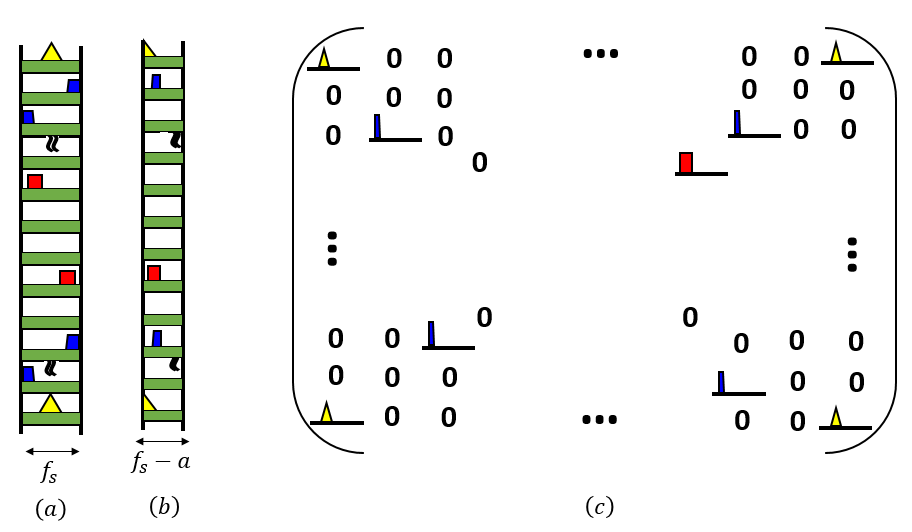}
    \caption{(a) Spectrum slices vector $\mathbf{x}(\tilde{f})$, (b) spectrum slices shifted vector $\mathbf{x}(\tilde{f}+a)$, for $a=f_s/2$, (c) correlated slices of $\mathbf{x}(\tilde{f})$ and $\mathbf{x}(\tilde{f}+a)$ in the matrix $\mathbf{R}_\mathbf{x}^a(\tilde{f})$, with $a =f_s/2$.}
    \label{fig:schema_a}
  \end{center}
\end{figure}

Figures~\ref{fig:schema_0} and~\ref{fig:schema_a} illustrate these correlations for $a=0$ and $a=f_s/2$, respectively.
First, in Fig.~\ref{fig:schema_x}, a sketch of the spectrum of $x(t)$, namely $X(f)$, is presented for the case of a sparse signal buried in stationary noise.
It can be seen that frequency bands of $X(f)$ may either appear in one $f_p$-slice or split between two slices at most since $f_p \geq B$.
The resulting vector of spectrum slices $\mathbf{x}(f)$ and the correlations between these slices without any shift, namely $\mathbf{R}_\mathbf{x}^0(\tilde{f})$, are shown in Fig.~\ref{fig:schema_0}(a) and (b), respectively.
In Fig.~\ref{fig:schema_0}(b), we observe that self-correlations appear only on the main diagonal since every frequency component is correlated with itself.
In particular, the main diagonal contains the noise's power spectrum (in green).
Cross-correlations between the yellow symmetric triangles appear in the $0$-anti diagonal, whereas those of the blue trapezes are contained in the $-1$ and $+1$ anti diagonals.
The red rectangles do not contribute any cross-correlations for $a=0$.

Figures~\ref{fig:schema_a}(a) and (b) show the vector $x(\tilde{f})$ and its shifted version $x(\tilde{f}+a)$ for $a=f_s/2$, respectively. The resulting correlation matrix $\mathbf{R}_\mathbf{x}^a(\tilde{f})$ appears in Figure~\ref{fig:schema_a}(c). Here, the self correlations of the yellow triangle appear in the main diagonal and that of the blue trapeze in the $-1$ diagonal. The cross-correlations all appear in the anti-diagonal, for the shift $a=f_s/2$. Note that since the noise is assumed to be wide-sense stationary, from (\ref{eq:papou}), a noise frequency component is correlated only with itself. Thus, $n(t)$ only contributes non-zero elements on the diagonal of $\mathbf{R}_\mathbf{x}^0(\tilde{f})$.

To recover $\mathbf{R}_\mathbf{x}^a(\tilde{f})$ from $\mathbf{R}_\mathbf{z}^a(\tilde{f})$, structured CS techniques are used in~\cite{cohen_cyclo} that aim at reconstructing a sparse matrix while taking into account its specific structure, as described above.
Once the cyclic spectrum is reconstructed, the number of transmissions and their respective carrier frequencies and bandwidths are estimated, as discussed in the next section. The detection performed on the cyclic spectrum is more robust to stationary noise than power spectrum based detection, at the expense of a slightly higher sampling rate, as shown in~\cite{cohen_cyclo}.
More precisely, in the presence of stationary noise, the cyclic spectrum may be reconstructed from samples obtained at $4/5$ of the Nyquist rate, without any sparsity assumption on the signal.
If the signal of interest is sparse, then the minimal sampling rate is further reduced to $8/5$ of the Landau rate~\cite{cohen_cyclo}.

\subsubsection{Carrier frequency and bandwidth estimation}

Many detection and classification algorithms based on cyclostationarity have been proposed (see reviews~\cite{Gardner_review, Napo_review}). We first survey several detection and classification approaches and then explain why they do not adhere to CR requirements. To assess the presence or absence of a signal, a first group of techniques requires a priori knowledge of its parameters and particularly of the carrier frequency~\cite{aparna, mismatch, cyclo_stats, cyclo_statsg, cyclo_symb, cyclo_class}, which is the information that CRs needs to determine in the first place. A second strategy focuses on a single transmission~\cite{cyclo_stats, cyclo_joint}, which does not fit the multiband model. Alternative approaches apply machine learning tools to the modulation classification of a single signal with unknown carrier frequency and symbol rate~\cite{peaks_table, cyclo_nn, cyclo_hmm, cyclo_svm}. Besides being only suitable for a single transmission, these methods require a training phase. Thus, these techniques can only cope with PUs whose modulation type and parameters were part of the training set.

\begin{figure}[tb]
  \begin{center}
    \includegraphics[width=1\columnwidth]{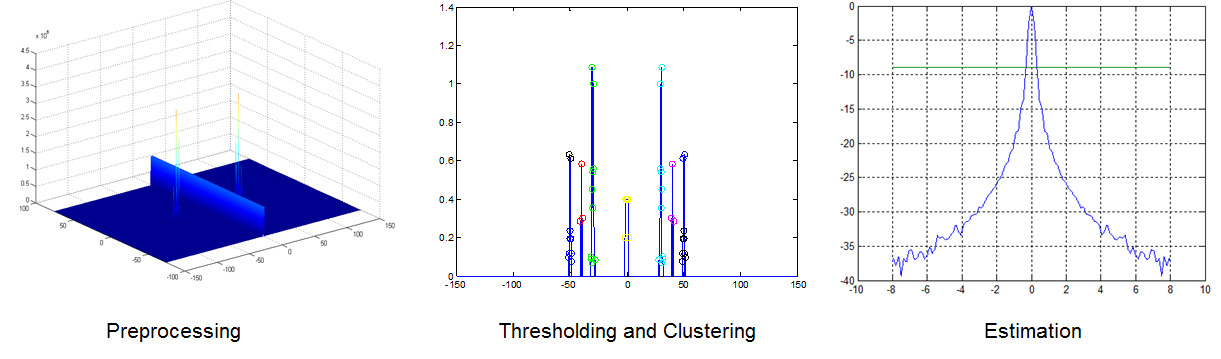}
    \caption{Carrier frequency and bandwidth estimation from the cyclic spectrum: preprocessing (left), thresholding and clustering (middle), parameter estimation (right).}
    \label{fig:cyclo_est}
  \end{center}
\end{figure}

For CR purposes, we seek a detector designed to comply with the following requirements:
(1) carrier frequency and bandwidth estimation rather than simple detection of the presence or absence of a signal;
(2) blind detection, namely without knowledge of the carrier frequencies, bandwidths and symbol rates of the transmissions;
(3) simultaneous detection of several transmissions;
(4) preferably a non-learning approach, i.e.\ with no training phase.
The parameter estimation algorithm, presented in~\cite{liad_cyclo}, is a simple parameter extraction method from the cyclic spectrum of multiband signals, that satisfies these requirements. It allows the estimation of several carriers and bandwidths simultaneously, as well as the number of transmissions, namely half the number of occupied bands $K/2$ for real-valued signals. The proposed parameter estimation algorithm can be decomposed into four main steps: preprocessing, thresholding, clustering and parameter estimation, as illustrated in Fig.~\ref{fig:cyclo_est}.

The preprocessing simply aims at compensating for the presence of stationary noise in the cyclic spectrum at the cyclic frequency $\alpha=0$, by attenuating the energy of the cyclic spectrum at this frequency. Thresholding is then applied to the resulting cyclic spectrum in order to find its peaks. The locations and values of the selected peaks are then clustered using k-means to find the corresponding cyclic feature, after estimating the number of clusters by applying the elbow method~\cite{elbow}. It follows that, apart from the cluster present in DC, the number of real signals, namely $N_{\text{sig}}=K/2$, is equal to half the number of clusters. Next, the carrier frequency $f_i$, which corresponds to the highest peak~\cite{GardnerBook}, is estimated for each transmission. The bandwidth $B_i$ is found by locating the edge of the support of the angular frequencies at the corresponding cyclic frequency $\alpha_i = 2f_i$.

Results presented in~\cite{cohen_cyclo} demonstrate that cyclostationary based detection as described in this section outperforms energy detection carried on the signal's spectrum or power spectrum, at the expense of increased complexity. We now show similar results obtained from hardware simulations, performed using the prototype from Fig.~\ref{fig:Prototype}.

\subsection{Hardware Simulations: Robustness to Noise}
\begin{figure*}
		\includegraphics[width=1\textwidth]{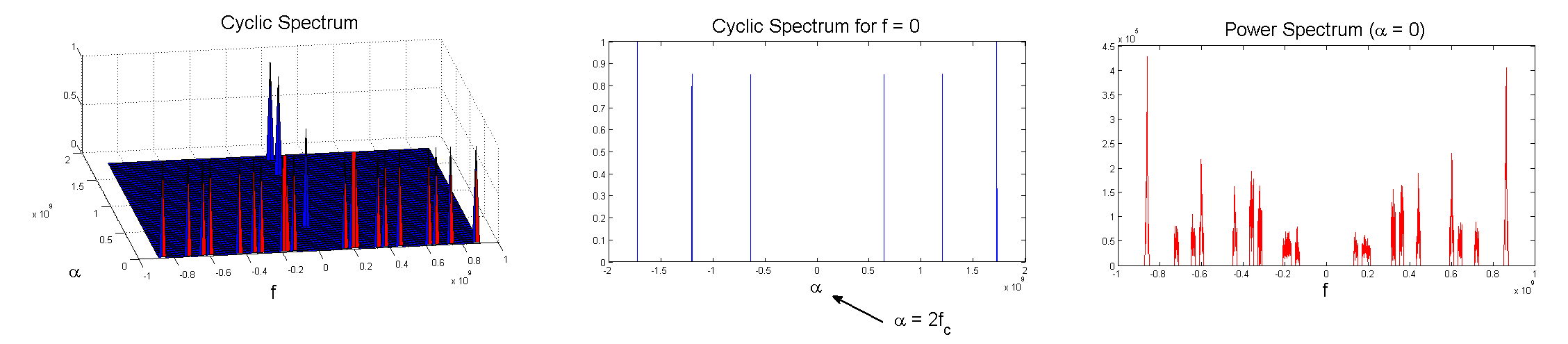}
		\caption{Screen shot from the MWC with cyclostationary detection. The input signal is composed of $N_{\text{sig}}=3$ transmissions (or $K=6$ bands) with carriers $f_1=320$MHz, $f_2=600$MHz and $f_3=860$MHz. Recovered cyclic spectrum from low rate samples (left), cyclic spectrum profile at the angular frequency $f=0$, where the cyclic peaks are clearly visible at twice the carrier frequencies (middle), power spectrum recovery failing in the presence of noise (right).}
		\label{fig:Cyclo1}
\end{figure*}

Cyclostationary detection has been implemented in the MWC CR prototype.
The analog front-end is identical to that of the original prototype and only the digital recovery part is modified since the cyclic spectrum is recovered directly from the MWC low rate samples.
Preliminary testing suggests that sensing success is still achievable at SNRs lower by $10\,\text{dB}$ than those allowed by energy detection performed on the recovered spectrum or power spectrum.
Representative results shown in Fig.~\ref{fig:Cyclo1} demonstrate the advantage of cyclostationary detection over energy detection in the presence of noise.
The figure presents the reconstructed cyclic spectrum from samples of the MWC prototype, as well as cross-sections at $f=0$ and $\alpha=0$, which corresponds to the power spectrum.
This increased robustness to noise comes at the expense of more involved digital processing on the low rate samples. The additional complexity stems from the higher dimensionality involved, since we reconstruct the $2$-dimensional cyclic spectrum rather than the $1$-dimensional (power) spectrum.

\begin{figure*}[t]\fboxsep1em
	\colorbox{BoxBackground}{\begin{minipage}{1\textwidth}\begin{multicols*}{2}

\section*{Channel Fading and Shadowing}

Analog signals transmitted over physical channels are affected by two main phenomena: Rayleigh fading, or small-scale fading, and log-normal shadowing, or large-scale fading~\cite{collaborative_brodersen, collaborative_ghasemi, fading_sklar}. The received signal is generally described in terms of the transmitted signal $s_i(t)$ convolved with the impulse response of the channel $h_{ij}(t)$, namely
\begin{equation}
\label{eq:gen}
r_{ij}(t) = s_i(t) \ast h_{ij}(t),
\end{equation}
where $r_{ij}(t)$ is the received signal corresponding to the $i$th transmission at the $j$th CR and $\ast$ denotes convolution. Fading and shadowing affect the channel response $h_{ij}(t)$.

\subsubsection*{Rayleigh fading}
For most practical channels, the free-space propagation model, which only accounts for path loss, is inadequate to describe the channel. A signal typically travels from transmitter to receiver over multiple reflective paths, which is traditionally modeled as Rayleigh fading. This implies that the amplitude and phase of the channel response $h_{ij}(t)=R(t)e^{j \phi(t)}$ are stochastically independent and identically distributed processes. The amplitude $R(t)$, for $t \in \mathbb{R}$, follows the Rayleigh distribution, given by
\begin{equation}
\label{eq:fad}
p_R(r) = \left \{ \begin{array}{ll} \frac{r}{\sigma^2} e^{-r^2/2\sigma^2} & r \geq 0 \\
0 & \text{otherwise},
\end{array} \right.
\end{equation}
where $2\sigma^2$ is the mean power~\cite{fading_sklar}.
The phase $\phi(t)$, for $t \in \mathbb{R}$, is uniformly distributed over the interval $[0, 2\pi)$.

\subsubsection*{Log-normal shadowing}
Large-scale fading represents the average signal power attenuation or path loss due to motion over large areas.
The resulting channel frequency response is therefore a constant.
This phenomenon is affected by prominent terrain contours between the transmitter and receiver.
Empirical measurements suggest that this type of fading, or shadowing, follows a normal distribution in dB units~\cite{shadowing}, or alternatively, the linear channel gain may be modeled as a log-normal random variable~\cite{collaborative_ghasemi}.
Therefore, the path loss (PL) measured in dB is expressed as
\begin{equation}
PL = PL_0+10 \gamma \log \frac{d}{d_0} + X_{\sigma}.
\end{equation}
Here, the reference distance $d_0$ corresponds to a point located in the far field of the antenna (typically 1 km for large cells).
The path loss to the reference point $PL_0$ is usually found through field measurements or calculated using free-space path loss.
The value of the path loss exponent $\gamma$ depends on the frequency, antenna heights, and propagation environment.
Finally, $X_{\sigma}$ denotes a Gaussian random variable (in dB) with variance $\sigma^2$ determined heuristically as well~\cite{fading_sklar}.
The shadowed received signal is thus given by
\begin{equation}
\label{eq:shad}
r_{ij}(t) = 10^{-PL_{ij}/20} \cdot s_i(t),
\end{equation}
where $PL_{ij}$ denotes the path loss between the $i$th transmitter and the $j$th receiver and the channel response $h_{ij}(t)=10^{-PL_{ij}/20} \delta(t)$.

\end{multicols*}\end{minipage}}\end{figure*}

\section{Collaborative Spectrum Sensing}

\subsection{Collaborative Model}

Until now, we assumed direct observation of the spectrum.
In practice, the task of spectrum sensing for CR is further complicated due to physical channel effects such as path loss, fading and shadowing~\cite{collaborative_balak, collaborative_brodersen, collaborative_ghasemi}, as described in the channel box.
To overcome these practical issues, collaborative CR networks have been considered, where different users share their sensing results and cooperatively decide on the licensed spectrum occupancy.



The different collaborative approaches can be distinguished according to several criteria~\cite{collaborative_balak}. First, cooperation can be either centralized or distributed. In centralized settings, the data is sent to a fusion center which combines the shared data to jointly estimate the spectrum or determine its occupancy. In the distributed approach, the CRs communicate among themselves and iteratively converge to a common estimate or decision. While centralized cooperation does not require iterations and can reach the optimal estimate based on the shared data, convergence to this estimate is not always guaranteed in its distributed counterpart. On the other hand, the latter is less power hungry and more robust to node and link failure, increasing the network survivability. An additional criterion concerns the shared data type; the CRs may share local binary decisions on the spectrum occupation (hard decision) or a portion of their samples (soft decision).

We consider the following collaborative model. A network of $N_{\text{rec}}$ CRs receives the $N_{\text{sig}}$ transmissions, such that the received signal at the $j$th CR is given by
\begin{equation}
x^{(j)}(t) = \sum_{i=1}^{N_{\text{sig}}} r_{ij}(t) = \sum_{i=1}^{N_{\text{sig}}} s_i(t) \ast h_{ij}(t).
\end{equation}
The channel response $h_{ij}(t)$ is determined by fading and shadowing effects.
Typical models are Rayleigh fading, or small-scale fading, and log-normal shadowing, or large-scale fading~\cite{collaborative_brodersen, collaborative_ghasemi, fading_sklar}, as described in ``Channel Fading and Shadowing''.
In the frequency domain, the Fourier transform of the $j$th received signal is given by
\begin{equation}
X^{(j)}(f)= \sum_{i=1}^{N_{\text{sig}}} S_i(f) H_{ij}(f).
\end{equation}
Therefore, the support of $x^{(j)}(t)$ is included in the support of the original signal $x(t)$.
Since the transmissions are affected differently by fading and shadowing from each transmitter to each CR, we can assume that the union of their respective supports is equivalent to the frequency support of $x(t)$.
The goal here is to assess the support of the transmitted signal $x(t)$ from sub-Nyquist samples of the received $x^{(j)}(t), 1 \leq j \leq N_{\text{rec}}$, by exploiting their joint frequency sparsity.

A simple and naive approach is to perform support recovery at each CR from its low rate samples and combine the local binary decisions, either in a fusion center for centralized collaboration or in a distributed manner.
In this hard decision strategy, the combination can be performed using several fusion rules such as AND, OR or majority rule.
Although this method is attractive due to its simplicity and low communication overhead, it typically achieves lower performance than its soft decision counterpart.
To mitigate the communication overhead, soft decision based methods may rely on sharing observations based on the low rate samples, rather than the samples themselves. In the next section, we review such techniques both in centralized and distributed contexts.

\subsection{Centralized Collaborative Support Recovery}

One approach~\cite{collaborative_leus3, collaborative_leus2} to centralized spectrum sensing considers a digital model based upon a linear relation between the $M$ sub-Nyquist samples $\mathbf{z}^{(j)}$ at CR $j$ and $N$ Nyquist samples $\mathbf{x}^{(j)}$ obtained for a given sensing time frame, namely
\begin{equation}
\label{eq:dig_eq}
\mathbf{z}^{(j)} = \mathbf{Ax}^{(j)},
\end{equation}
where $\bf A$ is the sampling matrix.
This model has been extensively studied by Leus et al.~\cite{collaborative_leus3, collaborative_leus1, collaborative_leus2}.
In all these works, the channel state information (CSI) is assumed to be known and the joint power spectrum of $\mathbf{x}$ is reconstructed, where $\bf x$ denotes the $N \times 1$ vector of Nyquist samples.
CSI is traditionally unknown by the CRs and should be estimated prior to detection to enable the use of this method.
The autocorrelation of the Nyquist samples is first related to that of the sub-Nyquist samples.
Then, in~\cite{collaborative_leus3}, the common sparsity of $\mathbf{r_x}^{(j)}$ is exploited in the frequency domain across all CRs to jointly reconstruct them at the fusion center, using a modified simultaneous orthogonal matching pursuit (SOMP)~\cite{CSBook} algorithm.
In~\cite{collaborative_leus1, collaborative_leus2}, besides exploiting joint sparsity, cross-correlations between measurements from different CRs are used, namely $\mathbf{R}_{\mathbf{z}_j \mathbf{z}_k}$, where $j$ and $k$ are the indices of two CRs.
These cross-correlations are related to the common power spectrum $\mathbf{s_x=Fr_x}$.
It is shown that if the total number of samples $N_{\text{rec}}M$ is greater than $N$ and these are suitably chosen to account for enough measurement diversity, then the power spectrum $\bf s_x$ of a non sparse signal can be recovered from compressed samples from a sufficient number of CRs. This shows that the number of receivers may be traded for the number of samples per CR.\@
However, increasing the number of samples per CR does not increase spatial diversity, as does increasing the number of receivers.

An alternative approach~\cite{coll_cent} relies on the analog model from (\ref{eq:multico}) and does not assume any a priori knowledge on the CSI.\@
This method considers collaborative spectrum sensing from samples acquired via multicoset sampling or the MWC at each CR.\@
In this approach, the $j$th CR shares its observation matrix $\mathbf{V}^{(j)}$, as defined in (\ref{eq:CTF}), rather than the sub-Nyquist samples themselves, and its measurement matrix $\mathbf{A}^{(j)}$, with a fusion center.
The sampling matrices are considered to be different from one another in order to allow for more measurement diversity. However, the same known matrix can be used to reduce the communication overhead. The underlying matrices $\mathbf{U}^{(j)}$ are jointly sparse since fading (\ref{eq:gen}), (\ref{eq:fad}) and shadowing (\ref{eq:shad}) do not affect the original signal's support.
Capitalizing on the joint support of $\mathbf{U}^{(j)}$, the support of the transmitted signal $x(t)$ is recovered at the fusion center by solving
\begin{eqnarray}
 \arg \min_{\mathbf{U}^{(j)}} & \bigcup_{i=1}^{N_{\text{rec}}} ||\mathbf{U}^{(j)}||_0 \\  \text{s.t. } & \mathbf{V}^{(j)} = \mathbf{A}^{(j)} \mathbf{U}^{(j)}, \text{ for all } 1 \leq j \leq N_{\text{rec}}. \nonumber
\end{eqnarray}

To recover the joint support of $\mathbf{U}^{(j)}$ from the observation matrices $\mathbf{V}^{(j)}$, both OMP and IHT, described in ``Compressed Sensing Recovery'', are extended to the collaborative setting~\cite{coll_cent}. Previously we considered support recovery from an individual CR, which boils down to an MMV system of equations (\ref{eq:CTF}).
CS algorithms have been extended to this case, such as SOMP from~\cite{tropp_somp} and simultaneous IHT (SIHT) presented in~\cite{valaee_iht}.
These now need to account for the joint sparsity across the CRs.

The distributed CS-SOMP (DCS-SOMP) algorithm~\cite{duarte_somp}, which extends the original SOMP to allow for different sampling matrices $\mathbf{A}^{(j)}$ for each receiver, is adapted to the CR collaborative setting in~\cite{coll_cent}.
The main modification appears in the computation of the index that accounts for the greatest amount of residual energy.
Here, the selected index is the one that maximizes the sum of residual projections over all the receivers. Once the shared support is updated, the residual matrices can be computed for each CR separately.
The resulting modified algorithm is referred to as block sparse OMP (BSOMP)~\cite{coll_cent}. Next, the block sparse IHT (BSIHT) algorithm~\cite{coll_cent} extends SIHT by selecting the indices of the common support though averaging over all the estimated $\mathbf{U}^{(j)}$ in each iteration.
Once the support is selected, the update calculations are performed separately for each receiver.
Both methods are suitable for centralized cooperation, in the presence of a fusion center.
As in the previous approach, if the CSI is known, then the sampling rate per CR can be reduced by a factor of $N_{\text{rec}}$ with respect to the rate required from an individual CR.\@



\subsection{Distributed Collaborative Support Recovery}

In the distributed approach, there is no fusion center and the CRs can only communicate with their neighbors.
Both the digital and analog centralized approaches have been extended to the distributed settings. In~\cite{collaborative_tian1, collaborative_tian2}, a digital model (\ref{eq:dig_eq}) is used where the spectrum is divided into $N$ known slots. Both unknown and known CSI cases are considered. In the first case, each CR computes its local binary decision for each spectral band by recovering the sparse spectrum using CS techniques. Then, an average consensus approach is adopted, with respect to the shared hard decision. If the CSI is known, similarly to~\cite{collaborative_leus1, collaborative_leus2}, then the spectrum is recovered in a distributed fashion. In~\cite{collaborative_tian1}, the proposed algorithm iterates through the following steps: local spectrum reconstruction given the support and consensus averaging on the support. In~\cite{collaborative_tian2}, a distributed augmented Lagrangian is adopted.

The centralized approach, based on the analog model (\ref{eq:multico}), presented in~\cite{coll_cent}, is modified in~\cite{coll_dist} to comply with distributed settings. The $i$th CR contacts a random neighbor $j$, chosen with some probability $P_{ij}$, according to the Metropolis-Hastings scheme for random transition probabilities,
\begin{equation}
\label{eq:prob}
P_{ij} = \left \{ \begin{array}{ll} \min\{\frac{1}{d_{i}},\frac{1}{d_{j}}\} & (i,j)\in E, \\
\sum_{(i,k)\in E} \max\{0,\frac{1}{d_{i}}-\frac{1}{d_{k}}\} & i=j, \\ 0 & \text{otherwise}.
\end{array} \right. 
\end{equation}
Here $d_i$ denotes the cardinality of the neighbor set of the $i$th CR and $E$ is the set of communication links between CRs in the network.

A single vector, computed from the low rate samples (and that will be defined below for each recovery algorithm), is passed between the CR nodes in the network, rather than the samples themselves, effectively reducing communication overhead. When a CR receives this vector, it performs local computation to update both the shared vector and its own estimate of the signal support accordingly. Finally, the vector is sent to a neighbor CR, chosen according to the random walk with probability (\ref{eq:prob}). Two distributed algorithms are presented in~\cite{coll_dist}. The first, distributed one-step greedy algorithm (DOSGA), extends the OSGA from~\cite{duarte_somp} to distributed settings. The second method, referred to as randomized distributed IHT (RDSIHT), adapts the centralized BSIHT~\cite{coll_cent} to the distributed case.

To describe the DOSGA algorithm, we first present its centralized counterpart OSGA.\@
Each CR computes the $\ell_2$-norm of the projections of the observation matrix $\mathbf{V}^{(j)}$ onto the columns of the measurement matrix $\mathbf{A}^{(j)}$, stored in the vector $\mathbf{w}^{(j)}$ of size $N$.
The fusion center then averages over all receivers' vectors, such that
\begin{equation}
\hat{\mathbf{w}}= \frac{1}{N_{\text{rec}}}\sum_{j=1}^{N_{\text{rec}}} \mathbf{w}^{(j)},
\end{equation}
and retains the highest values of $\hat{\mathbf{w}}$ whose indices constitute the support of interest.
In the absence of a fusion center, finding this average is a standard distributed average consensus problem, also referred to as distributed averaging or distributed consensus.
DOSGA~\cite{coll_dist} then uses a randomized gossip algorithm~\cite{Boyd06randomizedgossip} for this purpose, with the Metropolis-Hastings transition probabilities.

Next, we turn to the RDSIHT algorithm, which adapts the centralized BSIHT algorithm~\cite{coll_cent} to the distributed scenario. The distributed approach from~\cite{coll_dist} was inspired by the randomized incremental subgradient method proposed in~\cite{johansson2009randomized}, and recent work on a stochastic version of IHT~\cite{Needell14}. A vector $\mathbf{w}$ of size $N$, that sums the $\ell_2$-norms of the rows of the estimates of $\mathbf{U}^{(j)}$ before thresholding, is shared in the network through random walk. The indices of its $k$ largest values correspond to the current estimated support. When the $i$th CR receives $\mathbf{w}$, it locally updates it by performing a gradient step using its own objective function that is then added to $\mathbf{w}$.  Next, it selects a neighbor $j$ to send the vector to with probability $P_{ij}$ (\ref{eq:prob}). The joint sparsity across the CRs is exploited by sharing one common vector $\bf w$ by the network. It is shown numerically in~\cite{coll_dist} that both DOSGA and RDSIHT converge to their centralized counterparts.

\subsection{Hardware Simulations: Collaborative vs. Individual Spectrum Sensing}

\begin{figure*}
	\begin{center}
		\includegraphics[width = 1\textwidth]{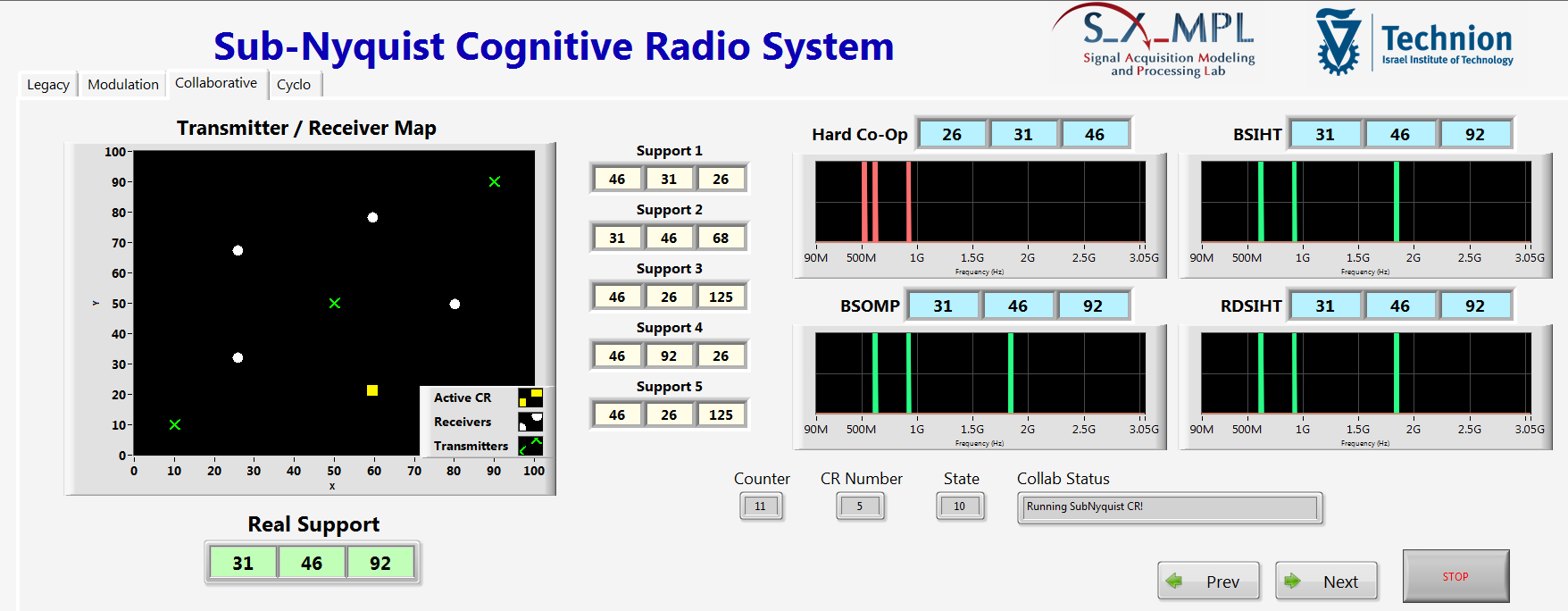}
	\end{center}
	\caption{Screen shot from the MWC CR collaborative hardware prototype.
    On the upper left side, we see the spatial map of the receivers in white, and transmitters in green.
    On the bottom left, the occupied band indices of the real spectral support are shown, while to the right of the transmitter/receiver map, the estimated indices by each CR individually are presented.
    On the right, we see the spectrum sensing results of 4 different algorithms: Hard Co-Op (hard decision collaboration that selects the most popular frequency band indices), BSIHT, BSOMP and RDSIHT.\@
    These results show both the superiority of collaborative spectrum sensing methods over individual detection and that of soft decision methods compared to the plain union of all CR results.}
	\label{fig:CollabSim}
\end{figure*}

We now confirm that the collaborative algorithms for spectrum sensing perform better than their individual counterparts by demonstrating a collaborative setting implemented using the MWC CR prototype, as can be seen in Fig.~\ref{fig:CollabSim}.
During the conducted experiments, $N_{\text{rec}}=5$ CR receivers, spread across different locations are emulated, denoted by white circles on the transmitter/receiver map.
The transmitters are also positioned in various locations depicted by green x-marks.
The transmitter positions and broadcasts mimic the effects of physical channel phenomena, i.e.\ fading and shadowing.
The frequency support recovered by each of the CRs is false, since they individually receive only a partial spectral image of their surroundings, as expected in a real-world scenario.

In all simulated scenarios, collaborative spectrum sensing outperforms detection realized by individual CRs.
This result is expected, since the soft collaborative methods take advantage of the spatial deployment of the receivers to reproduce the exact spectral map of the environment.
Moreover, the centralized and distributed algorithms BSOMP, BSIHT and RDSIHT, based on soft decisions, showed superior results in comparison with a hard decision method.
The same result can be seen in Fig.~\ref{fig:CollabSim}, where the hard decision support algorithm (Hard Co-Op) fails to recover the entire active frequency support (depicted by red bins).

\section{Joint Carrier Frequency and Direction Estimation}
\label{sec:doa}

Many signal processing applications, including CR, may require or at least benefit from joint spectrum sensing and DOA estimation. In order for CRs to map vacant bands more efficiently, spatial information about the PUs' locations can be of great interest. Consider the network of CRs presented in Fig.~\ref{fig:doa_model} and focus on CR1. Now, picture a scenario where PU2, with angle of arrival (AOA) $\theta_2$ with respect to CR1, is transmitting in a certain frequency band with carrier $f_2$. Assuming that CR2 does not receive PU2's transmission, CR1 could transmit in the same frequency band in the opposite direction of PU2 towards CR2. DOA estimation can thus enhance CR performance by allowing exploitation of vacant bands in space in addition to the frequency domain.

\subsection{Model and System Description}

\begin{figure}[ht]
\begin{center}
\includegraphics[width = 0.5\textwidth]{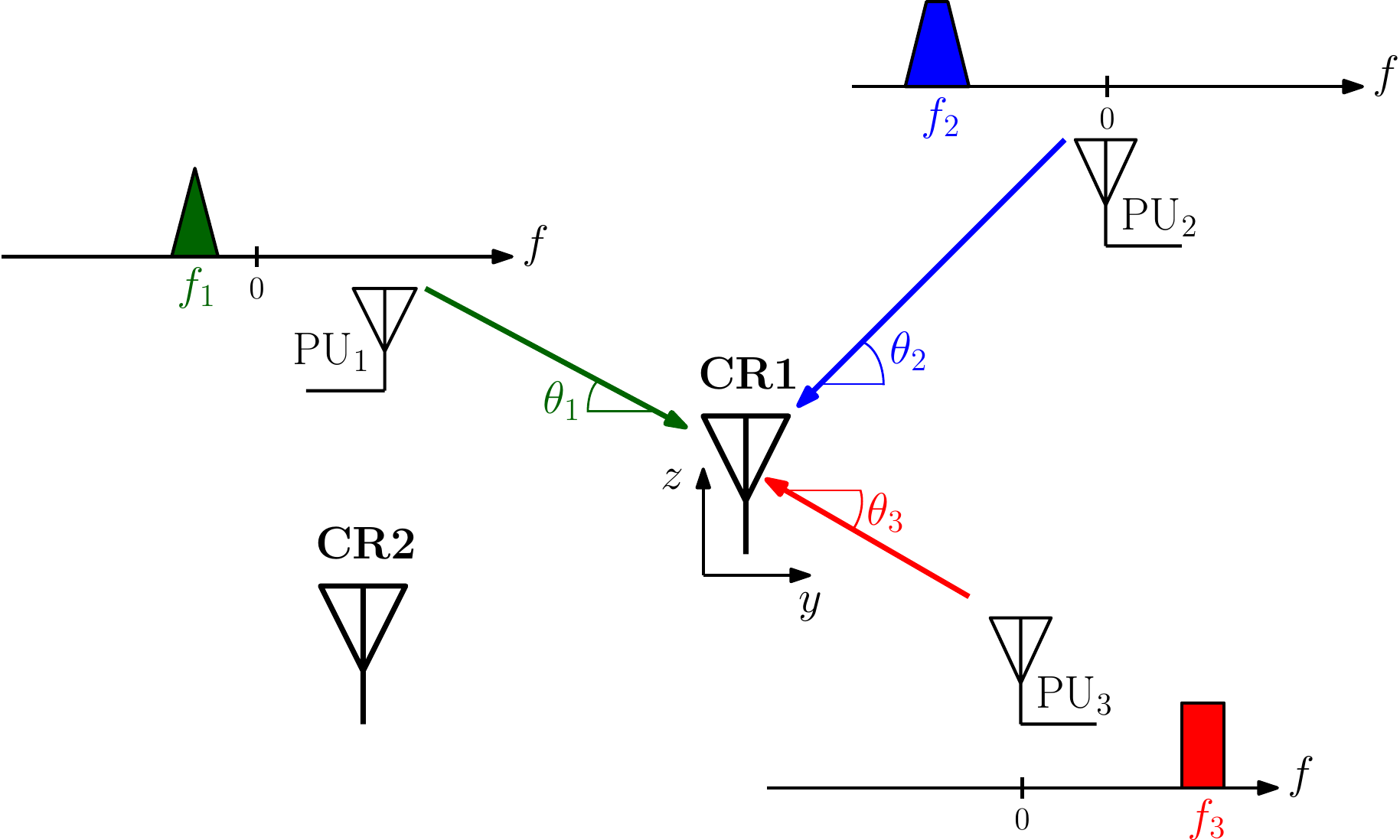}
\caption{Illustration of $N_{\text{sig}}=3$ source signals in the $yz$ plane. Each transmission is associated with a carrier
frequency $f_{i}$ and AOA $\theta_{i}$.}
\label{fig:doa_model}
\end{center}
\end{figure}

To formulate our problem mathematically, assume that the input signal $x(t)$ is composed of $N_{\text{sig}}$ source signals $s_{i}(t)$ with both unknown and different carrier frequencies $f_i$ and AOAs $\theta_i$. The main difference between this scenario and the one that has been discussed in the previous sections is the additional unknown AOAs $\theta_i$. Figure~\ref{fig:doa_model} illustrates this signal model. To recover the unknown AOAs, an array of sensors is required.
A similar problem thoroughly treated in the literature, is the 2D-DOA recovery problem, where two angles are traditionally recovered and paired. In our case, the second variable is the signal's carrier frequency instead of an additional angle.

\subsection{Multicoset Approach}
A few works have recently considered joint DOA and spectrum sensing of multiband signals from sub-Nyquist samples. In~\cite{leus_DOA} and~\cite{kumar_DOA}, low rate samples are obtained using the multicoset sampling scheme.
In~\cite{leus_DOA}, which considers the digital model (\ref{eq:dig_eq}), both time and spatial compression are applied by selecting samples from the Nyquist grid and receivers from a uniform linear array (ULA), such that
\begin{equation}
\mathbf{Z}[n]=\mathbf{C}_s \mathbf{X}[n] \mathbf{C}_t.
\end{equation}
Here, $\mathbf{X}[n]$ is the matrix of Nyquist samples from all receivers in the ULA and the selection matrices $\mathbf{C}_s$ and $\mathbf{C}_t$ operate on the spatial and time domain, respectively, to form the matrix of compressed samples $\mathbf{Z}[n]$. The 2D power spectrum matrix of the underlying signal is then reconstructed from the samples, where every row describes the power spectrum in the frequency domain for a given AOA and every column corresponds to the power spectrum information in the angular domain for a given frequency.

In~\cite{kumar_DOA}, an L-shaped array with two interleaved, or multicoset channels, with a fixed delay between the two, samples the signal below the Nyquist rate. Then, exploiting correlations between samples from the direct path and its delayed version, the frequencies and their corresponding AOAs are estimated using MUSIC~\cite{pisarenko73, schmidt86}.
However, the pairing issue between the two, that is matching each frequency with its corresponding angle, is not discussed.

In the next section, we describe the compressed carrier and DOA estimation (CaSCADE) system, presented in~\cite{cohen_doa}, that utilizes the sampling principles of the MWC.\@
This technique addresses the pairing issue and avoids the hardware issues involved in multicoset sampling.

\subsection{The CaSCADE System}

\begin{figure}
\begin{center}
\includegraphics[width = 0.4\textwidth]{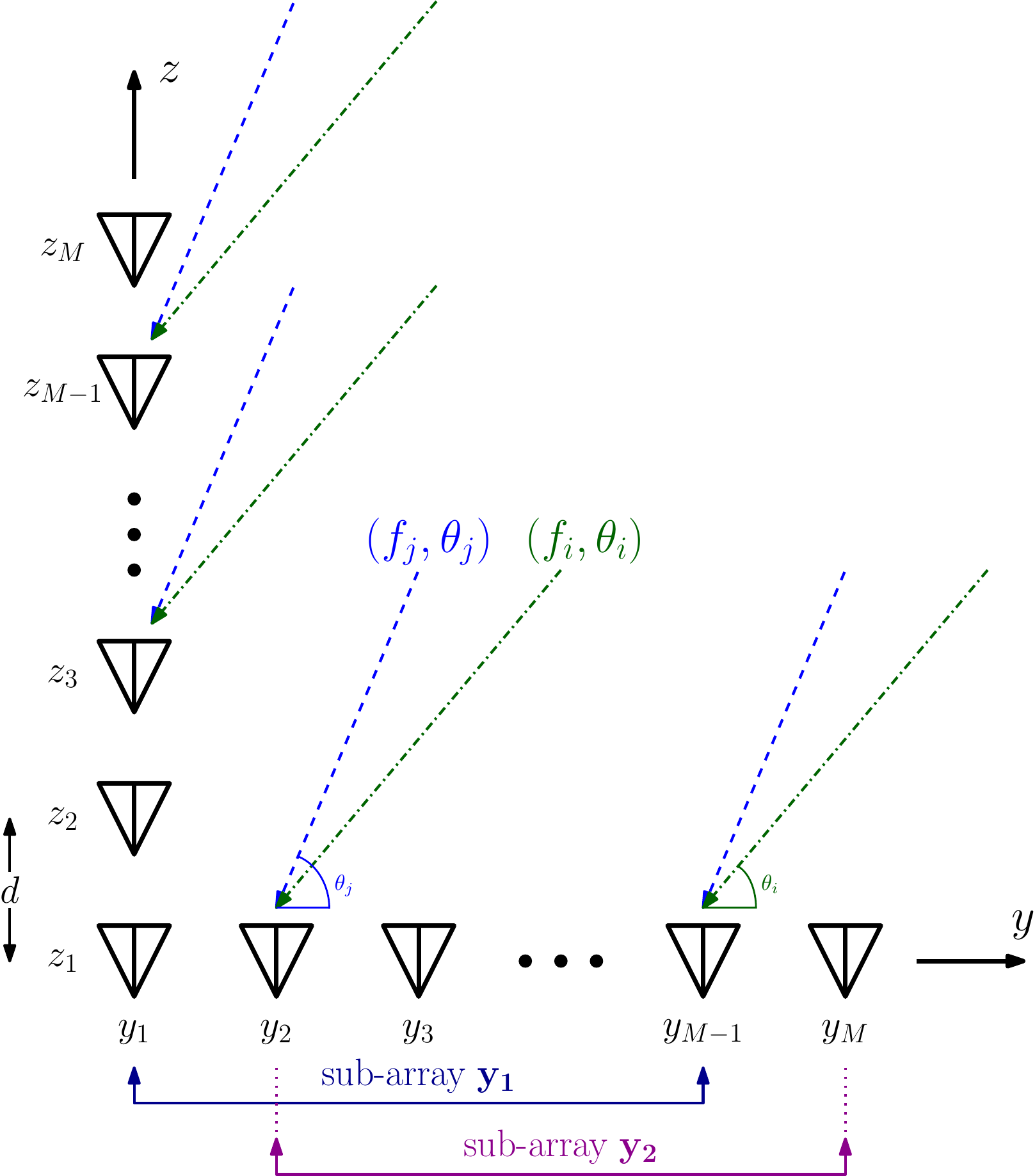}
\caption{CaSCADE system: L-shaped array with $M$ sensors along the $y$ axis and $M$ sensors along the $z$ axis including a common sensor at the origin. The sub-arrays $\mathbf{y}_1$ and $\mathbf{y}_2$ (and similarly $\mathbf{z}_1$ and $\mathbf{z}_2$) are defined in the derivation of the 2D-ESPRIT algorithm.}
\label{fig:cascade}
\end{center}
\end{figure}

The CaSCADE system implements the modified, or ULA based, MWC over an L-shaped array with $2M-1$ sensors ($M$ sensors along the $y$ axis and $M$ sensors along the $z$ axis including a common sensor at the origin) in the $yz$ plane.
Each transmission $s_{i}\left(t\right)$ impinges on the array with its corresponding AOA $\theta_{i}$, as shown in Fig.~\ref{fig:cascade}.
The array sensors have the same sampling pattern as the alternative MWC.\@
Each sensor is composed of an analog mixing front-end, implementing one physical branch of the MWC, that includes a mixer, a LPF and a sampler.

By treating the L-shaped array as two orthogonal ULAs, one along the $y$ axis and the other along the $z$ axis,
two systems of equations similar to (\ref{eq:alternative_mwc}) can be derived.
For the ULA along the $y$ axis, we obtain
\begin{equation}
\label{eq:Xeq}
\mathbf{y}(f)=\mathbf{A}_{y} \mathbf{x}(f), \quad f \in \mathcal{F}_s,
\end{equation}
where
\begin{equation}
\mathbf{A}_{y} =
\left[\begin{matrix}e^{j2\pi f_{1}\tau_{1}^{y}(\theta_{1})} & \cdots & e^{j2\pi f_{K}\tau_{1}^{y}(\theta_{K})}\\
\vdots &  & \vdots\\
\\
e^{j2\pi f_{1}\tau_{M}^{y}(\theta_{1})} & \cdots & e^{j2\pi f_{K}\tau_{M}^{y}(\theta_{K})}
\end{matrix}\right],
\end{equation}
and $
\tau_{m}^{y}\left(\theta\right)=\frac{dm}{c}\cos\left(\theta\right)$
denote the delays at the $m$th sensors with respect to the first sensor. Along the $z$ axis, the samples $\mathbf{z}(f)$, sampling matrix $\mathbf{A}_z$ and delays $\tau_m^z$ are similarly defined.

Two joint recovery approaches for the carrier frequencies and DOAs of the transmissions are proposed in~\cite{cohen_doa}.
Note that once the carriers and AOAs are estimated, the signals can be reconstructed, as shown for the alternative MWC.\@
For the sake of simplicity, a statistical model where $x(t)$ is wide-sense stationary is considered.
The first recovery approach is based on CS techniques and allows recovery of both parameters assuming the electronic angles $f_{i}\cos\theta_{i}$ and $f_{i}\sin\theta_{i}$ lie on a predefined grid.
The CS problem is formulated in such a way that no pairing issue arises between the carrier frequencies and their corresponding DOAs.
To that end, the time-domain samples from both ULAs are concatenated into one vector $\mathbf{v}[n] = {\left[ \mathbf{y}^T[n] \,\mathbf{z}^T[n] \right]}^T$, whose correlation matrix,
\begin{equation}
\mathbf{R}=\mathbb{E}\left[\mathbf{v}[n] \mathbf{v}^{H}[n]\right]=\mathbf{A}\mathbf{R}_{x}\mathbf{A}^{H},\label{eq:rvsp}
\end{equation}
is computed. Here, $\mathbf{A}= {[\mathbf{A}_y^T \, \mathbf{A}_z^T]}^T$ and the autocorrelation matrix $\mathbf{R}_x = \mathbb{E}\left[\mathbf{x}[k] \mathbf{x}^{H}[k]\right]$ is sparse and diagonal, from the stationarity of $x(t)$.
From the grid assumption, (\ref{eq:rvsp}) can be discretized with respect to the possible values taken by the electronic angles. The resulting sparse matrix derived from $\mathbf{R}_{x}$ is diagonal as well, and its sparse diagonal can be recovered using traditional CS techniques, similarly to (\ref{eq:rzrx}).

The second recovery approach, inspired by~\cite{Jgu07, Jgu15}, extends the ESPRIT algorithm to the joint estimation of carriers and DOAs, while overcoming the pairing issue. The 2D-ESPRIT algorithm presented in~\cite{cohen_doa} is directly applied to the sub-Nyquist samples, by considering cross-correlation matrices between the sub-arrays of both axis. Dropping the time variable $n$ for clarity, the samples from the sub-arrays can be written as
\begin{eqnarray}
\mathbf{y}_{1} = \mathbf{A}_{y_{1}}\mathbf{x}, & \quad & \mathbf{y}_{2} =  \mathbf{A}_{y_{2}}\mathbf{x} \nonumber \\
\mathbf{z}_{1} = \mathbf{A}_{z_{1}}\mathbf{x}, & \quad & \mathbf{z}_{2} = \mathbf{A}_{z_{2}}\mathbf{x},
\end{eqnarray}
where $\mathbf{y}_{1}$, $\mathbf{y}_{2}$, $\mathbf{z}_{1}$, $\mathbf{z}_{2}$ are samples from the sub-arrays shown in Fig.~\ref{fig:cascade}. The matrices $\mathbf{A}_{y_{1}}$ and $\mathbf{A}_{y_{2}}$ are the first and last $M-1$ rows of $\mathbf{A}_{y}$, respectively and $\mathbf{A}_{z_1}$ and $\mathbf{A}_{z_2}$ are similarly defined.
Each couple of sub-array matrices along the same axis are related as
\begin{eqnarray}
\mathbf{A}_{y_{2}} & = & \mathbf{A}_{y_{1}}\mathbf{\Phi} \nonumber \\
\mathbf{A}_{z_{2}} & = & \mathbf{A}_{z_{1}}\mathbf{\Psi},
\end{eqnarray}
where
\begin{eqnarray}
\label{eq:phipsi}
\mathbf{\Phi} & \triangleq & \mbox{diag}\left[\begin{matrix}e^{j2\pi f_{1}\tau_{1}^{y}(\theta_{1})} & \cdots & e^{j2\pi f_{K}\tau_{1}^{y}(\theta_{K})}\end{matrix}\right] \nonumber \\
\mathbf{\Psi} & \triangleq & \mbox{diag}\left[\begin{matrix}e^{j2\pi f_{1}\tau_{1}^{z}(\theta_{1})} & \cdots & e^{j2\pi f_{K}\tau_{1}^{z}(\theta_{K})}\end{matrix}\right].
\end{eqnarray}
We see from (\ref{eq:phipsi}) that the carrier frequencies $f_i$ and AOAs $\theta_i$ are embedded in the diagonal matrices $\bf \Phi$ and $\bf \Psi$. Applying the ESPRIT framework on cross-correlations matrices between the subarrays of both axis, allows to jointly recover $\bf \Phi$ and $\bf \Psi$~\cite{cohen_doa}. This leads to proper pairing of the corresponding elements $f_{i}\tau_{1}^{y}(\theta_{i})$ and $f_{i}\tau_{1}^{z}(\theta_{i})$. The AOAs $\theta_i$ and carrier frequencies $f_i$ are then given by
\begin{equation}
\theta_{i} = \tan^{-1}\left(\frac{\angle\Psi_{ii}}{\angle\Phi_{ii}}\right), \quad
f_{i}= \frac{\angle\Phi_{ii}}{2\pi\frac{d}{c}\cos\left(\theta_{i}\right)}. \label{eq:DOA_solution}
\end{equation}
It is proven in~\cite{cohen_doa} that the minimal number of sensors required for perfect recovery is $2K+1$. This leads to a minimal sampling rate of $(2K+1)B$, which is slightly higher than the minimal rate $2KB$ required for spectrum sensing in the absence of DOA recovery. These ideas can be extended to jointly recovery the transmissions' carrier frequencies, azimuth and elevation angles in a 3D framework.

\section{Conclusion and Future Challenges}

In this paper, we reviewed several challenges imposed on the traditional task of spectrum sensing by the new application of CR.\@
We first investigated sub-Nyquist sampling schemes, enabling sampling and processing of wideband signals at low rate, by exploiting their a priori known structure.
A possible extension of these works is to include adaptive updating of the detected support, triggered by a change in a PU's activity, either starting a transmission in a previously vacant band or withdrawing from an active band.
To increase efficiency, this should be performed by taking the current detected support as a prior and updating it with respect to the new acquired samples, without going through the support recovery process from scratch.
Additional preliminary assumptions on the structure or statistical behavior of the potentially active signals, such as statistics on channel occupancy, can also be exploited.

We then considered detection challenges in the presence of noise, where second-order statistics recovery, and in particular cyclostationary detection, are shown to perform better than simple energy detection.
Next, fading and shadowing channel effects were overcome by collaborative CR networks.
Finally, we addressed the joint spectrum sensing and DOA estimation problem, allowing for better exploitation of frequency vacant bands by exploiting spatial sparsity as well.
All these methods should be combined in order to map the occupied spectrum, in frequency, time and space, thus maximizing the CR network's throughput.
This requires an adequate spectrum access protocol as well, that translates the data acquired by spectrum sensing into transmission opportunities for the CRs.
Spectrum access challenges and algorithms were outside the scope of this review.

An essential part of the approach adopted in this survey is the relation between the theoretical algorithms and practical implementation, demonstrating real-time spectrum sensing from low rate samples using off-the-shelf hardware components.
We believe that prototype development is key to enabling sub-Nyquist sampling as a solution to the task of spectrum sensing in CR platforms.
A natural next step in that direction is the implementation of a complete CR network, collaboratively performing joint spectrum sensing and DOA estimation followed by spectrum access.
This prototype should then be tested on real data in order to assess its true capabilities.
We believe that a sub-Nyquist digital to analog interface can alleviate many of the bottlenecks currently hindering the development of CR systems, allowing fast deployment of low rate, simple and efficient CR devices, using currently available hardware.

In order for CRs to become a viable solution to spectrum shortage, other main challenges need to be addressed, as discussed earlier.
The issue of coexistence with existing communication links from PUs is crucial and is very particular to the CR scenario.
Here, coexistence is not a symmetric requirement as CRs are prohibited from interfering with PUs.
Another challenge is the establishment of a communication channel for CRs to be able to exchange the locations of their current transmission bands.
Finally, the issue of security against attacks to the CR networks still has numerous unresolved questions.

\bibliographystyle{IEEEtran}
\bibliography{references/IEEEabrv,references/CR_ref}

\end{document}